\documentclass[pra,aps,twocolumn,showpacs,superscriptaddress,amsmath,amssymb,floatfix]{revtex4}
\usepackage{color}
\usepackage{graphicx}

\newcommand{\be}{\begin{equation}}
\newcommand{\ee}{\end{equation}}

\newcommand{\rr}{{\mathbf r}}
\newcommand{\kk}{{\mathbf k}}
\newcommand{\qq}{{\mathbf q}}
\newcommand{\pp}{{\mathbf p}}
\newcommand{\uu}{{\mathbf u}}
\newcommand{\eee}{\mathbf{e}}

\newcommand{\QQ}{{\mathbf Q}}
\newcommand{\RR}{{\mathbf R}}

\newcommand{\jj}{{\mathbf j}}
\newcommand{\AAA}{{\mathbf A}}

\newcommand{\Psihd}{{\hat \Psi}^\dagger}
\newcommand{\Psih}{{\hat \Psi}}
\newcommand{\phih}{{\hat \phi}}
\newcommand{\phihd}{{\hat \phi^\dagger}}
\newcommand{\chih}{{\hat \chi}}

\newcommand{\ket}[1]{|#1\rangle}
\newcommand{\eq}[1]{(\ref{eq:#1})}
\newcommand{\eqname}[1]{\label{eq:#1}}

\newcommand{\bea}{\begin{eqnarray}}
\newcommand{\eea}{\end{eqnarray}}

\begin{document}
\title{Non-equilibrium and local detection of the normal fraction of a trapped
two-dimensional Bose gas} 
\author{Iacopo Carusotto}
\affiliation{INO-CNR BEC Center and Dipartimento di Fisica, Universit\`a
di Trento, I-38123 Povo, Italy}
\author{Yvan Castin}
\affiliation{Laboratoire Kastler Brossel, \'Ecole normale sup\'erieure, UPMC and CNRS, 24
rue Lhomond, 75231 Paris Cedex 05, France}

\begin{abstract}
We propose a method to measure the normal fraction of a two-dimensional Bose gas, a quantity that generally differs from the non-condensed fraction. 
The idea is based on applying a spatially oscillating artificial gauge field to the atoms. 
The response of the atoms to the gauge field can be read out either mechanically from the deposited energy into the cloud, or optically from the macroscopic optical properties of the atomic gas. 
The local nature of the proposed scheme allows one to reconstruct the spatial profile of the superfluid component; furthermore, the proposed method does not require having established 
thermal equilibrium in the gas in the presence of the gauge field.
The theoretical description of the system is based on a generalization of the Dum-Olshanii theory of artificial gauge fields to the interacting
many-body context. The efficiency of the proposed measurement scheme is assessed by means of classical field numerical simulations. An explicit atomic level scheme minimizing disturbing effects such as spontaneous emission and light-shifts
is proposed for $^{87}\textrm{Rb}$ atoms. 
\end{abstract}
\maketitle

\section{Introduction and summary of main results}

One of the most striking features of degenerate Bose gases in two dimensions is the possibility of having a superfluid behavior in the absence of a macroscopically populated Bose-Einstein condensate. The transition to the superfluid state is of the Berezinskii-Kosterlitz-Thouless (BKT) type, characterized by a sudden jump of the superfluid density from $0$ to the universal value $4$ (in units of the inverse square of the de Broglie thermal wavelength), independent from the details of the system~\cite{BKT,Minnhagen}. At the transition point, the asymptotic behavior of the field correlation function changes from an exponential to a power-law decay at large distances. In contrast to the three-dimensional case, superfluidity is then not related to the appearance of a macroscopically occupied Bose-Einstein condensate in the thermodynamic limit.

Pioneering experiments have addressed the mechanical properties of
two-dimensional layers of liquid Helium adsorbed on a
substrate~\cite{2Dexp_helium} and have characterized the universal jump
of the superfluid fraction at the BKT critical point. On the other hand, liquid Helium experiments have limited
access to the momentum distribution and the correlation functions of the fluid. The situation of ultracold atom experiments is almost the opposite: evidence of the BKT transition has been obtained
from the coherence functions~\cite{Dalibard2D}, the number of observed
vortices~\cite{2Dexp_cornell}, and the density profile after time-of-flight~\cite{2Dexp_phillips}, while the macroscopic mechanical properties of the fluid have not been characterized yet.  

Quite some effort has been recently devoted to the conceptual problem of how to experimentally detect genuine superfluidity in a quantum gas of ultracold atoms 
 and not simply Bose-Einstein condensation~\cite{IC_physics}. 
A possibility explored in~\cite{CH} is to look at the response of a gas in a toroidal trap to a static azimuthal artificial gauge field: a spectroscopic signature is proposed which should provide direct information on the total superfluid mass of the system.  A different strategy proposed in~\cite{Ho} consists of looking at the evolution of the 
equilibrium density profile of a trapped gas when it is set into rotation.

In the present paper we propose two experimental protocols to measure the normal fraction of a gas in a {\em local} way, so to extract its spatial dependence in a trapped geometry. This feature is most relevant for atomic samples, as the superfluid core coexists with an external ring of normal gas~\cite{dalib_profile}. In particular, the proposed diagnostic technique does not require to relate experimental observations after time of flight to in-trap quantities. Furthermore, in contrast to~\cite{Ho} our technique does not require thermodynamic equilibrium in the gas in presence of rotation \cite{foot_theq} 
and may be applied to more general, non-equilibrium conditions.

The basic idea of our proposal is based on the definition of normal and superfluid fractions of a quantum fluid in terms of its current response to a transverse gauge field in the low-frequency and long-wavelength limit~\cite{PN,dalfovo}. A spatially oscillating artificial gauge field~\cite{olshanii,GerbierDalibard,spielman_th,spielman_exp} with a spatially localized envelope can be applied to the atomic gas using a suitable combination of laser beams. The response of the fluid to the gauge field can be detected either mechanically or optically. In the former case, one has to measure the amount of energy that is deposited in the atomic gas at the end of a suitable temporal sequence of gauge field. In the latter case, one can observe e.g. the phase shift that is experienced by the laser fields while crossing the atomic cloud.

The structure and the main results of the paper can be summarized as follows.
In Sec.\ref{sec:principle} we review the definition of the normal and superfluid fractions that we adopt throughout the whole paper. 
A strategy to generate the artificial gauge field with the suitable spatial geometry is presented in Sec.\ref{sec:gauge} using three laser beams, namely a coupling beam and two probe ones. 

The first method to measure the normal fraction is by mechanical means. It is discussed in Sec.\ref{sec:deposited}: a pulse of spatially modulated gauge field is suddenly applied to the gas and then slowly switched-off according to an exponential law in time. An analytical calculation within the linear response theory and local density approximation
shows that the energy that is deposited in the gas at the end of the gauge field pulse is indeed proportional to the normal (total) density 
in the small fraction of the gas where the probe beam is focused, if the spatial modulation of the gauge field is orthogonal (parallel) to the coupling beam direction. Some of the complications that naturally occur in experiments are then numerically investigated: a Bogoliubov theory is used to assess the conditions to be imposed to the geometry and the temporal duration of the pulse; a classical field model is used to assess the conditions on the amplitude of the gauge field for the linear response theory to be valid. A main difficulty appears to be the relatively small amount of energy that can be deposited in the gas before nonlinear couplings become difficult to extrapolate out: the resulting figure is on the order of 1\% of the total energy of the gas, which is however
not far from the sensitivity of
state-of-the-art thermodynamic measurements of the energy 
\cite{Salomon_eos_unit,Dalibard_eos_2D,Zwierlein_eos,Hannaford1,Hannaford2}.

The second method to measure the normal fraction is by optical means.
It is discussed in Sec.\ref{subsec:eftl}. The spatially modulated gauge field is imposed using the same laser beam configuration in a continuous wave regime to generate a stationary current pattern in the gas. This current pattern can be read out from the phase shift accumulated by the same probe laser beams after crossing the atomic cloud. Analytical calculations show that this phase shift is indeed proportional to the normal (total) density if the spatial modulation of the gauge field is orthogonal (parallel) to the coupling beam direction. For realistic configurations, the amount of the phase shift is anticipated to be of the order of a fraction of $10^{-3}$, i.e. small but still appreciable with present-day optical techniques. 

Another optical set-up that is able to provide quantitative information on the normal fraction is discussed in Sec.\ref{subsec:cfatadol}: analytical calculations are used to relate the angular distribution of the scattered light off a single probe beam to the normal fraction of the gas. The main advantage of this last configuration is the rapidity of the measurement but the limiting factor is the relatively small number of photons (of the order of a few units) that are expected to be scattered in the useful directions.

In Appendix \ref{app:experimental} we discuss how to cope with all those spurious effects that arise from a realistic configuration of atomic levels and laser fields, in particular spontaneous emission and the mechanical effect of undesired light-shifts; even though the underlying concepts are general, the discussion is mostly focused on the most promising case of $^{87}$Rb atoms. 
The generalization of the Dum-Olshanii theory of artificial gauge fields to the many-body context is reported in Appendix \ref{app:adiab}: this 
development is required to put the gauge field concepts on firm ground in presence of atomic interactions
and to evaluate in a rigorous way the optical response of the atoms to the combined coupling and probe beams.
Appendix \ref{appendix:LDA} gives more details on the analytical derivation of the deposited energy and clarifies some issues related to the local density approximation.
The framework for calculating the deposited energy using the Bogoliubov theory within the linear response regime is discussed in Appendix \ref{app:bogo}.
The last Appendix \ref{app:noise} discusses issues related to the statistical noise on the deposited energy both in the numerical calculation and in an actual experiment. In particular, it shows how useful information on the normal fraction could be extracted from the noise if a sufficiently precise determination of the initial energy was possible.
We conclude in section \ref{sec:conclu}.

\section{Definition of superfluid and normal fractions}
\label{sec:principle}

Our proposal to quantitatively assess the superfluidity of the two-dimensional atomic gas is based 
on the traditional definition of the normal fraction $f_n$ in terms of the response to a transverse gauge field coupling to the atomic current operator~\cite{PN,dalfovo}.
The Hamiltonian giving the coupling of the matter current to an arbitrary vector potential $\AAA(\rr)$ is
\begin{equation}
V=-\int\!d^2\rr\,\AAA(\rr)\cdot\jj(\rr)
\end{equation}
with the current operator defined as usual as
\begin{equation}
\jj(\rr)=\frac{\hbar}{2im}\left[\phihd(\rr)\nabla\phih(\rr) - \textrm{h.c.} \right]
\end{equation}
in terms of the bosonic field operator $\hat{\phi}$ for the two-dimensional gas.
For a spatially homogeneous system, the linear response susceptibility relating the average
current~\cite{foot_phys_cur}
to the applied gauge field can be easily written in momentum space and frequency domain as
\begin{equation}
\langle \jj \rangle (\qq,\omega) =\chi(\qq,\omega)\,\AAA(\qq,\omega).
\eqname{chi}
\end{equation}
If the system is also invariant under reflection with respect to the direction of $\qq$, the susceptibility tensor $\chi(\qq,\omega)$ turns out to be diagonal in the longitudinal/transverse basis with respect to $\qq$, with diagonal matrix elements $\chi_{L,T}(\qq,\omega)$, respectively.

For a system of surface density $\rho$, the normal fraction $f_n$ of the system is then defined as the low-momentum, low-frequency limit of the susceptibility to transverse gauge fields:
\begin{equation}
f_n=\lim_{q\rightarrow 0} \lim_{\omega\rightarrow 0} \frac{m}{\rho}\chi_{T}(\qq,\omega).
\eqname{f_n}
\end{equation}
Note that the order of the limits is here important. A well-known sum
rule based on gauge invariance imposes that the same limit for the
longitudinal susceptibility $\chi_L(\qq,\omega)$ gives exactly unity,
\begin{equation}
1=\lim_{q\rightarrow 0} \lim_{\omega\rightarrow 0} \frac{m}{\rho}\chi_{L}(\qq,\omega).
\eqname{L=1}
\end{equation}
The definition \eq{f_n} can then be extended to large but finite systems using the standard local density approximation.
\\


\section{How to generate the gauge field}
\label{sec:gauge}

We consider a three-dimensional gas of bosonic atoms in a strongly anisotropic, pancake-shaped trap. The axial confinement frequency $\omega_z$ is much higher than the one $\omega_\parallel$ along the $xy$ plane; both the temperature $T$ (times the Boltzmann constant $k_B$) and the chemical potential $\mu$ of the gas are assumed to be smaller 
than $\hbar \omega_z$. In this regime, the gas will be eventually described in terms of a two-dimensional Hamiltonian.

Building on an idea originally introduced in~\cite{olshanii}, an artificial gauge field coupling to the atomic current can be obtained by illuminating the atoms with several laser beams with suitably chosen frequencies, wavevectors, and waist profiles.
Several schemes to generate artificial gauge fields for neutral atoms have been proposed in the last years~\cite{olshanii,GerbierDalibard,spielman_th}. The last proposal 
\cite{spielman_th} was recently implemented on an atomic Bose-Einstein condensate: for sufficiently strong gauge fields, 
 a disordered ensemble of vortices appeared in the gas~\cite{spielman_exp}.

\begin{figure}[htbp]
\includegraphics[width=0.24\columnwidth,clip]{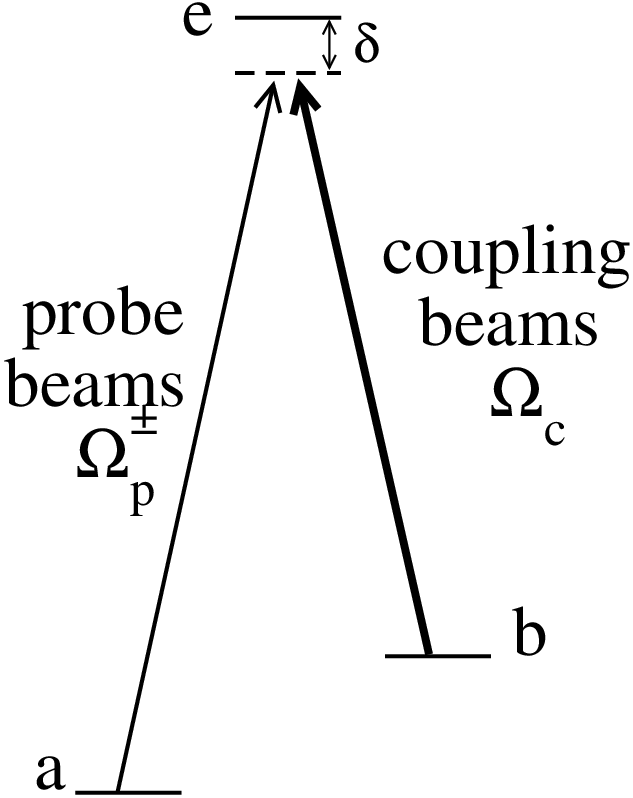}
\hspace{0.02\columnwidth}
\includegraphics[width=0.70\columnwidth,clip]{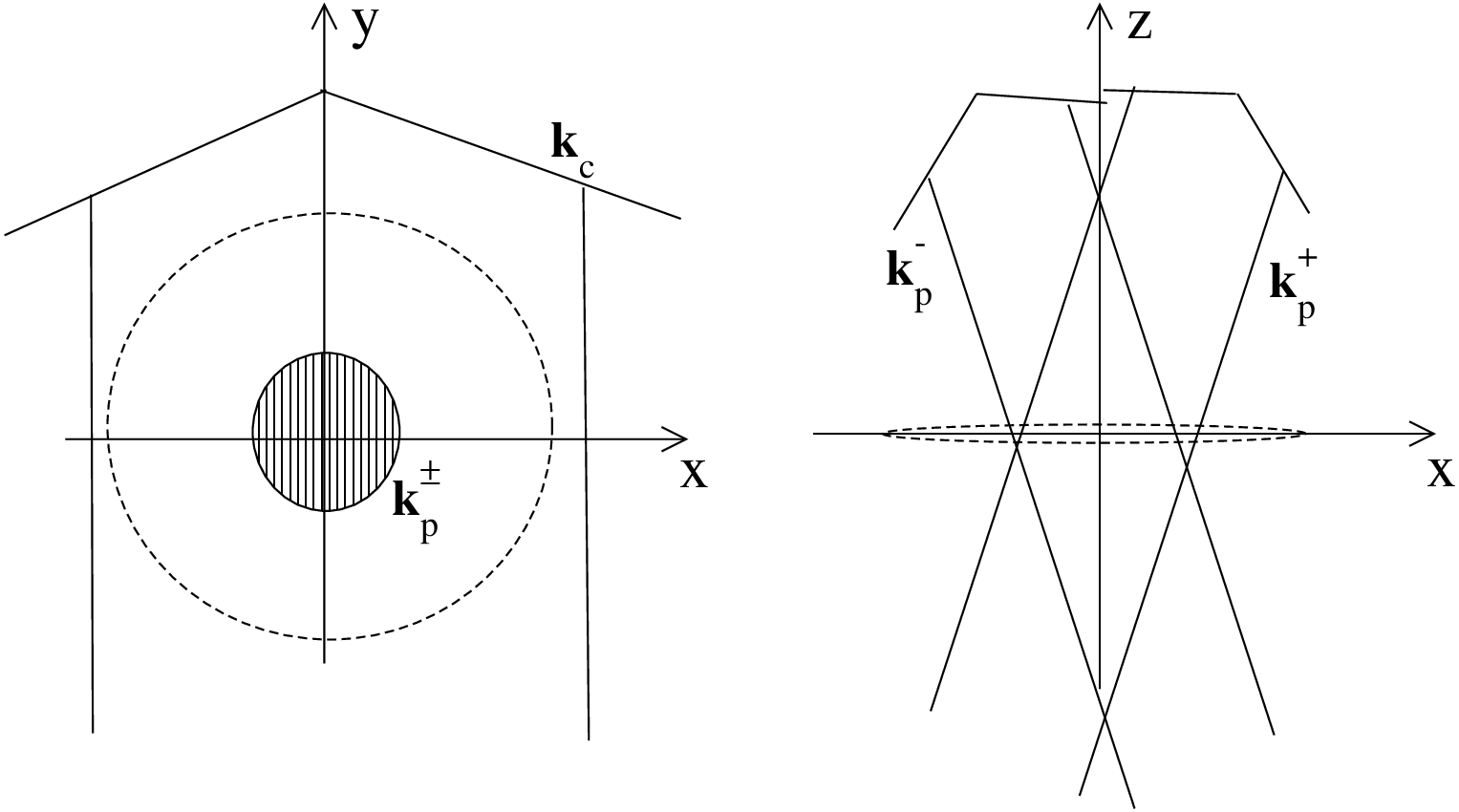}
\caption{Scheme of the set-up under consideration. Left panel: generic sketch of the $\Lambda$ configuration of atomic levels and laser beams involved in the optical processes. Center panel: view from above of the two-dimensional atomic pancake (lying within the dashed circle)
and of the laser beams (the hatched disk is the spot of the probe beams in the $xy$ plane).
 Right panel: side view. $\kk_c$ and $\kk_p^\pm$ are the wavevectors of the coupling and probe beams, respectively.
In practice, $\kk_p^{\pm}=(k_p^2-q^2/4)^{1/2}\, \eee_z \pm (q/2)\, \eee_x$ with $q\ll k_p$,
where $\eee_i$ is the unit vector along axis $i$.
}
\label{fig:scheme}
\end{figure}

In the present paper, we shall focus our attention on the level configuration shown in Fig.\ref{fig:scheme}. Three internal atomic levels in a $\Lambda$ configuration are connected by three laser fields according to the sketch given in the left panel of Fig.\ref{fig:scheme}: a coupling beam resonantly drives the $|b\rangle \rightarrow |e\rangle$ atomic transition,
while a pair of probe beams resonantly drive the $|a\rangle\rightarrow |e\rangle$ transition with
essentially the same detuning. The artificial gauge field originates from the spatial and temporal dependence of the 
resulting optically dark state \cite{olshanii}. All other atomic states are assumed to be far-off resonance. 
A discussion of their effect in the specific case of $^{87}$Rb atoms is given in the Appendix \ref{app:experimental}.
A summary of the suggested experimental parameters for this specific atomic species is given in Table~\ref{tab:val}.

The geometrical arrangement of the laser beams is sketched in the central and right panels of Fig.\ref{fig:scheme}. The continuous-wave control beam propagates along the $y$ direction with a wavevector $\kk_c$ and a carrier frequency $\omega_c$ close to resonance with the $|b\rangle\rightarrow |e\rangle$ transition, with a detuning
$\delta$, and has a peak Rabi frequency $\Omega^o_c$. Its waist profile is much wider than the size of the atomic cloud, so that it can be safely approximated by a plane wave.

The two probe beams share the same carrier frequency $\omega_p$ close to resonance with the $|a\rangle\rightarrow |e\rangle$ transition. 
The carrier frequencies of the coupling and 
probe beams are chosen exactly on resonance with the Raman transition $|a\rangle\rightarrow |e\rangle \rightarrow |b\rangle$, 
that is $\omega_p-\omega_c=\omega_b-\omega_a$.
The probe beams impinge on the atomic cloud with wavevectors $\kk^\pm_p\simeq k_p \eee_z\pm \qq/2$ close to the $z$ direction and symmetrically located with respect to it. The difference $\qq=\kk_p^+-\kk_p^-$ lies along the $xy$ plane and is in magnitude $q\ll k_p=\omega_p/c$. The probe beams spatial profile is taken to be a Gaussian with a waist $w$, centered 
at $\rr_0$ in the $z=0$ plane.

In what follows, we will need the condition $w\gg q^{-1}$, where $q$ is small enough for the limit in Eq.(\ref{eq:f_n}) to be almost reached.
In the numerical examples to come, this requires $q \xi < 1$, where $\xi$ is the healing length of the gas.  This automatically
shows that the concept of normal fraction can be used only when the portion of the gas within the waist is ``macroscopic" \cite{Toennies},
here $w \gg \xi$, which was physically expected.
At the same time, the waist $w$ is assumed to be much smaller than the atomic cloud radius, so to allow for a local measurement of the normal fraction.
The peak Rabi frequencies of the two probe beams are $\Omega_p^\pm(t)$, respectively (see Appendix \ref{app:adiab} for the precise definition of the Rabi frequencies). The spatial dependence of the Rabi frequencies of both the coupling and the probe beams is then summarized by the following expressions \cite{foot_3dgeom_beam},
\begin{eqnarray}
\label{eq:omegac_expli}
\Omega_c(\rr,t)&=&\Omega_c^{(0)} e^{-i \Delta_c t} \,e^{i\kk_c\cdot \rr} \eqname{omega_c} \\
\Omega_p(\rr,t)&\simeq&[\Omega_p^+(t)\,e^{i\kk_p^+\cdot
 \rr}+\Omega_p^-(t)\,e^{i\kk_p^-\cdot \rr} ] \nonumber
 \\
&&\times e^{-[(x-x_0)^2+(y-y_0)^2]/w^2}. \eqname{omega_p}
\end{eqnarray}
where we have allowed for the coupling beam to have a small detuning
$\Delta_c$ from Raman resonance on top of its carrier frequency at
$\omega_c$. 
On the contrary, $\Omega_p$ does not have a time dependent phase factor, but only contains the square root of a purely real non-negative
switch-on and switch-off function $f(t)$, common to the two probe beams:
\be
\Omega_p(\rr,t) = \Omega_p^{0}(\rr) [f(t)]^{1/2}.
\label{eq:def_ft}
\ee
In what follows, we shall restrict our attention to the  non-saturating regime $|\Omega_c|, |\Omega_p^\pm| \ll |\delta+i\Gamma/2|$, where
$\Gamma$ is the decay rate of $|e\rangle$ due to spontaneous emission, and $\delta$ is the common detuning of the probe
and control beam carrier frequencies from the transitions $|a\rangle\to |e\rangle$ and $|b\rangle\to|e\rangle$ respectively.
We shall also concentrate on the limit
$|\Omega_p^\pm| \ll |\Omega_c|$ where the structure of the gauge field is the simplest.

As the transitions driven by the probe and control beams share the excited state $|e\rangle$, 
there exists for each spatio-temporal coordinates $(\rr,t)$ an internal {\em non-coupled state}, for which the two excitation channels interfere destructively.
In terms of the local Rabi frequencies $\Omega_p(\rr,t)$ and $\Omega_c(\rr,t)$, this non-coupled state reads
\begin{equation}
|NC(\rr,t)\rangle=\frac{|a\rangle-{\Omega_p(\rr,t)}/{\Omega_c(\rr,t)}\,|b\rangle}{(1+|\Omega_p(\rr,t)|^2/|\Omega_c(\rr,t)|^2)^{1/2}}.
\eqname{NC}
\end{equation}
Adiabatically eliminating the excited state $|e\rangle$, one
sees that the bright orthogonal state, the so-called coupled state,
\be
|C(\rr,t)\rangle= \frac{[{\Omega_p(\rr,t)}/{\Omega_c(\rr,t)}]^* |a\rangle
+|b\rangle}{(1+|\Omega_p(\rr,t)|^2/|\Omega_c(\rr,t)|^2)^{1/2}}
\eqname{C}
\ee
is separated
from $|NC(\rr,t)\rangle$ by a (complex) energy gap  
\begin{equation}
\hbar[\delta'(\rr,t)-i\Gamma'(\rr,t)/2]\equiv \frac{\hbar [ |\Omega_c(\rr,t)|^2 +
 |\Omega_p(\rr,t)|^2]}{4(\delta+i\Gamma/2)},
\label{eq:def_gamma}
\end{equation}
where $\delta'$ and $\Gamma'$ are the light-shift and the decay rate of the coupled state.
If the energy gap is large enough as compared to both the motional coupling between $|NC\rangle$ and $|C\rangle$ due to the spatio-temporal dependence of $\Omega_c$ and $\Omega_p^\pm$ \cite{aspect}, 
and to the quantum of oscillation $\hbar\omega_z$ along the tightly confined $z$ direction, we can restrict the dynamics to the $|NC\rangle$ internal state.

Generalizing the single-particle theory of~\cite{olshanii} to the many-body context in Heisenberg picture, 
one gets to an effective Hamiltonian for the component $\phih_{3D}(\rr,t)$ of the three-dimensional atomic field operator in the (spatially and temporally-dependent) non-coupled state $|NC(\rr,t)\rangle$,
\begin{equation}
\phih_{3D}(\rr,t)=\langle NC(\rr,t)|a\rangle\,\Psih_a(\rr,t)+\langle NC(\rr,t)|b\rangle\,\Psih_b(\rr,t)
\label{eq:def_phi3D}
\end{equation}
in the simple form~\cite{foot_firstq}:
\begin{multline}
\mathcal{H}=\int\!d^3\rr\,\left\{\phihd_{3D}\left[-\frac{\hbar^2\nabla^2}{2m}+U_{3D}(\rr) +W_{3D}(\rr,t)\right]\phih_{3D} \right.\\
\left. -\jj_{3D}(\rr)\cdot\AAA_{3D}(\rr,t)+\frac{1}{2} g(\rr,t)\,\phihd_{3D}\phihd_{3D}\phih_{3D}\phih_{3D} \right\}
\eqname{H3}
\end{multline}
where $U_{3D}$ is the three-dimensional trapping potential (supposed to be common to the internal states $|a\rangle$ and $|b\rangle$)
and the vector gauge potential 
\begin{equation}
\AAA_{3D}(\rr,t) =   \frac{i\hbar}{2} \Big[\langle NC(\rr,t)|[\nabla |NC(\rr,t)\rangle] - \mbox{c.c.}\Big]
\eqname{A}
\end{equation}
couples to the three-dimensional atomic current operator
\begin{equation}
\jj_{3D}(\rr)=\frac{\hbar}{2im}\left[\phihd_{3D}(\rr)\nabla\phih_{3D}(\rr) - \textrm{h.c.} \right],
\eqname{j}
\end{equation}
and the scalar potential 
\begin{multline}
W_{3D}(\rr,t) = -\frac{i \hbar}{2} \Big[\langle NC(\rr,t)| \,[\partial_{t}  |NC (\rr,t)\rangle] - \mbox{c.c.}\Big]+ \\
+\frac{\hbar^2}{2m} \sum_{i=x,y,z} [\partial_{r_i} \langle NC(\rr,t)|] [\partial_{r_i} |NC(\rr,t)\rangle]
\eqname{W}
\end{multline}
couples to the three-dimensional  density
\begin{equation}
n_{3D}(\rr)=\phih_{3D}^\dagger(\rr)\,\phih_{3D}(\rr) .
\eqname{n3d}
\end{equation}
The derivation of the Hamiltonian \eq{H3} is based on the Quantum Stochastic Differential Equations formalism~\cite{QuantumNoise}; the details are given in the Appendix \ref{app:adiab}.

The spatial and temporal dependence of the weights of the non-coupled state \eq{NC} 
in the internal states $|a\rangle$ and $|b\rangle$
reflects into a similar variation of the coupling constant
describing the atomic interaction within the internal state $|NC\rangle$: 
\begin{equation}
g_{3D}(\rr,t) = \frac{|\Omega_c|^4\,g_{aa} + 2|\Omega_p(\rr,t)|^2\,|\Omega_c|^2\,g_{ab} + |\Omega_p|^4\,g_{bb}}
{(|\Omega_c|^2+|\Omega_p(\rr,t)|^2)^2},
\eqname{g}
\end{equation}
 where the coupling constants $g_{aa}$, $g_{ab}$  and $g_{bb}$ originate from the $a-a$, $a-b$ and $b-b$
elastic $s$-wave interactions.
In what follows, we shall be interested in isolating the response of the system to the gauge field $\AAA_{3D}$. 
To this purpose, it will be useful to minimize the effect of all unwanted couplings to the density introduced by the scalar potential  $W_{3D}$ and by the 
spatio-temporal dependence of the interaction constant $g_{3D}$. This latter effect is minimized if one chooses states $a,b$ with similar scattering properties $g_{aa}\simeq g_{ab} \simeq g_{bb}$. 
 In the limit $|\Omega_p/\Omega_c|\ll 1$, one simply needs to have 
$g_{aa}\simeq g_{ab}$, as it is assumed from now on.

The atomic field along $z$ is assumed to be frozen in the ground state of the harmonic confinement of wavefunction $\phi_0(z)$.
This allows to express the three-dimensional bosonic field $\hat{\phi}_{3D}$  in terms of the bosonic field for a two-dimensional gas, setting
\be
\hat{\phi}_{3D}(x,y,z) = \phi_0(z) \hat{\phi}(x,y).
\label{eq:def_phi}
\ee
Correspondingly, the two-dimensional coupling constant $g$ has the expression 
\be
g=\frac{\hbar^2}{m}\tilde{g}=\frac{g_{3D}}{\sqrt{2\pi}\,a_{\rm ho}^z}
\eqname{g2D}
\ee
in terms of the three-dimensional coupling constant $g_{3D}=4\pi \hbar^2 a_{3D}/m$, where
$a_{3D}$ is the $s$-wave scattering length in the state $|a\rangle$, and the size $a_{\rm ho}^z=\sqrt{\hbar/m\omega_z}$ of the ground state along $z$. 
Note that the dimensional reduction \eq{def_phi} does not require to be in the Lamb-Dicke limit $k_{p,c} a_{\rm ho}^z\ll 1$, 
it is sufficient
that the typical energy per particle corresponding to the various terms in $\mathcal{H}$ is $\ll\hbar\omega_z$.

The effective two-dimensional gauge and scalar potentials $\AAA$ and $W$ then result from an average of the Hamiltonian \eq{H3} over the motional ground state along $z$.
Including in \eq{A} and \eq{W} the explicit form of the beam profiles and restricting ourselves to zeroth order in the small parameters $q/k_{p,c}$, $1/wk_{p,c}$,
  and to second order in $\Omega_p^\pm/\Omega_c$, the resulting two-dimensional gauge potential turns out to be directed along the $y$ axis and to have the form
\begin{multline}
A_y(\rr) \simeq  \hbar k_c \frac{\left|\Omega_p^+ e^{i \qq\cdot \rr/2}+\Omega_p^- e^{-i \qq\cdot \rr/2}\right|^2} {|\Omega_c|^2} e^{-2 |\rr-\rr_0|^2 /w^2}  \\
 =\hbar k_c \frac{ |\Omega_p^+|^2 + |\Omega_p^-|^2 + [\Omega_p^+ \Omega_p^{-*} e^{i\qq \cdot \rr} + \textrm{c.c.} ] } {|\Omega_c|^2}\, \\ \times e^{-2 |\rr-\rr_0|^2 /w^2}.
\eqname{A_y}
\end{multline}
To the same level of approximation, the scalar potential has the form 
\begin{multline}
W(\rr)=\left[\frac{\hbar^2(k_c^2+k_p^2)}{2m}+\hbar \Delta_c\right]\\ 
\times \frac{\left|\Omega_p^+ e^{i\qq\cdot \rr/2} +\Omega_p^- e^{-i \qq\cdot \rr/2}\right|^2}{|\Omega_c|^2}\,
e^{-2|\rr-\rr_0|^2/w^2},
\end{multline}
which can be made to vanish by choosing a detuning $\Delta_c$
that exactly compensates the recoil of the atoms after the Raman 
process $a\rightarrow e \rightarrow b$:
\be
W \equiv 0 \ \ \ \mbox{for} \ \ \ \Delta_c=-\frac{\hbar (k_c^2+k_p^2)}{2m}.
\label{eq:magic_choice}
\ee
It is worth pointing out that the temporal derivative $df(t)/dt$  of the real switch-on and switch-off function $f(t)$ of the probe beam
Eq.(\ref{eq:def_ft}) has an exactly vanishing contribution to the $\partial_t$ term in the right-hand side of Eq.(\ref{eq:W}),  
so it does not contribute
to the scalar potential $W$ to all orders in $\Omega_p/\Omega_c$
\cite{foot_order_q2}. 
The same conclusion holds for the gauge field, since there is no temporal derivative in (\ref{eq:A}).

After expansion of the squared modulus as done in the second line of \eq{A_y}, 
two kinds of terms are immediately identified: (i) a slowly varying Gaussian term of size $w$ and peak amplitude $|\Omega_p^+|^2+|\Omega_p^-|^2$ that follows the laser envelopes, and (ii) an oscillating term at wavevector $\qq$ with a Gaussian envelope of size $w$ and peak amplitude $|\Omega_p^+\Omega_p^{-}|$. This spatially modulated term is indeed the one that we need to probe the normal fraction of the gas according to the definition \eq{f_n}: when $\qq$ is taken along the $x$ axis ($y$ axis), it provides an almost purely transverse (longitudinal) contribution to the gauge field $\AAA$. On the other hand, the slowly varying term always includes both longitudinal and transverse vector field components. Experimental procedures to subtract the effect of this unwanted term will be discussed in the next sections.

\section{Deposited energy measurement}
\label{sec:deposited}

\subsection{General idea}
\label{sec:gauge_gen}


In this section we shall present a method to extract the value of the normal fraction from a measurement of the energy that is deposited in the system by a suitably designed gauge field sequence. The coupling beam is assumed to be always switched on. On the other hand, both probe beam intensities
$|\Omega_p^\pm|^2$ are varied in time according to the (dimensionless) real envelope function $f(t)$. This function
is chosen to be $0$ for $t<0$ and to rapidly tend back to $0$ at long times.

As already mentioned, we assume that the atomic interaction constants satisfy $g_{aa}\simeq g_{ab}$~\cite{foot_sptpdep}.
As soon as $qw\gg 1$, the deposited energy is the sum of two independent contributions $\Delta E_{1,2}$ corresponding to the decomposition \eq{A_y} of the gauge field as the sum of a non-modulated term and a modulated one at the wavevector $\qq$. Using standard linear response theory within the local density approximation as discussed in the Appendix \ref{appendix:LDA}, the contribution $\Delta E_2$ of the modulated term can be written in the simplified form 
\begin{equation}
\Delta E_2 \simeq \frac{\pi}{4} w^2\,
\left(\frac{\epsilon_{\rm gauge}}{2}\right)^2
 \int_{-\infty}^{+\infty}\!\frac{d\omega}{\pi}\,\omega\,|f(\omega)|^2\,\textrm{Im}[\chi_{yy}(\qq,\omega)]
\eqname{deltaE2}
\end{equation}
 where we have introduced the amplitude of the spatially modulated part of the gauge field,
\begin{equation}
\epsilon_{\rm gauge}=2\,\hbar
 k_c\,\frac{|\Omega_p^+\,\Omega_{p}^{-}|(t=0^+)}{|\Omega_c|^2}.
\eqname{eps_gauge}
\end{equation}
A similar expression for the contribution $\Delta E_1$ of the non-modulated term is given in the Appendix
\ref{appendix:LDA} as \eq{deltaE1_app}.
The susceptibility $\chi_{yy}$ appearing in (\ref{eq:deltaE2}) is the one (as defined in section \ref{sec:principle})
of the fictitious spatially homogeneous two-dimensional Bose gas that approximates the state of the trapped
gas around the center $\rr_0$ of the probe beams' spot.

The expression \eq{deltaE2} for the deposited energy involves the imaginary part of that susceptibility, while the normal fraction \eq{f_n} involves the real part. To relate the two, one can make use of the well-known Kramers-Kronig relation of linear response theory,
\begin{equation}
\lim_{\omega\rightarrow 0}\textrm{Re}[\chi(\qq,\omega)]=\int_{-\infty}^{+\infty}\frac{d\omega'}{\pi}\,\frac{\textrm{Im}[\chi(\qq,\omega')]}{\omega'}.
\eqname{KK}
\end{equation}
For a suitably chosen envelope of the form 
\be
f(t)=e^{-\gamma t}\,\Theta(t),
\label{eq:choix_ft}
\ee
where $\Theta(t)$ is the Heaviside step function, 
the Fourier transform is equal to $f(\omega)=i/(\omega+i\gamma)$, and the integral in \eq{deltaE2} indeed reduces to the real part of the susceptibility \eq{KK} in the 
$\gamma\to 0$ limit.

As a consequence, the deposited energy $\Delta E_2$ for small $\qq$ perpendicular (parallel) to $\kk_c$ can be related to the normal (total) density $\rho_n$ ($\rho$) 
of the trapped gas at position $\rr_0$ by
\begin{equation}
\Delta E_2\simeq \frac{\pi}{4} \frac{w^2}{m}\, \left(\frac{\epsilon_{\rm gauge}}{2}\right)^2
\times \left\{\begin{array}{c} \rho_n(\rr_0) \hspace{0.5cm} \textrm{for $\qq \perp \kk_c$} \\ \rho(\rr_0) \hspace{0.5cm} \textrm{for $\qq \parallel \kk_c$} \end{array}
 \right. .
\eqname{Delta2} 
\end{equation}
For intermediate angles $\alpha$ between $\qq$ and $\kk_c$, $\Delta E_2$ is proportional to $\rho \cos^2\alpha + \rho_n \sin^2\alpha$
\cite{apropos_alpha}.

In an actual experiment, the undesired contribution $\Delta E_1$ can be eliminated by noting its independence on the relative orientation of $\qq$ and $\kk_c$ [see Eqs.\eq{deltaE1_app} and \eq{deltaE2_app}], as well as its different dependence on the probe amplitudes $\Omega_p^\pm$, proportional to $[|\Omega_p^+|^2 + |\Omega_p^-|^2]^2$ rather than $|\Omega_p^+\Omega_p^-|^2$. By measuring the deposited energy for at least two different values of the $\Omega_p^+/\Omega_p^-$ ratio, one is able to isolate the relevant contribution \eq{Delta2}. 

\subsection{How fast is the $q,\gamma\to 0$ limit reached~?}
\label{subsec:bogo}

\begin{figure}[htbp]
\includegraphics[width=0.95\columnwidth,clip]{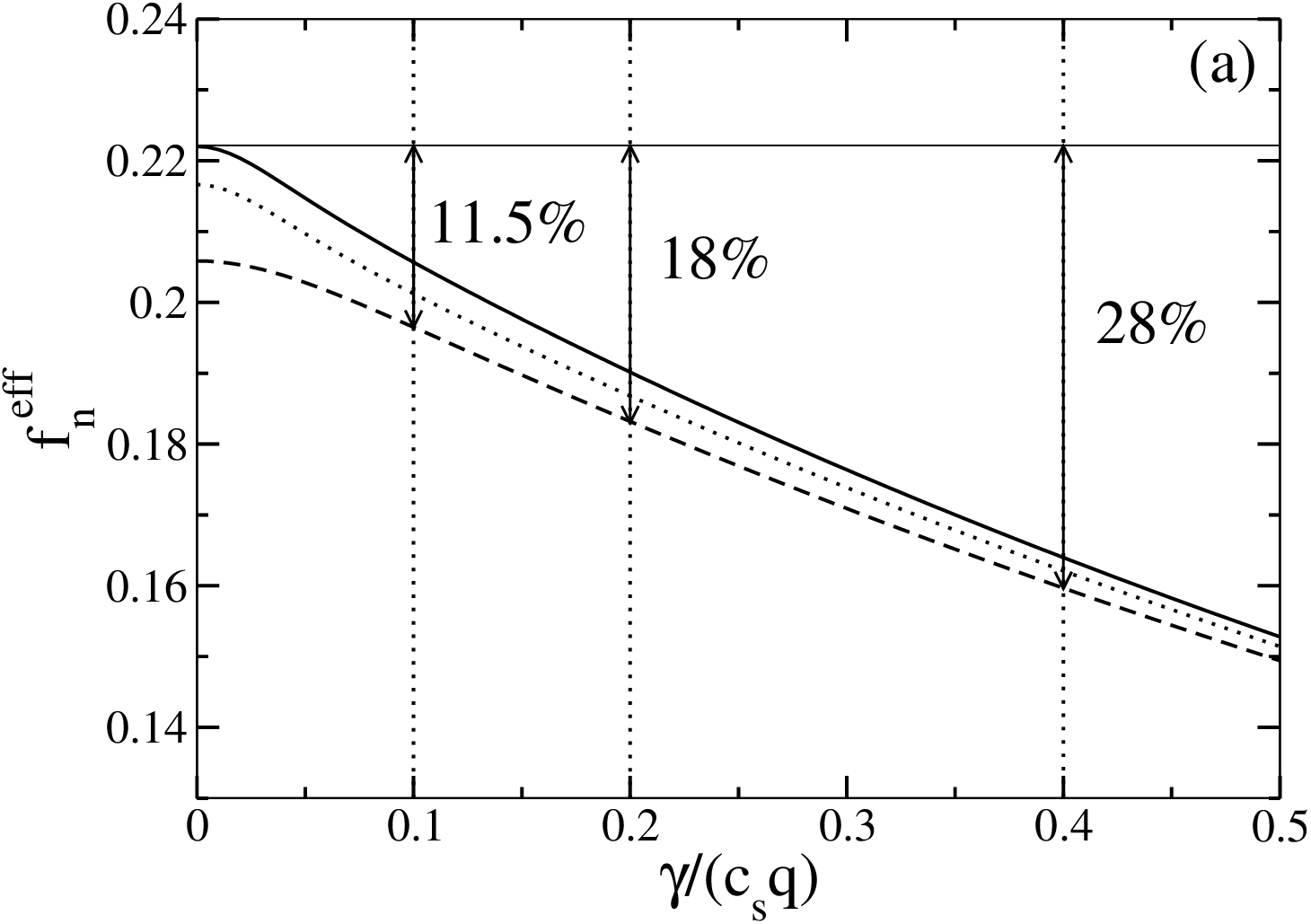}
\includegraphics[width=0.95\columnwidth,clip]{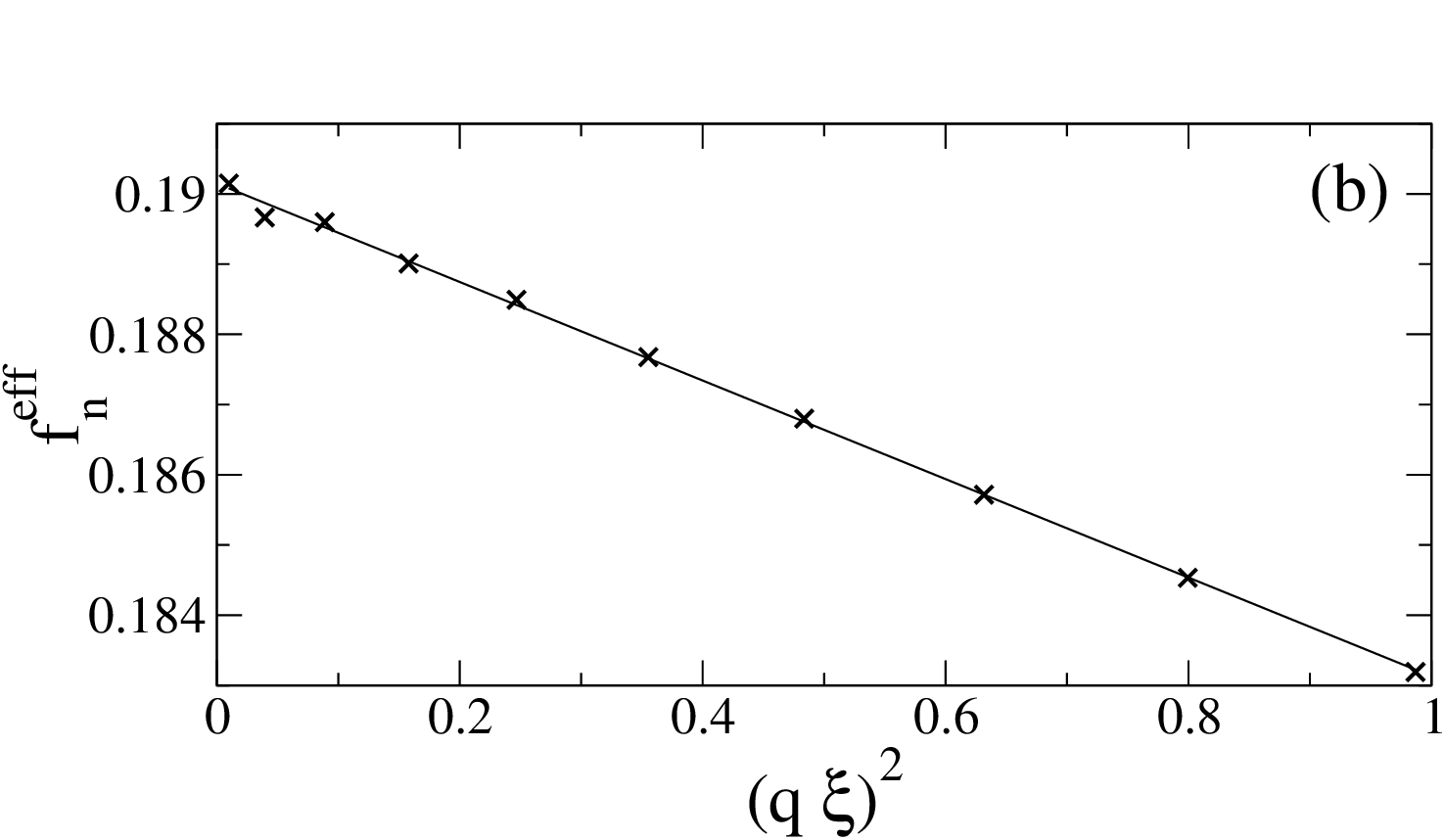}
\caption{
Quantum Bogoliubov prediction \eq{f_n_eff} for the effective normal fraction $f_n^{\rm eff}$.
 In (a), dependence of $f_n^{\rm eff}$ on the switch-off rate $\gamma$ of the gauge field, for 
three different values of the 
 gauge field spatial modulation wavevector $\qq=(2\pi/L)\, \eee_x$ (solid, $q\xi\simeq 0.1$), 
$\qq=7\times (2\pi/L)\, \eee_x$ (dotted, $q\xi\simeq 0.7$) and $\qq=10\times (2\pi/L)\, \eee_x$
 (dashed, $q\xi\simeq 1$).
System parameters: square box (with periodic boundary conditions) of size $L/\xi\simeq 63$ containing
 $N\simeq 40000$ particles with interaction constant $g=0.1\,\hbar^2/m$
 at a temperature $T/T_d=0.1$. $k_B T_d=2\pi \hbar^2 \rho/m$ is the degeneracy
 temperature. The healing length $\xi$ of the gas is defined by ${\hbar^2}/(m \xi^2) = \rho g$
and the Bogoliubov sound velocity is defined as $c_s=(\rho g/m)^{1/2}$.
The dashed curve corresponds to the same value of
 $q\xi\simeq 1$ as in Fig.\ref{fig:numeric2} and the vertical dotted lines indicate
 the values of $\gamma/(c_sq)$ considered in that figure.
The thin horizontal line is the prediction \eq{f_n_thermo_quant} of the quantum Bogoliubov theory
in the thermodynamic limit. Interestingly, for the considered atom number, the finite-size thermodynamic
formula Eq.(\ref{eq:f_n_thermo1}) evaluated with the quantum Bogoliubov theory gives already the same value at the scale of the figure~\cite{foot1}.
The quadratic rather than linear dependence of $f_n^{\rm eff}$ on $\gamma$
 for small values of $\gamma$ is a finite-size effect.
The fact that in (a), the values of $f_n^{\rm eff}$ for the dotted curve are close to the arithmetic mean of the 
solid and dashed curves, for common values of $\gamma/(c_s q)$, suggests
that $f_n^{\rm eff}$ is roughly an affine function of $(q\xi)^2$ for fixed $\gamma/(c_s q)$. 
This guess is substantiated in panel (b) where the dependence of the effective normal fraction 
on the modulation wavevector $\qq=q\,\eee_x$ is shown for $\gamma/(c_s q)=0.2$ 
[the crosses are the Bogoliubov predictions and the line is a linear fit of $f_n^{\rm eff}$
as a function of $q^2$].
}
\label{fig:bogo}
\end{figure}

An important point in view of experiments is to characterize how small the wavenumber $q$ and the switch-off rate $\gamma$ of the gauge field spatial modulation
have actually to be taken to obtain a quantitatively accurate measurement of the normal fraction $f_n$. 
To answer this question, we consider in this subsection the simplest case of a spatially homogeneous two-dimensional
system in a square box of size $L$ with periodic boundary conditions, excited by the gauge field in a plane wave form, i.e. in the limit 
$w\to \infty$. We also limit ourselves to the case of a transverse gauge field with $\qq=q\,\eee_x$ perpendicular to $\kk_c=k_c \, \eee_y$,
\begin{equation}
\AAA_{\rm ideal}(\rr,t)=e^{-\gamma t}\Theta(t) \eee_y
 \frac{\epsilon_{\rm gauge}}{2}\left( e^{i q x} + e^{-iqx} \right).
\end{equation}
The deposited energy at the end of the gauge field sequence can be evaluated by means of the Bogoliubov theory of dilute Bose gases. 
The main steps of the calculation are sketched in the Appendix \ref{app:bogo}. The final result reads
\begin{equation}
 \Delta E_2 = \left(\frac{\epsilon_{\rm gauge}}{2}\right)^2
		   \frac{N}{m}\,f_n^{\rm eff} 
\end{equation}
in terms of the wavevector- and $\gamma$-dependent effective normal fraction $f_n^{\rm eff}$:
\begin{multline}
f_n^{\rm eff}=\frac{1}{N}
\sum_{\kk\neq \mathbf{0},-\qq}
\frac{\hbar^2 k_y^2}{m} \times  \\
\textrm{Re}\Big[\frac{n_\kk - n_{\kk+\qq}}{\epsilon_{\kk+\qq}-\epsilon_\kk-i\hbar\gamma}
\left(U_\kk U_{\kk+\qq}- V_\kk V_{\kk+\qq} \right)^2 \\
+
\frac{1+n_\kk + n_{\kk+\qq}}{\epsilon_{\kk+\qq}+\epsilon_\kk -i\hbar\gamma}
\left(U_\kk V_{\kk+\qq} - V_\kk U_{\kk+\qq}\right)^2
\Big].
\eqname{f_n_eff}
\end{multline}
Here $N$ is the total particle number, $\rho=N/L^2$ is the surface density,
\begin{equation}
\epsilon_{\kk}=\left[\frac{\hbar^2k^2}{2m}
		\left(\frac{\hbar^2k^2}{2m}+2\rho g\right)\right]^{1/2}
\end{equation}
is the usual Bogoliubov dispersion relation and the amplitudes of the Bogoliubov modes satisfy
\begin{equation}
U_\kk+V_\kk=\frac{1}{U_{\kk}-V_{\kk}}=\left(\frac{\hbar^2k^2/2m}{\hbar^2k^2/2m+2\rho g}\right)^{1/4}.
\end{equation}
The $n_{\kk}$ are the thermal mean occupation numbers of Bogoliubov modes, $n_{\kk}=1/[\exp(\epsilon_{\kk}/k_BT)-1]$.

The thermodynamic limit $L\to \infty$ at fixed density $N/L^2$ can be worked out analytically by first taking the $\gamma\to 0$ limit and then the $q\to 0$ limit in the expression \eq{f_n_eff} for $f_n^{\rm eff}$. In this way one recovers the usual Bogoliubov expression for the normal fraction, which in dimension two reads:
\begin{equation}
f_n = \frac{1}{\rho} \int \frac{d^2k}{(2\pi)^2} \frac{\hbar^2 k_y^2 }{m}
(-\partial_{\epsilon_{\kk}} n_{\kk}).
\label{eq:f_n_thermo_quant}
\end{equation}
For a finite size system, the dependence of $f_n^{\rm eff}$ on $\gamma$ and $q$ is explored in Fig.\ref{fig:bogo}. For the smallest non-zero wavevector value allowed by the chosen 
quantization box, the relative error on $f_n^{\rm eff}$ is already as small as $10\%$ for $\gamma/c_sq=0.15$,
where $c_s=(\rho g/m)^{1/2}$ is the Bogoliubov sound velocity.
For $\gamma/(c_s q)$ fixed, in particular to the relevant value $0.2$, and over the relevant range
$q\xi \leq 1$, we have also found that 
$f_n^{\rm eff}$ is to a good approximation an affine function of $(q\xi)^2$, see Fig.\ref{fig:bogo}b, which was not obvious.

Another interesting result of Bogoliubov theory applied to our system is a sufficient condition on the amplitude of the gauge field to be within the linear response regime. To this purpose, we can write the equations of motion for the Bogoliubov mode operators $b_\kk$ in the interaction picture in the presence of the time-dependent gauge field, and impose that the amplitude change is small
as compared to the initial value. The expression of this amplitude change, that we shall not give here, involves as usual
energy denominators such as $\epsilon_{\kk}-\epsilon_{\kk\pm\qq} +i\hbar\gamma$.

In the limit $\gamma\to 0$, the condition of linear response is thus most stringent for modes of wavevector $\kk$
such that $\epsilon_\kk=\epsilon_{\kk\pm\qq}$, where the real part of the energy denominator can vanish. For the maximal value of $k$ set by the thermal occupation, this leads to the sufficient condition
\begin{equation}
\frac{\epsilon_{\rm gauge}}{(mk_BT_d)^{1/2}} \lesssim \left(\frac{T_d}{T}\right)^{1/2}
\frac{2\hbar\gamma}{k_B T_d},
\eqname{bound_gauge_amplitude}
\end{equation}
in terms of the degeneracy temperature $k_B T_d = 2\pi {\hbar^2 \rho}/{m}$.
This naive argument is however not able to determine to which extent this condition is actually necessary. This would require a higher order calculation which falls beyond the scope of the present work \cite{foot_ideal_gas}.

\subsection{Numerical investigation}
\label{subsec:num_inv}

To further assess the validity and accuracy of our proposed scheme we have performed full scale
numerical simulations of the response  of a two-dimensional Bose gas at finite temperature to the complete gauge field \eq{A_y}, including the Gaussian envelope of the gauge field and a circular well trapping potential. 
A very useful tool to this purpose is the classical field model developed and applied in a number of recent works~\cite{vortex_lattice_simul,CFT}. For this model both the thermal equilibrium state and the temporal dynamics can in fact be easily addressed with standard numerical tools and provide reliable results for the physics of the degenerate Bose gas.

We consider a classical (c-number) complex field defined on a square grid. 
The real space lattice constant $b$ is chosen in terms of the thermal de Broglie wavelength $\lambda=\sqrt{2\pi\hbar^2/mk_B
T}$ as $b/\lambda=\sqrt{\pi/(4\zeta)}$. The specific value $\zeta \simeq 0.80$~\cite{foot_zeta}
of the numerical coefficient is such that the classical field model correctly reproduces the total number of non-condensed particles for an ideal gas at zero chemical potential in the thermodynamic limit. This choice corresponds to setting the ultra-violet momentum cut-off $k_{\rm max}=\pi/b$ at $\hbar^2 k_{\rm max}^2/2m = \zeta k_B T$
in the classical field theory.

In the canonical ensemble, the thermal probability distribution for the interacting classical field 
follows a Boltzmann law $\delta(\|\Psi\|^2-N) \exp(-E[\psi]/k_B T)$ with the norm-squared
$\|\Psi\|^2=b^2 \sum_\rr |\Psi(\rr)|^2$ fixed to the total atom number, and the discrete Gross-Pitaevskii energy functional~\cite{foot_dis_lap} given by:
\begin{equation}
E[\Psi]=b^2 \sum_\rr \Psi^*\left[-\frac{\hbar^2}{2m}\Delta +U(\rr)+\frac{g}{2}\,|\Psi|^2\right]\Psi,
\eqname{EGP}
\end{equation}
where $U(\rr)$ is the trapping potential seen by the bidimensional gas.
This probability distribution can be sampled by the long time limit of a Ito stochastic differential
equation with drift terms and a noise term \cite{Mandonnet,AliceIto},
including projectors in order to keep the norm-squared constant, $\|\Psi\|^2=N$:
\begin{multline}
d\Psi=-\frac{1}{2}\,d\tau\,\mathcal{Q}_\Psi\left[-\frac{\hbar^2}{2m}\Delta\Psi
 + U\,\Psi+ g
 |\Psi|^2 \,\Psi\right] \\ 
+\frac{\sqrt{k_B T}}{b}\,\mathcal{Q}_\Psi d\xi -
 \frac{M-1}{2N}k_B T\,d\tau\,\Psi,
\end{multline}
where $M$ is the number of grid points, $\mathcal{Q}_\Psi$ is the projector onto the subspace orthogonal to the classical field $\Psi$,
$d\xi$ is a complex Gaussian, zero-mean, delta-correlated and temporally white noise such that $d\xi d\xi =0$ and
\begin{equation}
d\xi^*(\rr_i)\,d\xi(\rr_j) = d\tau\,\delta_{\rr_i,\rr_j}.
\end{equation}
Our numerical procedure simply consists in first generating a number $n_{\rm real}$ of independent wavefunctions distributed according to the thermal Boltzmann law with energy functional \eq{EGP} and then to let them evolve in real time according to the discrete time-dependent Gross-Pitaevskii equation including the gauge potential 
\eq{A_y} \cite{foot_real_time}.

\begin{figure}[htbp]
\includegraphics[width=0.95\columnwidth,clip]{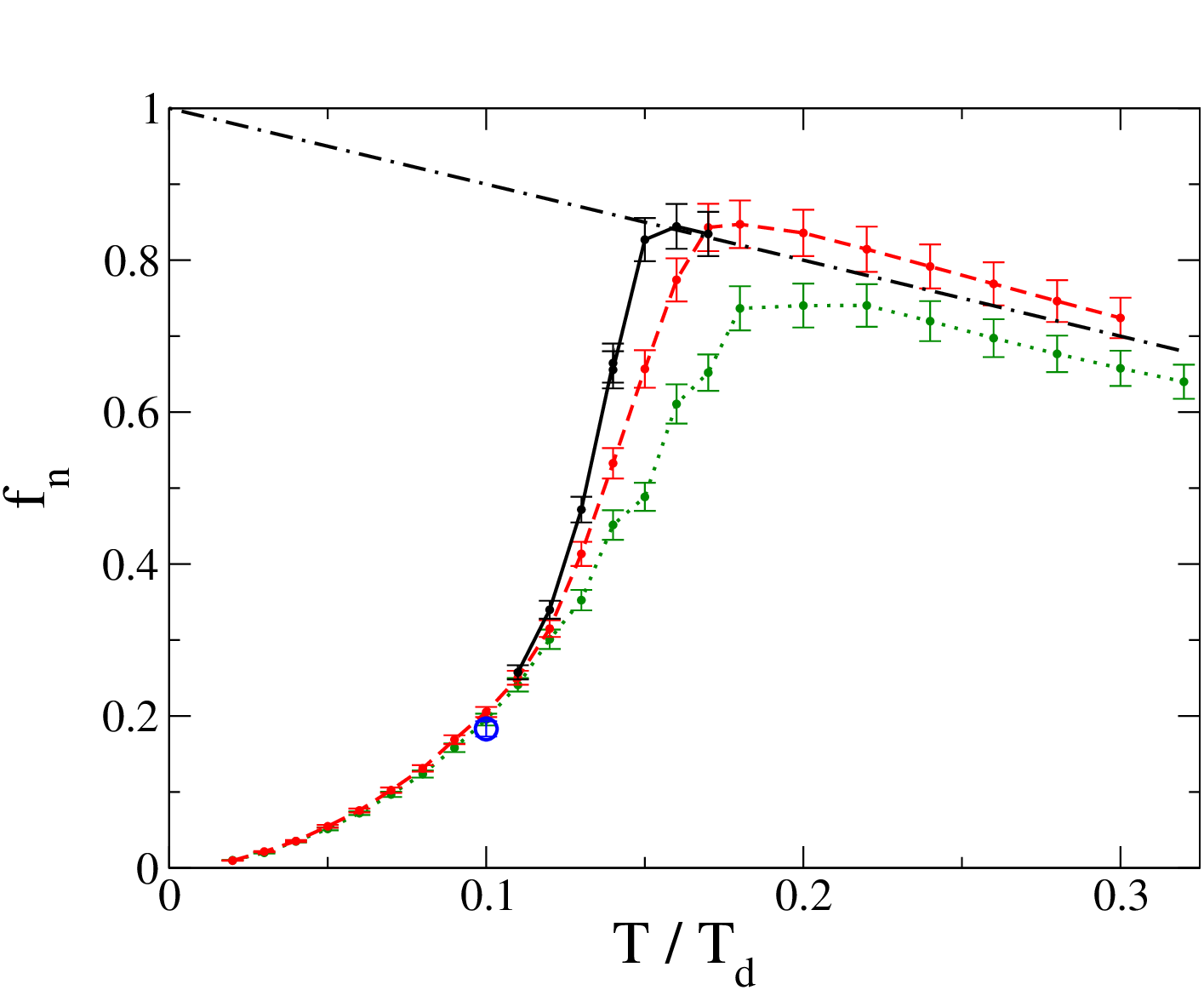}
\caption{Classical field simulation of the normal fraction $f_n$ for a finite size
 two-dimensional, spatially homogeneous  interacting Bose gas as a
 function of temperature. The calculation has been performed using the thermodynamic expression \eq{f_n_thermo1} for the normal fraction. The coupling constant is $g=0.1\, \hbar^2/m$. The different curves refer to simulations performed with different 
system sizes $L$, resulting in different numbers of grid points $M=16^2$ (green, dotted), $32^2$ (red, dashed), $64^2$ (black, solid) 
[the grid spacing is $b\simeq 0.99\lambda$, where $\lambda$ is the thermal de Broglie wavelength,
see text]. The number of classical field realizations is $n_{\rm real}=1000$.
The dot-dashed line is the classical field prediction for the normal fraction $f_n$ of an ideal gas in the thermodynamic limit: as discussed in the text, the decrease at high temperatures is an artifact of the classical field  model. 
The blue circle indicates the result of a numerical simulation of the deposited energy scheme 
as in Fig.\ref{fig:numeric2} for a homogeneous system with an infinite beam waist $w=\infty$
(in which case we simply dropped the non-modulated term in the gauge field), a number of grid points
$M=32^2$, a gauge field modulation wavevector $\qq=(2\pi/L)\, \eee_x \simeq (0.2/\xi)\, \eee_x$, and including a careful extrapolation of
$\epsilon_{\rm gauge}\to 0$ (numerics down to $\epsilon_{\rm gauge}/(m k_B T_d)^{1/2}=0.01$) and 
of $\gamma\to 0$ (numerics down to $\gamma/(c_s q)=0.05$ where $c_s=(\rho g/m)^{1/2}$ is the Bogoliubov sound velocity).
}
\label{fig:numeric}
\end{figure}
As a first application of the classical field method, we have determined the normal fraction of a spatially homogeneous two-dimensional interacting gas at thermal equilibrium. This is done using the
thermodynamic formula, applicable to finite size systems with periodic boundary conditions,
\begin{equation}
f_n^{\rm thermo}=\frac{\langle P_y^2 \rangle}{Nmk_B T}
\eqname{f_n_thermo1}
\end{equation}
which involves the thermal variance of the total momentum $P_y$ of the gas~\cite{Leggett,CRAS}.
The temperature dependence of the normal fraction is shown in Fig.\ref{fig:numeric} for increasing system sizes. The sudden variation around $T/T_d\simeq 0.13$ becomes sharper and sharper as the system size is increased and should eventually correspond to  a discontinuous jump in the superfluid fraction at the BKT transition~\cite{BKT,Minnhagen,amherst}. The slow decrease for larger values of $T/T_d$ is instead an artifact of the ultraviolet cut-off that has to be imposed to the classical field model in any dimension
$d\geq 2$. Indeed, the same decrease is visible also in the case of an ideal gas, for which one can show that $f_n=1-T/T_d +O( e^{-T_d/T} T/T_d )$ in the thermodynamic limit.

\begin{figure}[htbp]
\includegraphics[width=0.95\columnwidth,clip]{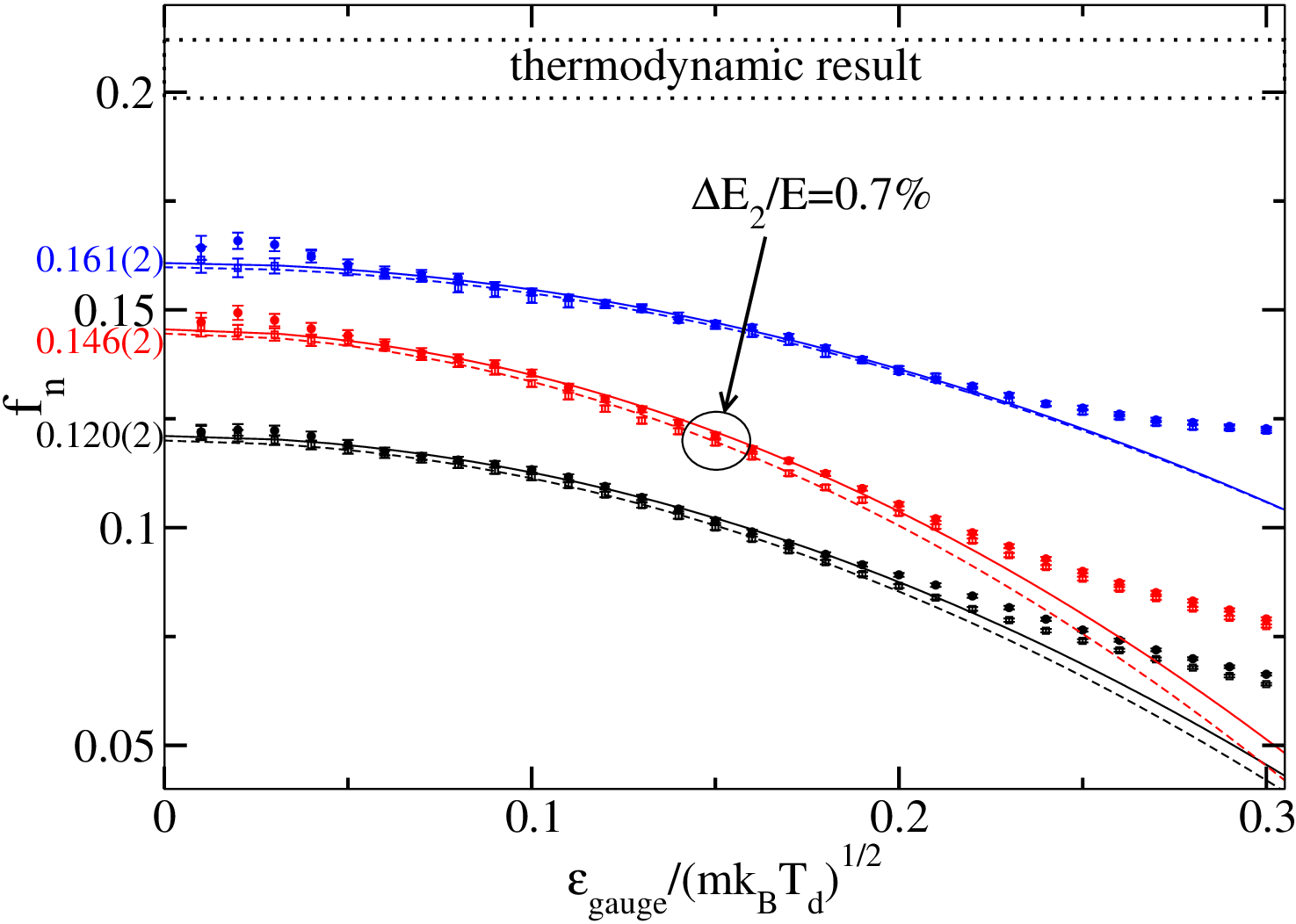}
\caption{
Classical field simulation of the deposited energy measurement scheme.
Initial thermal equilibrium state at $T/T_d=0.1$ with
 $g=0.1\,\hbar^2/m$, and a system size $L/\xi\simeq 63$ (the number of grid points is $M=64^2$,
and $n_{\rm real}=1000$ independent realizations of the classical field are used). 
Gauge field modulation wavevector $\qq\simeq (1/\xi)\, \eee_x$.
The real time evolution is followed during $\tau = 3/\gamma$.
The complete form Eq.(\ref{eq:A_y}) of the gauge field is considered, with $\rr_0$ located in the center of the system, 
and a beam waist $w=30\xi$.
For each value of the gauge field intensity $\epsilon_{\rm gauge}$ of Eq.\eq{eps_gauge},
two calculations are performed to extract the energy change $\Delta E_2$ due to the spatially modulated component,
from which an approximate value of the normal fraction $f_n$ is obtained
(see text): the first calculation with 
$|\Omega_p^+|^2(0^+)/|\Omega_c|^2=\epsilon_{\rm gauge}/(\hbar k_c)$ and
$\Omega_p^-\equiv 0$, the second one with $|\Omega_p^\pm|^2(0^+)/|\Omega_c|^2=
\epsilon_{\rm gauge}/(2\hbar k_c)$.
Empty squares with error bars: spatially homogeneous system of size $L$ (with periodic
 boundary conditions), which corresponds to $N\simeq 40000$
 atoms. Filled circles with error bars: 
 system in a circular potential well (see text), 
with a total atom number $N\simeq 30000$
 adjusted to have the same central density $\rho$, hence
 same degeneracy temperature $T_d$ and healing length $\xi$ as in the homogeneous case.
The suggested experimental values of Table~\ref{tab:val}
correspond to the circled point indicated by the oblique arrow and
require the measurement of a relative energy change of $\simeq 0.7\%$.
Thin lines: quadratic fit (with no linear term, see text) of the effective $f_n$ as a function
of $\epsilon_{\rm gauge}$, over the interval $0.05 \leq \epsilon_{\rm gauge}/(m k_B T_d)^{1/2} < 0.2$
(dashed lines for the spatially homogeneous system, solid lines for the system in the potential well).
Black, red and blue lines and points (from bottom to top) correspond to an excitation sequence with
 $\gamma/(c_s q)=0.4, 0.2$ and $0.1$ respectively,
with $c_s=(\rho g/m)^{1/2}$ the Bogoliubov sound velocity. 
The extrapolated zero-$\epsilon_{\rm gauge}$ values of the effective normal fraction are indicated on the
figure. 
The region between the horizontal dotted lines indicates the confidence interval of the thermodynamic prediction 
shown in Fig.\ref{fig:numeric}.
The residual $14\%$ deviation of $f_n$ from the thermodynamic value
(after linear extrapolation to $\gamma=0$) is a finite $q \xi$ effect (see text).
}
\label{fig:numeric2}
\end{figure}

The experimental estimation of the normal fraction obtained by the deposited energy method discussed in Sec.\ref{sec:gauge_gen} is simulated in Fig.\ref{fig:numeric2}. 
The value of the deposited energy is extracted from the classical field
simulation by taking the energy difference at the end of two evolutions
using the same value of
$|\Omega_p^+|^2 + |\Omega_p^-|^2$ but different relative magnitudes of
$\Omega_p^\pm$, $\Omega_p^-=0$ and $|\Omega_p^+|=|\Omega_p^-|$ respectively. This protocol aims at isolating the effect of the spatially modulated gauge potential: in the linear response limit, it is able to provide the exact value of $\Delta E_2$ alone. The effective normal fraction is then extracted from the deposited energy 
{\sl via} \eq{Delta2}, and it is plotted
in Fig.\ref{fig:numeric2}, as a function of the gauge field amplitude, for different values of the switch-off rate $\gamma/(c_s q)=
0.4,0.2,0.1$ (in black, in red,  in blue, from bottom to top) 
[$c_s=(\rho g/m)^{1/2}$ is the Bogoliubov sound velocity] and 
different geometries (empty squares and dashed lines for the spatially homogeneous case with periodic boundary conditions,
filled circles and solid lines for the circular potential well).

The $\epsilon_{\rm gauge}$-dependence in that figure allows to estimate the interval where the linear response approximation is reasonable, 
e.g. for $\epsilon_{\rm gauge}/\sqrt{mk_B T_d}\lesssim 0.08$ the deviation due to non-linear
effects is less than five percent. As expected, the estimate \eq{bound_gauge_amplitude} gives a more pessimistic bound around $0.04$
for $\gamma/(c_s q)=0.4$. 
From an experimental point of view, we expect that values of the gauge field amplitude as high as $\epsilon_{\rm gauge}/(m k_B T_d)^{1/2}=0.15$ should be well achievable, see Table~\ref{tab:val}.
As shown in the figure, for $\epsilon_{\rm gauge}/(m k_B T_d)^{1/2}\lesssim 0.2$, the effective normal fraction $f_n$ 
empirically has, within the error bars, a linear dependence with $\epsilon_{\rm gauge}^2$. 
This could be expected from the fact that
the first order correction to the linear response theory scales as $\epsilon_{\rm gauge}^4$
in the deposited energy $\Delta E$, the term of order $\epsilon_{\rm gauge}^3$ in $\Delta E$ 
being zero due to the odd parity of the matter-wave coupling to the gauge field. 
Assuming this linear dependence of the effective normal fraction,
a fit was performed over the range $(0.05)^2 \leq \epsilon_{\rm gauge}^2/(m k_B T_d) < (0.2)^2$, to extrapolate
to $\epsilon_{\rm gauge}=0$, see the solid and dashed lines in Fig.\ref{fig:numeric2}.  
Note that this extrapolation procedure also eliminates the contribution to the effective normal fraction
of the energy $\Delta E_{U^p}$ deposited by the undesired light-shifts, see the end of Appendix~\ref{app:experimental}.
For larger values of the gauge field amplitude $\epsilon_{\rm gauge}$, more serious non-linear effects set in
in the effective normal fraction  and it becomes less clear how
to extrapolate them down to $\epsilon_{\rm gauge}=0$. 

For the sake of completeness, it is important to note that for the weak gauge field amplitude that are required to be in the linear regime, 
or at least in the regime of the numerically suggested linear dependence of the effective normal fraction with
$\epsilon_{\rm gauge}^2$, 
the deposited energy $\Delta E_2$ is at most a few percent of the total energy of the system, 
which may be experimentally challenging to measure~\cite{foot_mesurer}.
This value is however larger than the statistical uncertainty of the energy in the canonical ensemble with $n_{\rm real}=1000$ realizations.
As explained in the Appendix \ref{app:noise},
such a large number of realizations turned out to be necessary in the small $\epsilon_{\rm gauge}$ regime to compensate the fluctuations 
of the current $\jj(t=0)$ at the initial time. The resulting statistical error for the effective normal
fraction is about $2\%$ for the parameters of Table~\ref{tab:val}. In the numerical results of 
Fig.\ref{fig:numeric2}, this issue was circumvented
by a simulation trick so that $n_{\rm real}=1000$ is a sufficiently large  number of realizations to keep error bars below
$2\%$ in the limit $\epsilon_{\rm gauge}\to 0$ \cite{foot_evolu}. 
It is also shown in the Appendix \ref{app:noise} that a measurement
of the {\sl variance} of the deposited energy (rather than its mean value) in principle also allows to access
the normal fraction, if one knows experimentally the initial energy of the gas.

After linear extrapolation to $\gamma=0$ from the lowest two values of $\gamma$ in Fig.\ref{fig:numeric2},
which gives $f_n\simeq 0.176$,
the residual disagreement of $f_n$ with the thermodynamic result
indicated by the dotted lines is about $14\%$, of the same order as the finite $q$  correction
predicted by Bogoliubov theory, see Fig.\ref{fig:bogo}a. We have thus performed simulations
for $q\xi=1/2$, for a spatially homogeneous system having the same parameters
as in Fig.\ref{fig:numeric2} and for a gauge field switch-off rate $\gamma/(c_s q)=0.2$ (not shown).
It is found that the range of linearity of the effective normal
fraction with $\epsilon_{\rm gauge}^2$ is of the same order, up to $\epsilon_{\rm gauge}/(m k_B T_d)^{1/2} \simeq 0.27$.
Repeating the zero-$\epsilon_{\rm gauge}$ extrapolation procedure as in Fig.\ref{fig:numeric2},
we find for $q\xi=1/2$ an effective normal fraction of $0.164(3)$, rather than $0.146(2)$ for $q\xi=1$.
Assuming a quadratic dependence of the effective normal fraction with $q$ at fixed $\gamma/(c_s q)$,
as predicted by Bogoliubov theory, see Fig.\ref{fig:bogo}b, this leads to an additive correction to $f_n$ 
close to $0.024$. We finally reach $f_n\simeq 0.20$, which is within the statistical error bars
of the thermodynamic value \cite{foot_finite_tau}.
On a smaller system with $M=32^2$ modes, we have reached a similar conclusion,  taking
even smaller values of $q \xi$ and $\gamma/(c_s q)$, 
see the blue circle in Fig.\ref{fig:numeric} and the figure caption. The deviation
observed in Fig.\ref{fig:numeric2} is therefore not a systematic error of the proposed method.

As a final check, we have tested the locality of the proposed measurement scheme by performing the simulation for two different geometries. 
Empty squares and dashed lines in Fig.\ref{fig:numeric2} correspond to a spatially homogeneous system with periodic boundary conditions, while the 
filled disks and solid lines correspond to a
system trapped in a circular well with steep walls of the form
$U(\rr)=\zeta k_B T\,\{\textrm{tanh}[(r-L/2)/(\xi/2)] +1\}$, where the numerical parameter
$\zeta$ determines the energy cut-off in the classical field model \cite{foot_zeta}. The probed
region is at the center of the potential well, $\rr_0=\mathbf{0}$.
Even in the non-linear regime, where linear response theory fails,
the effective normal fractions are almost the same in both
geometries.

\begin{table}[htbp]
\begin{tabular}{|c|c|c|}
\hline
2D density & $\rho \lambda_c^2 =9 $ & $\rho=14 \mu\textrm{m}^{-2}$ \\ \hline
\parbox[t]{3cm}{degeneracy \\ temperature} & $k_B T_d \equiv 2\pi \hbar^2 {\rho}/{m}$ & $T_d=500$ nK \\ \hline
temperature & $T=0.1\,T_d$ & $T=50$ nK \\  \hline
\parbox[t]{3cm}{2D interaction \\ constant} & $\tilde{g}=m g / \hbar^2$ 
& $\tilde{g}=0.1$ \\  \hline
\parbox[t]{3cm}{transverse \\ confinement} & \parbox[t]{3cm}{$\hbar \omega_z=0.23\,{\hbar^2 k_c^2}/{m}$\\$=0.16\,k_B T_d$} & $\frac{\omega_z}{2\pi}=1.65$ kHz \\  \hline
healing length & $\xi\equiv (\rho \tilde{g})^{-1/2}$ & $\xi=0.84 \mu m$ \\
\hline 
\end{tabular}

\vspace{1cm}
\begin{tabular}{|c|c|c|}
\hline 
\parbox[t]{3cm}{reduced gauge \\ field amplitude} & $\tilde{\epsilon}_{\rm gauge}\equiv \frac{\epsilon_{\rm gauge}}{(m k_B T_d)^{1/2}}$ &
$\tilde{\epsilon}_{\rm gauge}=0.15$ \\ \hline
\parbox[t]{3cm}{probe beam \\ Rabi frequencies} & $\frac{(|\Omega_p^+|^2+|\Omega_p^-|^2)_{t=0^+}}{2|\Omega_c|^2}$ & 
$0.09$ \\ \hline 
\parbox[t]{3cm}{gauge field\\switch-off rate} & \parbox[t]{3cm}{$\gamma=0.2\, c_s q$ \\ with $q=1/\xi$} & 
 \parbox[t]{2cm}{$1/\gamma=4.8$ ms \\ $q=0.15k_c$} \\ \hline
\end{tabular}

\vspace{1cm}
\begin{tabular}{|c|c|c|}
\hline level schemes of Fig.\ref{fig:Rb_levels}: & first choice & second choice \\ \hline
level $|a\rangle$  & $|F=1,-1\rangle$ & $|F=2,-2\rangle$ \\
\hline
level $|b\rangle$ & $|F=2,-2\rangle$ & $|F=1,-1\rangle$ \\
\hline
level $|e\rangle$ & $|F'=2,-1\rangle$ & $|F'=2,-2\rangle$ \\
\hline
\parbox[t]{4cm}{squared coupling Rabi frequency  ${|\Omega_c|^2}/{\Gamma^2}$} &  $0.21$ & 
$0.5$ \\
\hline \parbox[t]{4cm}{minimum detuning $\delta/\Gamma$} & $1$ & $1.5$ \\
\hline \parbox[t]{4cm}{fluorescence probability \\ per atom $P_{\rm fluo}$} & $0.22$ & $0.045$ \\
\hline \parbox[t]{4cm}{spurious deposited energy \\ ${\Delta E_U}/{\Delta E_2}$
(for $f_n=0.2$)}  & $33$
& $0.16$ \\
\hline
\end{tabular}
\caption{Suggested values of the physical parameters for an experimental measurement of the normal fraction of a two-dimensional
Bose gas of $^{87}\textrm{Rb}$ using an artificial gauge field produced by laser (coupling and probe beam) excitation on the D1 line 
with an optical wavelength $\lambda_c=795$ nm.
The first block characterizes the thermal equilibrium of the gas.
The second block determines the gauge field. The third block deals with the issues of spontaneous emission 
and spurious light-shift, for two atomic level schemes: the ``first choice"
is Fig.\ref{fig:Rb_levels}a, the ``second choice"
is Fig.\ref{fig:Rb_levels}b.
For this second choice, the indicated value $|\Omega_c|^2/\Gamma^2=0.5$ corresponds to a compromise
between minimization of the fluorescence and of the spurious light-shift, the truly minimal $P_{\rm fluo}$ being
$< 0.01$, see (\ref{eq:Pfluo_min}).
Note that $P_{\rm fluo}$ and ${\Delta E_U}/{\Delta E_2}$ are basically unchanged
if one takes  $q\xi=1/2$ and $\gamma/(c_s q)=0.4$.
The three-dimensional scattering length $a_{3D}\simeq 100$ Bohr radii 
is related to the two-dimensional coupling constant by Eq.\eq{g2D}. 
A useful relation is $m\Gamma/(\hbar k_c^2)\simeq 792$.
$c_s=(\rho g/m)^{1/2}$ is the Bogoliubov sound velocity.
\label{tab:val}
}
\end{table}

\section{Optical measurement}
\label{sec:optical}
The proposal that we have illustrated in the previous section was based on the measurement of atomic quantities, namely the deposited energy in the atomic cloud at the end of the gauge field sequence.
The present section is devoted to the presentation and the characterization of an alternative, all-optical route to measure the normal fraction $f_n$: information on the response of the atomic cloud to the gauge field can be retrieved from the transmitted probe beams once they have crossed the atomic cloud. Recent works have in fact pointed out that the strong frequency-dependence of the dielectric constant of an optically dressed medium in the electromagnetically induced transparency (EIT),  already used experimentally to strongly reduce the light group velocity \cite{Imamoglu_rev,Hau,Ketterle},
can be exploited for velocimetry experiments: Information on the current profile of an atomic cloud was predicted to be imprinted onto the phase of the transmitted probe beam~\cite{ohberg,ma-ic}. 

In the present case, the gas is illuminated by probe light of angular frequency $\omega_p$
and coupling light of angular frequency $\omega_c$. As in Sec.\ref{sec:deposited} the coupling light is
a plane wave propagating along $y$ axis.
In the subsection \ref{subsec:eftl}, where it is proposed to measure the transmitted mean electric field amplitude,
 the probe light consists, as in Sec.\ref{sec:deposited}, of two beams with wavevectors $\kk_p^{\pm}$ at a small angle with the $z$ axis, 
as sketched in Fig.\ref{fig:scheme}; differently from~\cite{ohberg,ma-ic}, the matter-wave
current pattern is generated by the same laser beams that are then used for probing.
The configuration is slightly different in the subsection \ref{subsec:cfatadol}, where it is proposed to measure the pattern of the scattered light intensity: in this case, the probe light consists of a single beam with a wavevector
$\kk_p$ directed along the $z$ axis, as sketched in Fig.\ref{fig:scheme_scatt}. Even though no spatially modulated matter-wave current is generated by the laser beams, still one can extract information on $f_n$ from the zero-mean  matter-wave currents induced by thermal fluctuations in the gas.

\subsection{Extracting $f_n$ from the amplitude of transmitted light}
\label{subsec:eftl}

The transmission and reflection of probe light from the two-dimensional atomic cloud can be described in terms of Maxwell equations. In particular, the dipole polarization of the atoms provides a source term for the 
probe electric field $\mathcal{E}_p$ at angular frequency $\omega_p$: for the positive frequency parts, one has
 in the paraxial approximation with respect to the $z$ axis:
\begin{equation}
(\Delta+k_p^2)\,\mathcal{E}_p=-\frac{k_p^2}{\epsilon_0}\mathcal{P}_p,
\eqname{maxw}
\end{equation}
where $k_p=\omega_p/c$ and the Laplacian operator is three-dimensional.
In the considered $\Lambda$ atomic configuration, see Fig.\ref{fig:scheme}, one may think that the atoms
occupy in each point of space the non-coupled state $|NC(\rr)\rangle$, leading to a vanishing
mean atomic polarization $\mathcal{P}_p$. This is actually not exactly the case: Due to the atomic
motion, the atomic internal state does not follow adiabatically and
there exists a small coupling between the position dependent coupled state $|C(\rr)\rangle$
and non-coupled state $|NC(\rr)\rangle$, the so-called motional coupling \cite{aspect}.
This leads to a small non adiabatic atomic polarization, that we now evaluate.

Within a perturbative picture, we simply need to calculate the mean atomic polarization $\mathcal{P}_p$ induced by the 
unperturbed laser fields. In terms of the three-dimensional atomic field operators, this reads 
\begin{equation}
\mathcal{P}_p(\rr)=d_{ae}\,\langle \Psihd_a(\rr)\,\Psih_e(\rr) \rangle.
\eqname{P_coh}
\end{equation}
With the usual adiabatic elimination of state $|e\rangle$, as shown in the Appendix \ref{app:adiab},
the atomic field in the excited state can be written
in terms of the atomic field operator $\chih_{3D}$ in the coupled $|C\rangle$ state,
\begin{equation}
\chih_{3D}(\rr,t)=\langle C(\rr,t)|a\rangle\,\Psih_a(\rr,t)+\langle C(\rr,t)|b\rangle\,\Psih_b(\rr,t)
\label{eq:def_chi3D}
\end{equation}
as  
\begin{equation}
\Psih_e \simeq 
\frac{\Omega_c}{2(\delta+i\Gamma/2)}\,[1+|{\Omega_p}/{\Omega_c}|^2]^{1/2}
\,\chih_{3D} +\Gamma^{1/2} \hat{B}_e.
\eqname{Psie}
\end{equation}
From the explicit form of the noise term $\hat{B}_e$ given in the Appendix \ref{app:adiab}, it is immediate to see that it gives a zero contribution to the mean in Eq.(\ref{eq:P_coh}). 
Since the atoms are mostly in the uncoupled $|NC\rangle$ state,  we can approximate the atomic field $\Psi_a$ in the $|a\rangle$ state by its $|NC\rangle$ component.
To lowest order in $\Omega_p/\Omega_c$ we then have
\begin{equation}
\mathcal{P}_p(\rr)\simeq \frac{d_{ae}\,\Omega_c}{2(\delta+i\Gamma/2)}\,\langle
 \phihd_{3D}(\rr)\,\chih_{3D}(\rr) \rangle.
\eqname{polar1}
\end{equation}
The next step is to perturbatively evaluate the field $\chih_{3D}$ that is created in the coupled $|C\rangle$ state by the 
motional coupling 
\cite{aspect} of $|C\rangle$ to $|NC\rangle$. The details of the procedure are given in the Appendix \ref{app:adiab}. 
To first order in $\Omega_p/\Omega_c$ and for the magic choice Eq.(\ref{eq:magic_choice}) of $\Delta_c$, one obtains after adiabatic elimination
\be
\hat{\chi}_{3D}\simeq 
-\frac{4\hbar(\delta+i\Gamma/2)}{m|\Omega_c|^2}
\nabla\,\phih_{3D}\cdot\nabla\,\frac{\Omega_p}{\Omega_c}
+\Gamma^{'1/2}\hat{B}_\chi.
\eqname{champ_chi}
\ee
An explicit expression for the noise term $\hat{B}_\chi$ is given in the Appendix \ref{app:adiab}: again, the noise term has a zero expectation value and does not contribute to the optical polarization. 
The final form of the optical polarization in terms of the three-dimensional 
atomic density and current operators \eq{n3d} and \eq{j} reads, to zeroth order in $q/k_p$:
\begin{multline}
\mathcal{P}_p(\rr)=-\frac{4\,|d_{ae}|^2}{\hbar\,|\Omega_c|^2}\,\times \\
\times (\kk_p-\kk_c)\cdot\left[ \langle \jj_{3D}(\rr) \rangle+
 \frac{\hbar}{2im}\,\nabla \langle n_{3D}(\rr) \rangle  \right] \,
\mathcal{E}_p(\rr),
\eqname{polar2}
\end{multline}
where the Rabi frequency of the probe beam has been eliminated in favor of the electric field using the 
definition $-d_{ea} \mathcal{E}_p=\hbar\Omega_p/2$ and the detuning $\delta$ has disappeared from the formula.
The first term proportional to the atomic current operator has a simple semi-classical interpretation in terms of the reduced group velocity in the Electromagnetically Induced Transparency regime, as anticipated in~\cite{ohberg,ma-ic}: in this regime, the refractive index strongly depends on the Raman detuning, which in turn depends on the atomic speed because of the Doppler effect.

However, the expression \eq{polar2} differs from the semi-classical one that was used in~\cite{ma-ic} in two ways. 
First, the current operator in Eq.\eq{polar2} differs from the physical current of atoms by the gauge field
\begin{equation}
\jj_{\rm phys} = \jj_{3D} - \frac{1}{m}\,n_{3D} \, \AAA_{3D}.
\end{equation}
As the proposal in~\cite{ma-ic} addressed a pre-existing current profile and the weak probe beam induced a vanishingly small gauge field, the difference was irrelevant in that case. Here, on the contrary, the mean current is itself proportional to the gauge field so that the difference between the two operators really matters.
Second, the expression \eq{polar2} contains an extra term proportional to the average density gradient. 
In contrast to the first term, this one is purely imaginary. As a result, it only affects the intensity of the transmitted light 
{\sl via} a combination of absorption and/or amplification effects. In particular, as it does not induce any phase shift on the light, it does not interfere with the proposal of~\cite{ma-ic}.

In order to calculate the modification $\delta\mathcal{E}_p$ induced by
the atoms on the transmitted electric field of the probe, one has to
insert the polarization \eq{polar2} as a source term into the Maxwell
equation \eq{maxw}.  
Within a standard approximation, we can neglect diffraction effects stemming from the in-plane part of the Laplace operator in \eq{maxw} and integrate the $z$ dependence across the atomic cloud. 
Taking into account the appropriate boundary conditions for $\delta\mathcal{E}_p$,
this leads to the expression
\begin{equation}
\delta\mathcal{E}_p(x,y,z)\simeq e^{ik_p z}\,\frac{ik_p}{2\epsilon_0}\,
\int_{-\infty}^{\infty} dz'\,e^{-ik_pz'}\,\mathcal{P}_p(x,y,z')
\end{equation}
for the transmitted field in the $z>0$ region. In order for the approximation to be accurate, $z$ has to be much larger than the thickness 
$a^z_{\rm ho}$ of the atomic pancake, but at the same time much smaller than the diffraction length $k_p/q^2$, 
where $q$ is the characteristic wavevector of the in-plane modulation of the atomic density and current.

Along $z$, the atomic field varies as the harmonic oscillator ground state wavefunction $\phi_0(z)$,
see Eq.(\ref{eq:def_phi}). Performing the integral over $z'$, this gives the final expression for the variation of the transmitted field 
\begin{equation}
\delta\mathcal{E}_p(\rr)= 
 \frac{2ik_p\,|d_{ae}|^2}{\hbar\epsilon_0\,|\Omega_c|^2}\,\kk_c
 \cdot\left[\langle \jj(\rr) \rangle+ \frac{\hbar}{2im}\,\nabla \langle
 n(\rr) \rangle  \right] \mathcal{E}_p(\rr)
\eqname{transm_var}
\end{equation}
in terms of the two-dimensional density $n(\rr)$ and current $\jj(\rr)$
operators. The first contribution proportional to the atomic current gives a phase
shift, while the second contribution proportional to the atomic density
gradient is responsible for absorption and amplification of the probe
beam.

The atomic current profile created by the gauge field is evaluated using the linear response formulas \eq{chi} and \eq{f_n} as discussed in detail in the previous sections.  
The gauge field is assumed to be switched on slowly enough as compared to the characteristic frequencies of all the excitation modes of the gas at wavevector $\qq$. 
Within the linear response regime, the contribution to the current due to the spatially modulated gauge field at $\qq$ may be isolated by a suitable combination of measurements with different values of $\Omega_p^\pm$, which gives 
\begin{multline}
\langle\jj\rangle_2(\rr)=\frac{\hbar \rho k_c}{m q |\Omega_c|^2} 
(\qq\cos\alpha + f_n \mathbf{e}_z\times \qq \sin \alpha) \\
\times \left( \Omega_p^+ \Omega_p^{-*}\,e^{i\qq\cdot \rr}
+\Omega_p^{+*} \Omega_p^- \,e^{-i\qq\cdot \rr}
\right) e^{-2|\rr-\rr_0|^2/w^2},
\end{multline}
where $\rr$ is now in the $xy$ plane and $\alpha$ is the oriented angle that $\kk_c$ makes with $\qq$ \cite{apropos_alpha}. 
Inserting this expression into \eq{transm_var} and recalling the form \eq{omega_p} of the incident probe field, 
one can extract the phase shift experienced by the central part of the probe beams after crossing the atomic pancake,

\begin{equation}
\Delta \phi_2^\pm=\frac{6\pi\hbar\rho}{m}\frac{|\Omega_p^\mp|^2}{|\Omega_c|^4}\,B\Gamma\,
[\cos^2\alpha+f_n\sin^2\alpha],
\eqname{deltaphi}
\end{equation}
where we have assumed $k_p\simeq k_c$. The total decay rate of the $e$ state by spontaneous emission is indicated by $\Gamma$ and $B$ is the branching ratio for the decay to the $a$ state, so that $B \Gamma=|d_{ae}|^2 k_c^3/(3\pi\hbar\epsilon_0)$. 
As we have already mentioned above, the density gradient term in \eq{transm_var} only introduces an intensity modulation and is not responsible for any phase shift.

From a nonlinear optics point of view, the phase shift \eq{deltaphi} can be interpreted as arising from a $\chi^{(3)}$ optical nonlinearity of opto-mechanical origin similar to the one that was demonstrated in the experiment~\cite{esslinger_cavity}: the nonlinear modulation of the optical response of the atoms is determined by the mechanical distortion of the cloud by the optical forces. 

Inserting into \eq{deltaphi} the values of Table~\ref{tab:val} for the $^{87}$Rb case,
for the first choice of level scheme reported in the Appendix \ref{app:experimental},
 one has a branching ratio $B=1/4$
and one finds a small, yet appreciable phase shift on the order of  
\begin{equation}
\Delta\phi_2^{\rm choice\ 1}\simeq 6\cdot 10^{-4}\times [\cos^2\alpha+f_n\sin^2\alpha].
\end{equation}
For the second choice of level scheme in the Appendix \ref{app:experimental}, the branching ratio is slightly larger, $B=1/3$, but
for the compromise choice Eq.(\ref{eq:compromise}) $|\Omega_c|^2/\Gamma^2$
is larger so that one finds a smaller phase shift
\be
\Delta\phi_2^{\rm choice\ 2}\simeq 3\cdot 10^{-4} \times [\cos^2\alpha+f_n\sin^2\alpha].
\ee

In addition to the phase shift of the transmitted beam that we have discussed so far, Bragg diffraction on the spatially modulated current profile produces a pair of additional beams of in-plane wavevector respectively $\pm 3\qq/2$ {\sl via} a sort of four-wave mixing process. 
The relative intensity of these beams as compared to the incident probe beams is on the order of $|\Delta\phi_2|^2$.
For transverse gauge fields such that $\qq\cdot \kk_c=0$ the contribution of the induced density gradient term of \eq{transm_var} vanishes by symmetry. In the case of longitudinal gauge fields, the relative correction is on the order of $q\xi$.


\subsection{Current fluctuations and the angular distribution of scattered light}
\label{subsec:cfatadol}

All the calculations presented in the previous subsection
aimed at evaluating the expectation value of the transmitted field amplitude.
At this level of the theory, we were allowed to describe the probe beam as a coherent, classical field and we could neglect the fluctuations around the expectation value of both the light field amplitude and the atomic current and density operators.
The formalism can be straightforwardly extended to quantum optical fields so to include the fluctuations of the atomic density and current. This is crucial when one aims at investigating the spontaneous scattering of light off the current fluctuations in the atomic gas. In this subsection, we shall in particular show how information on the normal fraction of the gas can be inferred from the angular distribution of scattered light. We shall make the approximation of replacing temporal derivatives of the electromagnetic field
$\partial_t \mathcal{E}$ by $-i c k_p \mathcal{E}$ in Maxwell's equation. In particular, this misses retardation effects in the expression
of the scattered fields in terms of the atomic dipoles, which is accurate since the system size is much smaller than $c/\Gamma$.

\begin{figure}[htbp]
\includegraphics[width=0.6\columnwidth,clip]{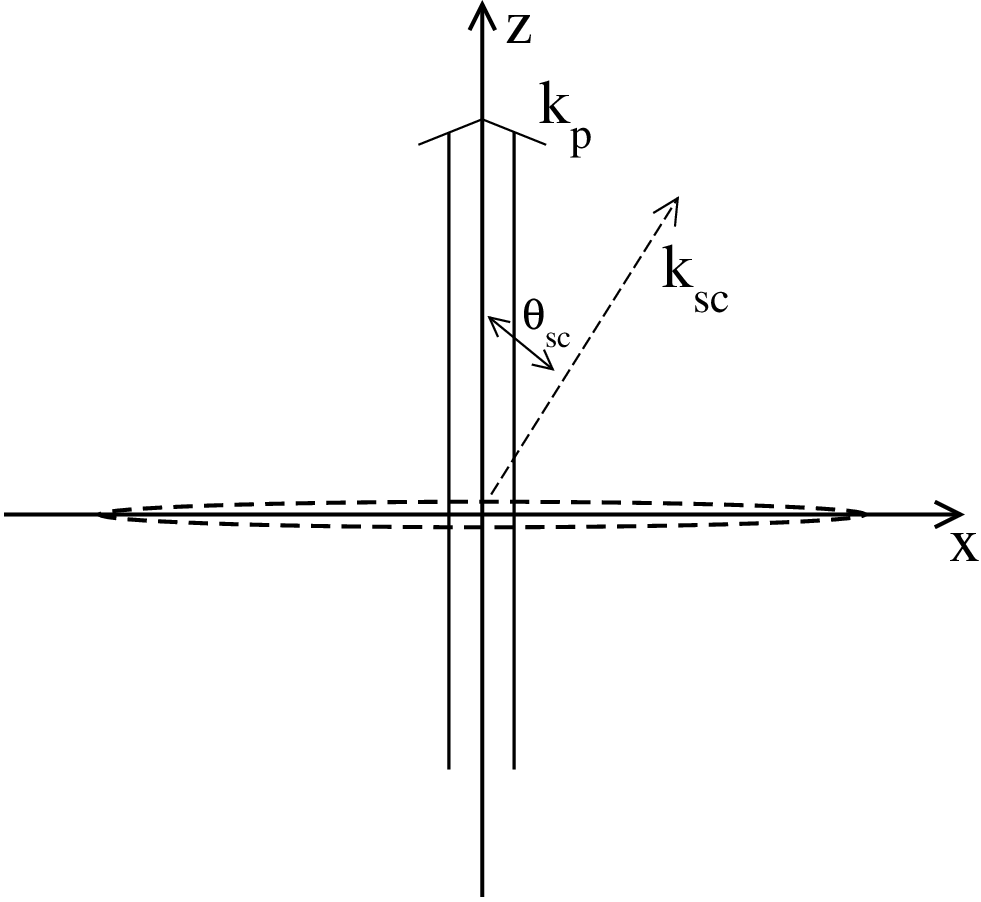}
\caption{Scheme of the scattering geometry under examination. Probe light is incident at wavevector $\kk_p=k_p\eee_z$ and the scattered light is collected at a wavevector 
$\kk_{\rm sc}=(k_p^2-Q^2)^{1/2}\,\eee_z+\QQ$.}
\label{fig:scheme_scatt}
\end{figure}

We consider the geometry sketched in Fig.\ref{fig:scheme_scatt}: a single Gaussian probe beam is incident onto the atoms with a wavevector $\kk_p$ exactly orthogonal to the atomic plane, waist $w$ 
centered at $\rr_0$ and a weak peak amplitude $\mathcal{E}_p^0$. 
Inserting this form into \eq{transm_var} and taking the Fourier transform along the $xy$ plane, one obtains the following operator equation for the scattered field component at in-plane wavevector $\QQ$ \cite{precision},
\begin{multline}
\mathcal{E}(\QQ,z)\simeq \frac{2i\, k_p\,|d_{ae}|^2}{\epsilon_0\,\hbar\,|\Omega_c|^2}\,\pi\,w^2\,\mathcal{E}_p^0\,\int\!\frac{d^2q}{(2\pi)^2} \,e^{-q^2 w^2 /4} \, e^{-i \qq\cdot\rr_0}
\\\times \left\{\kk_c \cdot\left[ \jj_{\QQ-\qq} + \frac{\hbar (\QQ-\qq) }{2m}\,n_{\QQ-\qq}\right]\right\}\,e^{ik_z(\QQ)\,z}.
\label{eq:EQQz}
\end{multline}
Here, $\jj_\QQ$ and $n_\QQ$ are the spatial Fourier transforms of the two-dimensional current $\jj(\rr)$ and density $n(\rr)$ operators; the Fourier transform of a product of two functions has been rewritten in terms of the convolution of their Fourier transforms.
The $z$ component of the propagation wavevector is determined by the photon dispersion as $k_z(\QQ)=(k_p^2-Q^2)^{1/2}$.

The intensity of the scattered light at in plane wavevector $\QQ$ is quantified by \cite{precision}
\begin{multline}
\langle \mathcal{E}^\dagger(\QQ) \mathcal{E}(\QQ)
 \rangle = 
\left[\frac{2\pi
 k_p\,|d_{ae}|^2\,w^2\,|\mathcal{E}_p^0|}{\epsilon_0\,\hbar\,|\Omega_c|^2}\right]^2\times \,\\
\int\!\frac{d^2q}{(2\pi)^2} \,\int\!\frac{d^2q'}{(2\pi)^2}
\left\langle 
\left\{\kk_c\cdot \left[\jj^\dagger_{\QQ-\qq} + \frac{\hbar (\QQ-\qq) }{2m}\,n^\dagger_{\QQ-\qq}\right]\right\} \right.\\ 
\times\left.\left\{\kk_c\cdot \left[\jj_{\QQ-\qq'} + \frac{\hbar (\QQ-\qq') }{2m}\,n_{\QQ-\qq'}\right]\right\}
\right\rangle \\
\times  e^{i(\qq-\qq')\cdot\rr_0} e^{-(q^2+{q'}^2)\, w^2 /4}.
\eqname{scatt}
 \end{multline}
Since the system size is much larger than the waist $w$ of the probe beam, we can for simplicity assume an effective translational symmetry along the $xy$ plane. As a consequence, the correlation function that appears in \eq{scatt} has a delta-distribution
shape around equal wavevectors $\QQ-\qq=\QQ-\qq'$ (see e.g. the next Eq.\eq{j-j-corr}).

In contrast to the schemes proposed in the previous sections, where the duration $1/\gamma$ of the experiment had to be at least on the order of $1/(c_s q)$, 
where $c_s=(\rho g/m)^{1/2}$ is the Bogoliubov sound velocity,
the light scattering experiment discussed here can be performed on a much faster time scale, only limited by the characteristic rate $\Gamma'$ of the internal atomic evolution time, Eq.\eq{dchidt}. 
As a result, the experiment can be performed in the small wavevector region $Q\xi \ll 1$ where the contribution to \eq{scatt} of the terms involving the density fluctuations $n_\QQ$ is negligible~\cite{foot_petitQ}.
Of course, efficient isolation of the scattered light from the incident beam requires that the scattering angle $\theta_{\rm sc}\simeq Q/k_p$ be much larger than the diffraction cone of the probe beam, i.e. $Q \gg 1/w$.

The instantaneous correlation function of the current in the $y$ direction parallel to $\kk_c$ can be evaluated applying the fluctuation-dissipation relation \eq{correl} to the current operator $\jj_\qq$ in an infinite space geometry. This gives 
\begin{multline}
\langle j_{\QQ,y}^\dagger\,j_{\QQ',y} \rangle=(2\pi)^2\,\delta^2(\QQ-\QQ')
 k_B T \,\textrm{Re}[\chi_{yy}(\QQ',\omega=0)] \\
\underset{Q\to 0}{\simeq}
(2\pi)^2\,\delta^2(\QQ-\QQ') \frac{k_B T \,\rho(\rr_0)}{m}\\
\times [\cos^2\phi_{\rm sc}+f_n\sin^2\phi_{\rm sc}],
\eqname{j-j-corr}
\end{multline}
where $\phi_{\rm sc}$ is now the azimuthal angle between $\kk_c$ and $\QQ$. 
Inserting this expression into \eq{scatt} and taking the thermodynamic limit, one gets to the final expression for the scattered intensity in the momentum $\QQ$-space
\cite{precision},
\begin{multline}
\langle \mathcal{E}_\QQ^\dagger \mathcal{E}_\QQ
 \rangle = 
\left[\frac{ k_p\,k_c\,|d_{ae}|^2}{\epsilon_0\,\hbar\,|\Omega_c|^2}\right]^2\,2\pi w^2\,|\mathcal{E}_p^0|^2\\
\times\frac{ k_B T \, \rho(\rr_0)}{m}\,
[\cos^2\phi_{\rm sc}+f_n\sin^2\phi_{\rm sc}].
\eqname{scatt3}
 \end{multline}
To estimate the relative intensity of scattered light, it is useful to rewrite the expression \eq{scatt3} for the momentum space intensity $\langle \mathcal{E}_\QQ^\dagger \mathcal{E}_\QQ\rangle$ in terms of physically more transparent quantities such as the angular distribution $I(\theta_{\rm sc},\phi_{\rm sc})$ of scattered intensity.
For small scattering angles $|\theta_{\rm sc}|\ll 1$, the infinitesimal solid angle and momentum space volume elements are related by 
$d\Omega=\sin\theta_{\rm sc}\,d\theta_{\rm sc}\,d\phi_{\rm sc}\simeq \theta_{\rm sc}\,d\theta_{\rm sc}\,d\phi_{\rm sc}\simeq d^2 Q/k_p^2$, 
so that 
\begin{equation}
I_{\rm sc}(\theta_{\rm sc},\phi_{\rm sc})\simeq \frac{k_p^2}{(2\pi)^2}\,
\langle \mathcal{E}_\QQ^\dagger \mathcal{E}_\QQ \rangle. 
 \eqname{conversion}
\end{equation}
This immediately leads to the final expression for the angular distribution of the scattering intensity \cite{precision}
\begin{multline}
\frac{I_{\rm sc}(\theta_{\rm sc},\phi_{\rm sc})}{I_{\rm inc}}=\left[\frac{k_p^2\,k_c\,|d_{ae}|^2}{\pi \epsilon_0\,\hbar\,|\Omega_c|^2}\right]^2 
 \frac{k_B T \rho(\rr_0)}{m}
\times \\ [\cos^2\phi_{\rm sc}+f_n\sin^2\phi_{\rm sc}]
  \eqname{relat_scatt_int}
\end{multline}
in units of the incident intensity,
\begin{equation}
I_{\rm inc}=\int d^2r\, |\mathcal{E}_p(\rr)|^2= \frac{\pi w^2}{2}\,|\mathcal{E}_p^0|^2.
\label{eq:incident_intensity}
\end{equation}
From this expression, it is immediate to see that information on the normal fraction of the gas can be retrieved from the azimuthal dependence of the scattered intensity. In terms of the total spontaneous emission decay rate $\Gamma$ of the $e$ state and the 
$|e\rangle \to |a\rangle$ branching ratio $B$
(we recall that $B \Gamma=|d_{ae}|^2 k_c^3/(3\pi\hbar\epsilon_0)$), Eq.\eq{relat_scatt_int} can be rewritten in the more transparent form \cite{precision},
\begin{multline}
\frac{I_{\rm sc}(\theta_{\rm sc},\phi_{\rm sc})}{I_{\rm inc}}=\frac{9 k_B^2 T T_d}{2\pi\hbar^2} 
\frac{B^2 \Gamma^2}{|\Omega_c|^4} [\cos^2\phi_{\rm sc}+f_n\sin^2\phi_{\rm sc}]. 
  \eqname{relat_scatt_int2}
\end{multline}

To estimate the feasibility of the proposed light scattering experiment, we now derive an upper bound on the number of useful scattered photons
in a single shot of duration $\tau$. Calculating the Poynting's vector of the probe beam, and using (\ref{eq:incident_intensity}), we find an incident flux
of probe photons $\Phi_{\rm inc}=(k_p w)^2 |\Omega_p^0|^2/(12 B \Gamma)$. Integrating in (\ref{eq:relat_scatt_int}) the term proportional to $f_n$ 
over solid angles in the cone $\theta_{\rm sc}\leq 1/(k_p \xi)$, we obtain the flux $\Phi_{\rm sc}^{\rm use}$ of useful scattered photons.
As a maximal duration, we take $\tau=1/\Gamma_{\rm fluo}^{\rm non-ad}$ where the fluorescence rate of the atoms
due to motional coupling between the non-coupled and the coupled states is given by Eq.(\ref{eq:gam_non_adiab})
(with $2|\Omega_p^+|^2$ replaced here with $|\Omega_p^0|^2$). The number of single shot useful scattered photons is thus bounded by
\be
N_{\rm ph}^{\rm use} \leq \frac{3\pi B}{16} \frac{k_B T}{\hbar\omega_z} \rho w^2 f_n \frac{1}{(k_c \xi)^2}.
\label{eq:upper_phot}
\ee
Remarkably the Rabi frequencies $\Omega_c$ and $\Omega_p^0$ have canceled out in the ratio of the scattered flux to  the fluorescence rate.
One recognizes in the right-hand side of (\ref{eq:upper_phot}) the effective mean number of atoms $N_{\rm at}^{\rm eff}=\pi\rho w^2 f_n/4$ in the normal component illuminated by the probe
beam, as in Eq.(\ref{eq:Delta2}). There is however a severe geometrical reduction factor, $1/(k_c \xi)^2$, due to the small aperture of the useful
scattering cone.
For the parameters of Table~\ref{tab:val}, with $B=1/3$, and taking a waist $w=30\xi$ and $f_n=0.2$ as in Fig.\ref{fig:numeric2}, 
we find $N_{\rm at}^{\rm eff}\simeq 1400$, $1/(k_c \xi)\simeq 0.15$, which leads to $N_{\rm ph}^{\rm use}\leq 5$. This remains
accessible to current quantum optics experiments. For $k_B T/\hbar \omega_z$
fixed, the upper bound in 
Eq.(\ref{eq:upper_phot}) scales as $\rho^2$, since $1/\xi^2$ scales as $\rho$,
so that larger values of photon numbers for a given waist 
may be obtained by increasing the density $\rho$ of the two-dimensional Bose gas.

\section{Conclusions and perspectives}
\label{sec:conclu}

In this paper we have proposed and validated two methods to measure the superfluid fraction of a quantum fluid of ultracold atoms.
The idea is to apply an artificial gauge field to the atoms with spatial oscillations within a localized envelope and to detect, within
the linear response regime,  
the matter-wave current pattern that is generated in the fluid. This can be done in a mechanical way by measuring the energy that is deposited in the fluid at the end of a gauge field sequence. This can also be done in an all-optical way by observing the phase shift experienced by the same laser beams that are used to generate the artificial gauge field or the angular pattern of scattered light.
We have shown that, by a careful choice of the parameters and of the atomic level scheme for $^{87}\textrm{Rb}$,
two experimental obstacles, the spontaneous emission and (in the case of the mechanical method) undesired light-shifts,
can be put to an acceptable level. The experimental challenge remains in the required high sensitivity
of the measurements, that is a detection of small energy changes in the mechanical method, and a detection
of small optical phase shifts or small photon numbers in the all-optical methods.

The interest of the proposed methods is twofold: they do not require that the gas reaches thermal equilibrium in presence of the gauge field,
and furthermore they give the possibility of reconstructing in a local way the spatial profile of the superfluid fraction of a trapped gas,
independently from the presence or the absence of a Bose-Einstein condensate. 
This last feature is attractive in the study of the Berezinskii-Kosterlitz-Thouless transition to a superfluid state in two-dimensional Bose gases and of the superfluidity properties of Bose gases in disordered environments.  It would also be interesting to
extend the method to the study of superfluidity in multi-component atomic fermionic gases, which may require identification of suitable level schemes. 

\acknowledgments

We acknowledge useful discussions with G. La Rocca, S. Stringari, F. Gerbier, C. Salomon and J. Dalibard.
The work of Y.C. was done as part of the ERC Project FERLODIM N.228177. 
Y.C. is a member of IFRAF and acknowledges financial support from IFRAF.
I.C. acknowledges financial support from ERC through the QGBE grant.

\appendix

\section{Experimental issues}
\label{app:experimental}

\begin{figure*}[htbp]
\includegraphics[width=0.9\columnwidth,clip]{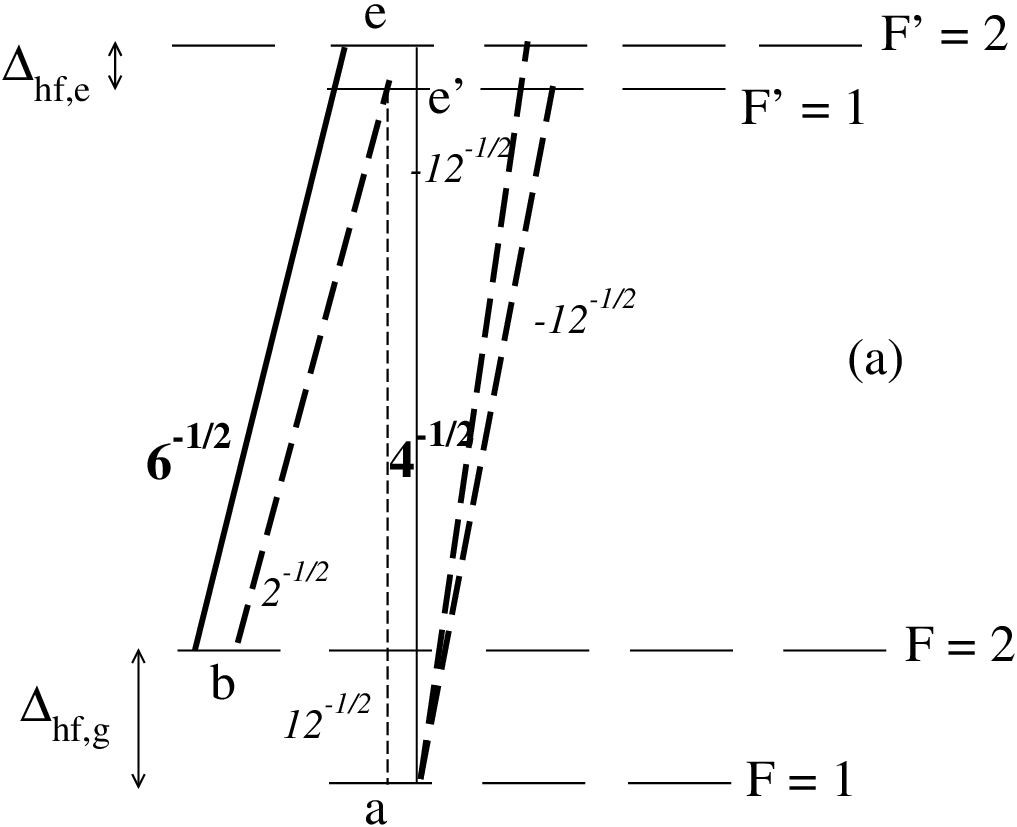}
\hspace{1cm}
\includegraphics[width=0.9\columnwidth,clip]{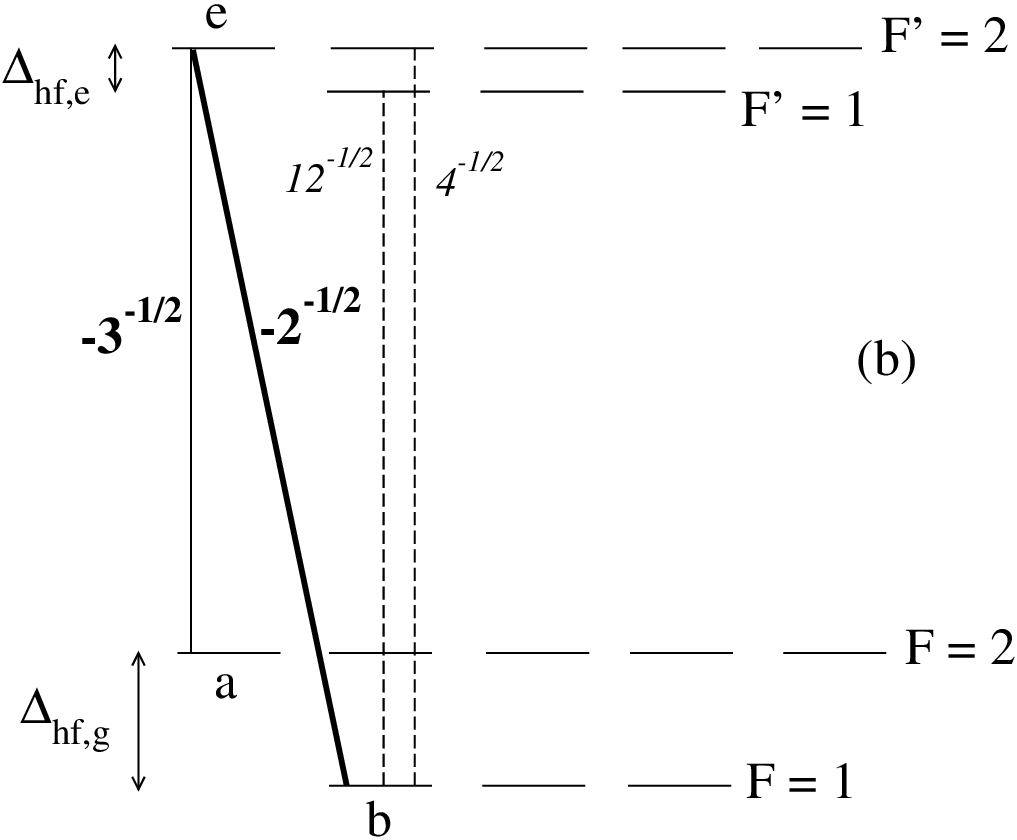}
\caption{Scheme of the internal levels of $^{87}\textrm{Rb}$ atom involved in the $D_1$ transition ($J=1/2\to J'=1/2$). 
The hyperfine splitting of the ground (excited) state is $\Delta_{{\rm hf,g}}=2\pi\times 6.834\,\textrm{GHz}$ 
($\Delta_{{\rm hf,e}}=2\pi\times 814\,\textrm{MHz}$). The natural linewidth of the excited state
is $\Gamma=2\pi\times 5.75\,\textrm{MHz}$. Thin (thick) lines indicate transitions that are induced by the 
probe (coupling) beams. Solid (dashed) lines indicate the desired (main undesired) transitions. 
The relevant dipole matrix elements are shown in units of the reduced dipole element $\langle J=1/2||er||J'=1/2\rangle$ of the D1 line, in boldface for the desired transitions and in italic for the undesired ones.
Our first choice for the $\Lambda$ system $|a\rangle,|e\rangle,|b\rangle$ states is shown in the left (a) panel, with
$|a\rangle=|F=1,m_F=-1\rangle$, $|e\rangle=|F'=2,m_{F'}=-1\rangle$, $|b\rangle=|F=2,m_F=-2\rangle$.
The second choice is shown in the right (b) panel, with $|a\rangle = |F=2,m_F=-2\rangle$, $|e\rangle=|F'=2,m_{F'}=-2\rangle$, $|b\rangle=|F=1,m_F=-1\rangle$. Note that we have taken the $y$ axis as the quantization axis of angular momenta.
}
\label{fig:Rb_levels}
\end{figure*}

In this Appendix we review some issues that may hinder an experimental implementation of our proposal.
Our attention will be concentrated onto the most relevant case of $^{87}\textrm{Rb}$ atoms considered in the experiment of \cite{Cornell1,Cornell2} and, in a two-dimensional context, in \cite{Dalibard2D}.  The level structure of this atomic species is sketched in Fig.\ref{fig:Rb_levels}. A possibly important advantage of this atom in view of an experimental implementation of the present proposal is that the singlet and triplet scattering lengths of ground state atoms are equal within a few percent \cite{theory_ratio}.
We thus expect that all scattering lengths between arbitrary $F=1$ or $F=2$ sublevels of the ground state have all almost the same values~\cite{RMP_Fesh}, which leads to $g_{aa}\simeq g_{ab}$ and therefore to a suppression of the spatio-temporal variation of the effective interaction constant $g_{3D}(\rr,t)$ defined in \eq{g}. 
In the deposited energy method proposed in Sec.\ref{sec:deposited}, this is important to reduce 
the emission of phonons in the atomic gas by the temporal modulation of the interaction constant. In the optical detection scheme of Sec.\ref{sec:optical}, this is also important to suppress the contribution of the interaction term to the field $\hat{\chi}_{3D}$ and then to the optical polarization.
Other atomic species such as Yb~\cite{GerbierDalibard} or metastable He \cite{aspect_manip,leduc_manip} and/or different laser beam configurations~\cite{spielman_th,spielman_exp,dalibard_rev} are expected to be useful for other purposes, e.g. to suppress spontaneous emission and/or generate artificial gauge fields with different geometries~\cite{foot_un_un}.

Two possible choices for the three states $|a\rangle, |b\rangle, |e\rangle$ forming the $\Lambda$ system on the $D1$ line
of $^{87}\textrm{Rb}$ are considered, as sketched in the two panels of Fig.\ref{fig:Rb_levels}.
For each choice, we determine the undesired effects (spontaneous emission, light-shifts, Raman leaks) stemming from deviations from the perfect adiabatic following of the non-coupled state by the moving atoms and from optical transitions to other levels not included in the $\Lambda$ system. An eye will also be kept on trying to maximize the $|e\rangle$ to $|a\rangle$ branching ratio so as to reinforce the optical signal of Sec.\ref{sec:optical}. To minimize spontaneous emission within the low saturation regime, we shall allow for a small detuning $\delta$ of both the probe and coupling beam carrier frequencies from the $|a\rangle\to |e\rangle$ and $|b\rangle\to |e\rangle$ transitions, respectively. The Raman detuning of the two beams is taken in a way to always fulfill the magic Raman condition \eq{magic_choice}.

\subsubsection{First choice}

The coupling beam propagates along the $y$ axis and is taken with a $\sigma_+$ polarization with respect to the $y$ quantization
axis. 
The probe beam is taken as linearly polarized along $y$. For the three atomic levels forming the $\Lambda$ system, we take $|a\rangle \equiv |F=1, m_F=-1\rangle$,  $|b\rangle\equiv |F=2,m_F=-2\rangle$ and $|e\rangle \equiv |F'=2,m_{F'}=-1\rangle$~\cite{foot_this_choice}.
This scheme of levels and lasers is illustrated in the left panel (a) of Fig.\ref{fig:Rb_levels}.
To estimate the importance of the non-adiabatic coupling between the $\ket{NC}$ and the $\ket{C}$ states due to the atomic motion, 
we can evaluate the ratio $\rho_C/\rho_{NC}$ of the two-dimensional densities in the two states close to the center of the laser spot.
This is done using the explicit formula \eq{champ_chi} for the field in the $\ket{C}$ state
\cite{foot_effet_ft}.
Using the fact that for quasi-2D samples the gradient is mostly along the harmonically trapped $z$ direction and assuming $\Omega_p^+=\Omega_p^-$, we obtain for
$k_p\simeq k_c$: 
\be
\frac{\rho_C}{\rho_{NC}}\simeq 
\left|\frac{4(\delta+i\Gamma/2) k_c\,\Omega_p^+}{\Omega_c^3} \right|^2\,\frac{\hbar\omega_z}{m}
.
\ee
Accuracy of the adiabatic approximation requires that this ratio is much smaller than unity.

The finite population that is present in the $\ket{C}$ state as a consequence of non-perfect adiabaticity is responsible for the spontaneous emission of photons 
at a  single atom rate:
\begin{equation}
\Gamma_{{\rm fluo}}^{{\rm non-ad}}=\Gamma' \frac{\rho_{NC}}{\rho_C} = 4\Gamma \frac{\,|\Omega_p^+|^2}{|\Omega_c|^4} \,\frac{\hbar k_c^2}{m}\omega_z,
\label{eq:gam_non_adiab}
\end{equation}
where the fluorescence rate $\Gamma'$ of the coupled state $|C\rangle$ is defined by Eq.(\ref{eq:def_gamma}) and a spatial average has been performed. Remarkably, $\Gamma_{{\rm fluo}}^{{\rm non-ad}}$ does not depend on the detuning $\delta$.

Other contributions to the fluorescence rate come from non-resonant excitation processes. 
Dominating among these are the excitation of the $|a\rangle$ state to the  states $\ket{F'=1\ \textrm{or}\ 2,m_{F'}=0}$ by the coupling beam 
at a rate 
\begin{equation}
\Gamma_{{\rm fluo}}^{c}=\Gamma\,\frac{|\Omega_c|^2}{4\Delta_{{\rm hf,g}}^2},
\end{equation}
and the excitation of the non-coupled state $|NC\rangle$ to $|e'\rangle=|F'=1,m_{F'}=-1\rangle$ by the total probe plus coupling field
with an effective Rabi frequency $-2\Omega_p/\sqrt{3}$, which results in the fluorescence rate on the parasitic $\Lambda'$ configuration
$|a\rangle\rightarrow |e'\rangle \rightarrow |b\rangle$:
\begin{equation}
\Gamma_{{\rm fluo}}^{\Lambda'}=\frac{2}{3} \Gamma\,\frac{|\Omega_p^+|^2}{\Delta_{{\rm hf,e}}^2}.
\end{equation}
In these expressions, we have taken into account the tabulated hyperfine dipole matrix elements of the various optical transitions, and we have introduced the hyperfine splittings $\Delta_{\rm hf,g}$ and $\Delta_{\rm hf,e}$ given for $^{87}\textrm{Rb}$ in the caption of Fig.\ref{fig:Rb_levels}.
Limiting ourselves to the most relevant regime where $|\Omega_p^+/\Omega_c|^2 > 1/100$, we see that in the present case of $^{87}\textrm{Rb}$ atoms, $\Gamma_{{\rm fluo}}^{\Lambda'} \gg \Gamma_{{\rm fluo}}^{c}$.

The total fluorescence rate can then be approximated as the sum of $\Gamma_{{\rm fluo}}^{\Lambda'}$ and $\Gamma_{{\rm fluo}}^{{\rm non-ad}}$. For a given value of the gauge field (proportional to $|\Omega_p^+/\Omega_c|^2$), the total fluorescence rate is minimized to
\begin{equation}
\Gamma_{{\rm fluo}}^{{\rm min}}\simeq 
\frac{4\sqrt{6}}{3} 
\left|\frac{\Omega_p^+}{\Omega_c} \right|^2
\frac{\Gamma}{\Delta_{{\rm hf,e}}}\,
\left(\frac{\hbar k_c^2\omega_z}{m}\right)^{1/2}
\end{equation}
by a careful choice of the coupling beam Rabi frequency
\begin{equation}
\left|\Omega_c^{\rm opt}\right|^2 =\Delta_{{\rm hf,e}}\,\sqrt{\frac{6\hbar k_c^2\,\omega_z}{m}}=\frac{\sqrt{3}\Delta_{\rm hf,e} }{2\pi^{1/2}} 
\frac{\hbar k_c}{ma_{3D}} \tilde{g}.
\eqname{Omega_c_opt}
\end{equation}
Here, we have expressed $\omega_z$ in terms of the three-dimensional scattering length $a_{3D}$ and the reduced two-dimensional coupling constant as given by Table~\ref{tab:val} and Eq.\eq{g2D}.
Inserting the actual parameters of the $^{87}\textrm{Rb}$ atom, 
and taking $\tilde{g}=0.1$, we obtain $|\Omega_c^{\rm opt}|^2/\Gamma^2\simeq 0.21$.
It remains to adjust the detuning $\delta$ to be in the weak saturation
regime,
\be
s \equiv \frac{|\Omega_c|^2/2}{|\delta + i\Gamma/2|^2} \lesssim \frac{1}{10}.
\ee
Let us take the same values of the gauge field sequence as in Fig.\ref{fig:numeric2}: $\gamma/(c_s q)=0.2$ and $q\xi=1$ and $\epsilon_{{\rm gauge}} \simeq 0.15 (m k_B T_{{\rm d}})^{1/2}$, 
$c_s=(\rho g/m)^{1/2}$ being the Bogoliubov sound velocity. 
This choice of $\epsilon_{{\rm gauge}}$ leads to $|\Omega_p^+(0^+)/\Omega_c|^2\simeq 0.03 (\rho\lambda_c^2)^{1/2}$: as this quantity has to be much smaller than 1 in order for the gauge field description of Sec.\ref{sec:gauge} to be valid, it is safe to impose $\rho \lambda_c^2 < 10$.
Integrating over the exponential switch-off ramp of $\Omega_p$ and eliminating $g$ in terms of $\omega_z$ and $a_{3D}$, this gives for the total fluorescence probability per atom,
\begin{equation}
P_{\rm fluo} = \frac{\sqrt{6}}{2} \frac{\Gamma}{\Delta_{{\rm hf,e}}}\,
\left(\frac{k_BT_d \hbar \omega_z}{\rho^2 g^2}\right)^{1/2}
=\frac{\sqrt{6}}{4} \frac{\Gamma}{\Delta_{\rm hf,e}} 
\frac{1}{(\rho a_{3D}^2)^{1/2}}.
\end{equation}
For $^{87}\textrm{Rb}$ with the choice $\rho \lambda_c^2=9$, we obtain the not very impressive result,
\be
P_{\rm fluo} \simeq 0.22\,.
\label{eq:proba_choice1}
\ee
For the sake of completeness, it is important to note that for this choice, $k_B T= 0.1 k_B T_d$ remains smaller than $\hbar\omega_z$, so that the Bose gas retains a two-dimensional character.

The existence of other atomic levels in addition to the ones strictly needed to create the gauge field is responsible not only for dissipative effects such as fluorescence, but also creates reactive effects such as a spatially and temporally-dependent light-shift of the non-coupled state $\ket{NC}$. 
Among the dominating processes, the parasitic $\Lambda'$ scheme creates a modulated light-shift potential 
\begin{equation}
U^{\Lambda'}(\rr)= \frac{\hbar\,|\Omega_p(\rr)|^2}{3\Delta_{{\rm hf,e}}}.
\eqname{U}
\end{equation}
A shift of the same order of magnitude arises from the coupling of the $|a\rangle$ 
state to the $\ket{F'=2\ \textrm{or}\ 1,m_{F'}=0}$ by the coupling beam.

An estimate of the energy deposited in the system by the $U^{\Lambda'}$ term as compared to the one $\Delta E_2$ due to the gauge field can be obtained 
with Bogoliubov theory: Using Eq.(\ref{eq:degen}) with $\mathcal{U}_0=\hbar (\Omega_p^{+}\Omega_p^{-*})(0^+)/(3\Delta_{{\rm hf,e}})$, $\eta=\gamma$ and $\QQ=\qq$, one gets
for $\hbar\gamma \ll \epsilon_q \ll \rho g$:
\begin{equation}
\frac{\Delta E_{U^{\Lambda'}}}{\Delta E_2}= \frac{|\Omega_c|^4}{9 f_n (k_c c_s)^2 \Delta_{{\rm hf,e}}^2}.
\end{equation}
For actual parameters, the energy change due to $\Delta U^{\Lambda'}$ turns out to be non-negligible.
For the optimal value of the Rabi frequency $\Omega_c^{\rm opt}$ from Eq.\eq{Omega_c_opt}, the ratio is 
\begin{equation}
\frac{\Delta E_{U^{\Lambda'}}}{\Delta E_2} =\frac{2\hbar \omega_z}{3f_n\,\rho g }.
\end{equation}
For $^{87}\textrm{Rb}$, one finds the discouraging result $\Delta E_{U^{\Lambda'}}/\Delta E_2\simeq 600\tilde{g}/(\rho \lambda_c^2 f_n)$, which remains much larger than unity even for $\rho\lambda_c^2=9$.

Even if the deposited energy by the spurious potential $U^{\Lambda'}$ is much larger than the desired one of the gauge field, 
a suitable extrapolation procedure may take advantage of the different dependence on the laser intensities to isolate the effect of the gauge potential. 
An alternative possibility is to exploit the fact that the $\Delta E_{U^{\Lambda'}}$ contribution does not depend on the direction of $\qq$: Within the regime of linear response, this contribution can therefore be eliminated by taking the difference of the energy changes for respectively longitudinal and transverse gauge fields.

Another possible nuisance is the existence of stimulated Raman processes that may out couple the non-coupled $|NC\rangle$ state to atomic ground state sublevels $|c\rangle$ other than $|a\rangle$ and $|b\rangle$, {\sl via} excited state sublevels other than $|e\rangle$
\cite{foot_raman_zero}.
One may however check that for the proposed scheme these leaky Raman couplings are detuned from resonance by a frequency amount at least $\Delta_{\rm hf,g}$ in absolute value and therefore harmless.

\subsubsection{Second choice}

Another possible choice for $^{87}$Rb atoms is to take a $\sigma_-$ polarization for the coupling beam propagating along the $y$ axis. The probe beam is again linearly polarized along $y$. The atomic levels forming the $\Lambda$ system are now $|a\rangle \equiv |F=2, m_F=-2\rangle$, $|b\rangle\equiv |F=1,m_F=-1\rangle$ and $|e\rangle \equiv |F'=2,m_{F'}=-2\rangle$. 
The strong two-body losses that are generally experienced by the upper hyperfine manifold of the ground state are here suppressed by the choice of a maximal $m_F$ state for $|a\rangle$: as collisions between ultra-cold atoms mostly occur in the $s$-wave scattering channel, conservation of the sum of the $m_F$'s then prevents transition to the lower hyperfine manifold.

The fluorescence rate per atom due to the motional coupling between $|NC\rangle$ and $|C\rangle$ is still given by $\Gamma_{{\rm fluo}}^{{\rm non-ad}}$ as defined in Eq.(\ref{eq:gam_non_adiab}). 
As there is no longer any parasitic $\Lambda'$ system, the fluorescence due to laser excitation
of $|a\rangle$ or $|b\rangle$ to excited state sublevels other than $|e\rangle$ is now dominated 
by the transitions $|b\rangle\rightarrow |F'=1\ \textrm{or}\ 2, m_{F'}=-1\rangle$ due to the
probe beam. Thanks to the reduced occupation probability $\simeq |\Omega_p|^2/|\Omega_c|^2$ of sublevel
$|b\rangle$ in the atomic state $|NC\rangle$ and to the larger hyperfine splitting $\Delta_{\rm hf, g}$ of the ground state, the fluorescence rate is strongly suppressed. After spatial averaging it amounts to
\be
\Gamma_{\rm fluo}^p = \frac{3}{2} \Gamma \frac{|\Omega_p^+|^4}{\Delta_{\rm hf,g}^2 |\Omega_c|^2}.
\ee
Other fluorescence processes on the D2 line (e.g. the transition $|a\rangle\rightarrow |J'=3/2,F'=3,m_{F'}=-3\rangle$ excited by the coupling beam) are several orders of magnitude weaker than $\Gamma_{\rm fluo}^p$ thanks to the huge fine structure splitting of $ 
2\pi\times 7$ THz.

Since the two terms in the sum $\Gamma_{\rm fluo}^{\rm non-ad}+\Gamma_{\rm fluo}^{p}$  experience different switch-off functions $e^{-\gamma t}$ and $e^{-2\gamma t}$, we integrate over time to calculate the total fluorescence probability: 
\be
P_{\rm fluo} = \frac{4\Gamma}{\gamma} \frac{|\Omega_p^+(0^+)|^2}{|\Omega_c|^4} \frac{\hbar k_c^2}{m} \omega_z
+ \frac{3\Gamma}{4\gamma} \frac{|\Omega_p^+(0^+)|^4}{|\Omega_c|^2 \Delta_{\rm hf,g}^2}.
\ee
After optimization over the coupling beam intensity, the minimal fluorescence probability
\be
P_{\rm fluo}^{\rm min}= \frac{2\sqrt{3}}{\gamma}  \frac{\Gamma}{\Delta_{\rm hf,g}} \left|\frac{\Omega_p^+(0^+)}{\Omega_c}\right|^3
\left(\frac{\hbar k_c^2}{m} \omega_z\right)^{1/2}
\ee
is obtained for a coupling beam Rabi frequency such that
\be
|\Omega_c^{\rm opt}|^4 =  \frac{16}{3} \frac{|\Omega_c|^2}{|\Omega_p^+|^2} \Delta_{\rm hf,g}^2 \frac{\hbar k_c^2}{m} \omega_z.
\eqname{Omega_c_opt2} 
\ee
Introducing the reduced quantities $\tilde{\gamma}=\hbar\gamma/(\rho g)$
and $\tilde{\epsilon}_{\rm gauge}=\epsilon_{\rm gauge}/(m k_B T_d)^{1/2}$, and eliminating $\omega_z$ in terms
of $g$ and $a_{3D}$, we finally obtain
\be
P_{\rm fluo}^{\rm min} = \left(\frac{9}{128\pi}\right)^{1/4} \frac{\tilde{\epsilon}_{\rm gauge}^{3/2}}{\tilde{\gamma}}
\frac{\Gamma}{\Delta_{\rm hf,g}} \frac{\lambda_c}{a_{3D}} (\rho\lambda_c^2)^{-1/4}.
\ee
For the parameters of Table~\ref{tab:val}, in particular $\rho \lambda_c^2=9$, one finds $|\Omega_c^{\rm opt}|^2/\Gamma^2\simeq 5.5$ so that a detuning $|\delta| > 5 \Gamma$ is required to remain in the weak saturation regime.
This resulting probability of spontaneous emission per atom in the deposited energy measurement is very small,
\be
P_{\rm fluo} \simeq 0.008.
\label{eq:Pfluo_min}
\ee
As compared to the first one, this second choice then provides a strong reduction of the spontaneous emission rate by a factor almost 30.

Another advantage of this second choice is that light-shift effects are potentially smaller 
thanks to the absence of the parasitic $\Lambda'$ scheme. 
The probe beam on the $|b\rangle\to |F'=1\ \textrm{or}\ 2, m_{F'}=-1\rangle$ transitions
produces a light-shift which, after average in the $|NC\rangle$ state, leads to the spurious potential
\be
U^p(\rr) = -\frac{\hbar |\Omega_p|^4}{4|\Omega_c|^2 \Delta_{\rm hf,g}}.
\ee
The amount of energy that is deposited in the gas by this spurious potential can be estimated using twice Eq.(\ref{eq:degen}), first with $\mathcal{U}_0=
-\hbar |\Omega_p^+(t=0^+)|^4/(|\Omega_c|^2 \Delta_{\rm hf,g})$, $\eta=2\gamma$, $\QQ=\qq$,  and second with $\mathcal{U}_0$ four times
smaller, $\eta=2\gamma$, $\QQ=2\qq$.
Neglecting $2\hbar \gamma$ with respect to $\epsilon_q$, and taking $q\xi=1$, we obtain the following estimate for the spurious deposited energy
\be
\Delta E_{U^p} = \frac{133}{160} \frac{N}{\rho g} 
\frac{\hbar^2 |\Omega_p^+(0^+)|^8}{|\Omega_c|^4 \Delta_{\rm hf,g}^2}.
\ee
For the  coupling beam Rabi frequency \eq{Omega_c_opt2} minimizing spontaneous emission, the ratio of the energies deposited by the gauge field and the spurious potential amounts to:
\be
\frac{\Delta E_{U^p}}{\Delta E_2} = \frac{133\sqrt{2}}{960\pi^{3/2}} \frac{\tilde{g}\,\tilde{\epsilon}_{\rm gauge}}{f_n} 
\frac{\lambda_c}{\rho^{1/2} a_{3D}^2}.
\ee
For $\tilde{g}=0.1$ and $\tilde{\epsilon}_{\rm gauge}=0.15$ and using the $^{87}$Rb parameters summarized in Table~\ref{tab:val}, one finds the still quite unfortunate result
\be
\frac{\Delta E_{U^p}}{\Delta E_2}\simeq \frac{4}{f_n} \gg 1. 
\ee
A possibility to overcome this difficulty and separate $\Delta E_2$ from $\Delta U^{p}$ is to use the same strategy proposed to separate $\Delta E_2$ from $\Delta E_1$ by exploiting the different dependence of the two quantities on the ratio $|\Omega_p^+/\Omega_p^-|(0^+)$.
This can be combined with the choice of a compromise value of $|\Omega_c|^2/\Gamma^2$ 
that allows to strongly suppress 
the light-shift potential without introducing a too large spontaneous emission rate.

To this purpose, we fix $\tilde{\epsilon}_{\rm gauge}=0.15$, $\gamma=0.2\rho g/\hbar$, $q\xi=1$, $\tilde{g}=0.1$ and we take as free parameters $X=\rho \lambda_c^2$ and $Y=|\Omega_c|^2/\Gamma^2$. 
Inserting the relevant parameters for $^{87}\textrm{Rb}$ as in Table~\ref{tab:val}, we obtain 
\bea
P_{\rm fluo} &=& \frac{6.86\cdot 10^{-2}}{X^{1/2}Y} + 7.46\cdot 10^{-4} Y, \\
\frac{\Delta E_{U^p}}{\Delta E_2}  &=& \frac{0.132 Y^2}{f_n}.
\eea
A reasonable compromise between the two competing effects is to choose $X=9$ and $Y=0.5$, which corresponds to
\be
\label{eq:compromise}
\rho \lambda_c^2 = 9 \ \ \ \mbox{and}\ \ \ \frac{|\Omega_c|^2}{\Gamma^2} = 0.5\quad .
\ee
As a result, for the same parameters $T/T_d=0.1$ and $f_n=0.2$ used in the classical field simulations of section \ref{sec:deposited}, 
we obtain the quite encouraging values
\bea
P_{\rm fluo} &\simeq& 0.045 \\
\frac{\Delta E_{U^p}}{\Delta E_2} &\simeq & 0.16\quad.
\eea
We have checked that at the resulting temperature $k_B T/\hbar \omega_z=0.62$ the Bose gas retains a two-dimensional character and that the validity of the gauge field model of Sec.\ref{sec:gauge} is guaranteed by the resulting probe beam Rabi frequency $|\Omega_p^+(0^+)|^2/|\Omega_c|^2=0.03 X^{1/2}=0.09\ll 1$.

To conclude, we remark that the energy deposited by the spurious light-shift is suppressed in the extrapolation procedure
proposed in subsection \ref{subsec:num_inv} to eliminate effects beyond the linear response regime to the gauge field, 
that is an extrapolation of the effective normal fraction that is linear in $\epsilon_{\rm gauge}^2$.
Since $\Delta E_{U^p}$ scales as $|\Omega_p^+(0^+)|^8$ with the probe beam Rabi frequency,
it varies as $\epsilon_{\rm gauge}^4$ with the amplitude of the gauge field, so it indeed contributes to the effective normal fraction as 
$\epsilon_{\rm gauge}^2$, which extrapolates to zero. This works of course for $\epsilon_{\rm gauge}$ small enough
for the mechanical effect of the spurious light-shift to be treatable in the linear response regime.

\section{Dum-Olshanii theory for many-body systems}
\label{app:adiab}

In a seminal work \cite{olshanii}, Dum and Olshanii have shown that an effective gauge field appears in the theoretical description of a three-level atom interacting with a laser field on a $\Lambda$ transition. Here we use the formalism of the Quantum Stochastic Differential Equations (see e.g.\S 8.3.2 of \cite{QuantumNoise}) to extend this idea to an interacting Bose gas in second quantized form. 

We start with the master equation for the density operator $\hat{\sigma}$ of the many-body system, assuming for simplicity that spontaneous emission corresponds to a net loss of atoms,
\begin{multline}
\frac{d}{dt}\hat{\sigma} = \frac{1}{i\hbar} [H,\hat{\sigma}]
\\
+\Gamma \int d^3r\, 
\left[\hat{\Psi}_e(\rr)  \hat{\sigma} \hat{\Psi}_e^\dagger(\rr)
-\frac{1}{2} \{ \hat{\Psi}_e^\dagger(\rr) \hat{\Psi}_e(\rr),\hat{\sigma}\}
\right],
\end{multline}
where $[,]$ ($\{,\}$) stands for the commutator (anticommutator).
The Hamiltonian $H$ is the sum of the single-particle kinetic and trapping terms, of the interaction terms (that we formally model as local Dirac-delta interactions), of the internal energy of the atomic excited state, and of the coupling terms of the atoms to the laser fields.

Since the loss rate $\hbar \Gamma$ greatly exceeds the kinetic, trapping and interaction
energies, we can neglect the external dynamics of the excited state and write
\begin{multline}
H \simeq \int d^3r \sum_{\alpha=a,b} 
\left[-\frac{\hbar^2}{2m} \hat{\Psi}_\alpha^\dagger
\Delta \hat{\Psi}_\alpha + U_{3D} \hat{\Psi}_\alpha^\dagger \hat{\Psi}_\alpha\right] 
\\
+ \int d^3r \left[\frac{g_{aa}}{2} \hat{\Psi}_a^{\dagger 2} \hat{\Psi}_a^2
+ \frac{g_{bb}}{2} \hat{\Psi}_b^{\dagger 2} \hat{\Psi}_b^2 +
g_{ab} \hat{\Psi}_a^{\dagger} \hat{\Psi}_b^{\dagger} \hat{\Psi}_b
\hat{\Psi}_a\right] \\
+ \int d^3r\,(-\hbar \delta) \hat{\Psi}_e^\dagger \hat{\Psi}_e \\
+ \int d^3r \left[\frac{\hbar \Omega_p}{2} \hat{\Psi}_e^\dagger 
\hat{\Psi}_a + \frac{\hbar \Omega_c}{2} \hat{\Psi}_e^\dagger \hat{\Psi}_b
+\mbox{h.c.}\right].
\end{multline}
As previously defined, $\delta$ is the common value of the detuning of the probe and coupling beams from the $|a\rangle \to |e\rangle$ and $|b\rangle \to |e\rangle$ transitions.

In a Heisenberg picture for the open atomic system, the ground state atomic field operators $\Psih_{\alpha=a,b}$ satisfy the usual evolution equations $i\hbar\partial_t \hat{\Psi}_\alpha=[\hat{\Psi}_\alpha,H]$. 
On the other hand, conservation of the canonical commutation relations of the fields and of the Hermitian conjugation relation between $\Psih_{\alpha}$ and $\Psihd_{\alpha}$ requires including a quantum Langevin term $\hat{F}_e$ in the evolution equation for the excited state field $\Psih_e$,
\be
\partial_t \hat{\Psi}_e= \frac{1}{i\hbar} [\hat{\Psi}_e, H] 
-\frac{1}{2} \Gamma \hat{\Psi}_e + \Gamma^{1/2} \hat{F}_e(\rr,t)~.
\label{eq:exact_psiepoint}
\ee
Here, the quantum noise term $\hat{F}_e$ is $\delta$-correlated in position and time, e.g.\ $[\hat{F}_e(\rr,t),\hat{F}_e^\dagger(\rr',t')]=\delta(\rr-\rr') \delta(t-t')$,
and we recall that the expectation value of normally-ordered products of noise operators vanish, e.g. $\langle \hat{F}^\dagger_e\,\hat{F}_e\rangle=0$,
since the bath does not provide an incoming flux of $e$ atoms. 

The only non-zero contributions to the commutator in Eq.(\ref{eq:exact_psiepoint}) originate from the excited state internal energy and from the atom-laser coupling term. This latter term can be expressed solely in terms of the atomic field operator $\hat{\chi}_{3D}$ in the coupled internal state $|C\rangle$, as defined in Eq.(\ref{eq:def_chi3D}).
Along the lines of~\cite{Gerbier_heating}, we formally integrate $[\partial_t +(-i\delta+\Gamma/2)] \hat{\Psi}_e = \hat{S}$ neglecting a transient of duration $1/\Gamma$ as 
\be
\hat{\Psi}_e(\rr,t) = \int_0^{+\infty} d\tau e^{-(-i\delta+\Gamma/2)\tau}
\hat{S}(\rr,t-\tau).
\ee
The Rabi frequencies $\Omega_{c,p}$ and the atomic field $\hat{\chi}_{3D}$
have a negligible variation during $1/\Gamma$ and may be replaced by their values at time $t$ in the integrand.
This leads to Eq.(\ref{eq:Psie}) of the main text, where the noise term is defined as 
$\hat{B}_e(\rr,t)= \int_0^{+\infty}
d\tau\, e^{-(-i\delta+\Gamma/2)\tau}\hat{F}_e(\rr,t-\tau).$

As explained in section \ref{sec:gauge}, we are in a regime where the atoms are mostly in the non-coupled state and the field
$\hat{\chi}_{3D}$ in the coupled state is small and a perturbation expansion in powers of $\hat{\chi}_{3D}$ can be performed. 
The gauge field formalism discussed in Sec.\ref{sec:gauge} for the evolution of the atomic field $\hat{\phi}_{3D}$ in the non-coupled state [defined in Eq.(\ref{eq:def_phi3D})] is already recovered at zeroth order in $\hat{\chi}_{3D}$. From this zeroth order approximation of $\hat{\phi}_{3D}$, it is then easy to obtain the first order contribution to the field $\hat{\chi}_{3D}$ that is required in Sec.\ref{sec:optical} to evaluate the optical polarization of the moving atoms.

From Eq.(\ref{eq:def_phi3D}) the equation of motion for $\hat{\phi}_{3D}$ is
\be
\partial_t \hat{\phi}_{3D} = \frac{1}{i\hbar} [\hat{\phi}_{3D},H] 
+\sum_{\alpha=a,b} \hat{\Psi}_\alpha \partial_t \langle NC|\alpha\rangle.
\label{eq:motphi}
\ee
By the very definition of non-coupled state, the excited state internal energy and the atom-laser coupling terms give an exactly vanishing contribution to the commutator. In the other terms of the Hamiltonian as well as in the last sum in Eq.(\ref{eq:motphi}), we can perform the approximation $\hat{\Psi}_\alpha\simeq \langle \alpha | NC\rangle  \hat{\phi}_{3D}$, which is accurate at zeroth order in $\hat{\chi}_{3D}$. After an integration by part and noting that $\langle NC|\partial_t|NC\rangle$ and $\langle NC| \nabla |NC\rangle$ are purely imaginary quantities, we find that  up to this order $\hat{\phi}_{3D}$
follows a purely Hamiltonian evolution governed by Eq.(\ref{eq:H3}).

The equation of motion of $\hat{\chi}_{3D}$ has the form
\be
\partial_t \hat{\chi}_{3D} = \frac{1}{i\hbar} [\hat{\chi}_{3D},H] 
+\sum_{\alpha=a,b} \hat{\Psi}_\alpha \partial_t \langle C|\alpha\rangle.
\label{eq:motchi}
\ee 
The commutator with the internal excited state energy term introduces a $\hat{\Psi}_e$ term, that we replace with Eq.(\ref{eq:Psie}): in this way, both a noise term and a complex, position dependent energy term $-(i\delta'+\Gamma'/2) \hat{\chi}_{3D}$ appear in the equation. 
The real quantities $\delta'$ and $\Gamma'$ are given by Eq.(\ref{eq:def_gamma}) and correspond to light-shift and damping effects, respectively.

Since $\hbar\Gamma'$ is much larger than the kinetic, trapping, interaction and
recoil energies of the atoms, we can neglect these latter terms in the evolution equation of the coupled state, 
and only keep the coupling to $\hat{\phi}_{3D}$.
This amounts to keeping in Eq.(\ref{eq:motchi}) only the contributions to the kinetic, trapping and interaction terms of the Hamiltonian $H$ that contain one single factor $\hat{\chi}_{3D}^\dagger$ and an arbitrary number of $\hat{\phi}_{3D}$ and $\hat{\phi}_{3D}^\dagger$ factors. 
In this way, we obtain
\begin{multline}
\partial_t \hat{\chi}_{3D} \simeq -(i\delta'+\frac{\Gamma'}{2}) \hat{\chi}_{3D} 
+ \hat{\phi}_{3D} \langle C| [-\partial_t+ \frac{i\hbar}{2m} \Delta]
|NC\rangle \\
 + \frac{i\hbar}{m} \nabla\hat{\phi}_{3D} \cdot
\langle C| \nabla |NC\rangle
+ \frac{1}{i\hbar} \mathcal{G} \hat{\phi}^\dagger_{3D} \hat{\phi}_{3D}^2 
+\Gamma^{'1/2} \hat{F}_\chi
\label{eq:dchidt}
\end{multline}
where we have introduced a complex position and time dependent coupling constant
\be
\mathcal{G} = \langle C|a\rangle \langle a|NC\rangle [(g_{aa}-g_{ab}) \langle NC|a\rangle \langle a | NC\rangle
-a\leftrightarrow b]
\ee
with the convention $g_{ba}=g_{ab}$. 
The noise term is defined by $\hat{F}_\chi = -i|\delta+i\Gamma/2| (\Omega_c^*/|\Omega_c|)\hat{B}_e$. Its correlation properties are determined by the commutation relation $[\hat{F}_\chi(\rr,t),\hat{F}_\chi^\dagger(\rr',t')] \simeq \frac{|\delta+i\Gamma/2|^2}{\Gamma} e^{i(t-t')\delta-\Gamma|t-t'|/2} \delta(\rr-\rr')$.
Since we are working in a low saturation regime in which $\Gamma\gg \Gamma'$, the time dependent factor in front of $\delta(\rr-\rr')$ may be replaced with a Dirac of $t-t'$, so that $\hat{F}_\chi$ is in practice a spatio-temporally delta-correlated noise.

The last step is to expand Eq.(\ref{eq:dchidt}) to first order
in $\Omega_p/\Omega_c$. We also limit ourselves to zeroth order in $q/k_p$ and in
$1/(k_p w)$, and we neglect the temporal derivative of the switch-off function $f(t)$. Then (i) for $g_{aa}\simeq g_{ab}$,
the fourth contribution in the right-hand side of Eq.(\ref{eq:dchidt})
vanishes~\cite{foot_si_petite_diff},
and (ii) for the magic choice Eq.(\ref{eq:magic_choice}),
the second contribution in the right-hand side of Eq.(\ref{eq:dchidt})
vanishes. 
With the same adiabatic elimination technique adopted for $\hat{\Psi}_e$ and taking into account the fact that $\delta'$ and $\Gamma'$ vary very slowly on the scale of $1/\Gamma'$, we are finally led to the final equation Eq.(\ref{eq:champ_chi}) with a noise term defined by
$\hat{B}_\chi(\rr,t) =  \int_0^{+\infty} d\tau e^{-[i\delta'+\Gamma'/2](\rr,t)\tau}
\hat{F}_\chi(\rr,t-\tau)$.

We complete the discussion by giving the back-action of the field $\hat{\chi}_{3D}$ on the field $\hat{\phi}_{3D}$,
a back-action that was already considered for a specific single atom geometry in \cite{aspect}.
The linear coupling of $\hat{\phi}_{3D}$ to $\hat{\chi}_{3D}$ originates from terms in the Hamiltonian that are linear in 
$\hat{\chi}_{3D}$, leading to
\begin{multline}
\left(\partial_t \hat{\phi}_{3D}\right)_{\rm back} = \hat{\chi}_{3D} \langle NC|[-\partial_t+ \frac{i\hbar}{2m} \Delta] |C\rangle
 \\ +\frac{i\hbar}{m} \nabla \hat{\chi}_{3D} \cdot \langle NC| \nabla |C\rangle
+\frac{1}{i\hbar} [\mathcal{G} \hat{\chi}_{3D}^\dagger \hat{\phi}_{3D}^2 + 2\mathcal{G}^* \hat{\phi}_{3D}^\dagger
\hat{\phi}_{3D} \hat{\chi}_{3D}].
\end{multline}
Expression of the back-action solely in terms of $\hat{\phi}_{3D}$ and noise operators is obtained by 
replacing $\hat{\chi}_{3D}$ in the resulting equations of motion with its adiabatic approximation derived from Eq.(\ref{eq:dchidt}).
This leads in general to a lengthy formula.
For simplicity, we give the result for $g_{aa}=g_{ab}$, to leading order in $\Omega_p/\Omega_c$,
we also neglect the contribution  
to $\langle C|\partial_t |NC\rangle$ of the time-derivative of the switch-off function $f(t)$, and we use the specific form 
$\Omega_p/\Omega_c$ considered in this paper, see Eqs.(\ref{eq:omegac_expli},\ref{eq:omega_p}), 
restricting to  zeroth order in $q/k_p$ and $1/(k_p w)$
so that
\begin{multline}
\left(\partial_t \hat{\phi}_{3D}\right)_{\rm back}\simeq
-\frac{i\hbar^2}{m^2} \frac{4(\delta+i\Gamma/2)}{|\Omega_c|^2}\,
\left|\frac{\Omega_p}{\Omega_c}\right|^2\times\\ \times [(\kk_p-\kk_c)\cdot\nabla]^2\,\hat{\phi}_{3D}+\textrm{noise terms}.
\end{multline}
After reduction to the $xy$ plane, the deterministic term gives rise to two corrections to the evolution
of $\hat{\phi}_{3D}$: (i) a complex position dependent energy shift, 
\be
\hbar(\delta''-i\Gamma''/2) = 
-2 (\delta+i\Gamma/2) \omega_z \frac{\hbar^2 k_c^2}{m} \frac{|\Omega_p|^2}{|\Omega_c|^4},
\ee
and (ii) a complex correction to the mass along $y$, $\delta m_y =8 \hbar k_c^2 |\Omega_p|^2(\delta + i \Gamma/2)/|\Omega_c|^4$. 
The quantity $\hbar\delta''$ is the light-shift potential experienced by the non-coupled bidimensional field.
The spatial average of the fluorescence rate $\Gamma''$ of the non-coupled field coincides with the $\Gamma_{\rm fluo}^{\rm non-ad}$ fluorescence rate 
previously discussed in \eq{gam_non_adiab}, as it should be.
For the parameters of Table~\ref{tab:val}, the reactive corrections $\delta''$ and $\delta m_y$ are small provided that the
detuning is not too large, $|\delta/\Gamma|< 5$. For instance, an estimate for the undesired 
energy deposited by the $\hbar\delta''$ 
potential can be obtained from the Bogoliubov theory for a homogeneous system, see Eq.(\ref{eq:degen}), leading to
\be
\frac{\Delta E_{\hbar\delta''}}{\Delta E_2} \approx \frac{5\cdot 10^{-5}}{f_n}  (\delta/\Gamma)^2
\ee
where $\Delta E_2$ is the desired deposited energy giving access to the normal fraction $f_n$.

\section{Derivation of the expression for the deposited energy}
\label{appendix:LDA}
We start from a two-dimensional system at thermal equilibrium with no average current and we apply a gauge field of the form
\be
\mathbf{A}(\rr,t) = f(t) \eee_y |c_+ e^{i\qq\cdot\rr/2} + c_- e^{-i\qq\cdot\rr/2} |^2
e^{-(\rr-\rr_0)^2/(2\sigma^2)},
\label{eq:pf}
\ee
where $\eee_\alpha$ is the unit vector along direction $\alpha$. The derivable dimensionless envelope function $f(t)$ is assumed to be zero for $t<0$ and to rapidly tend to zero for $t\to +\infty$. The time-independent coefficients $c_\pm$ have the dimension of the square root of a momentum.

We are interested in evaluating the energy change of the system from $t=0^-$ to $t=+\infty$ at the lowest order in $c_\pm$. We work in Schr\"odinger picture and we first use the exact relations:
\bea
\label{eq:dE_exact1}
\Delta E &\equiv&  \int_{-\infty}^{+\infty} dt \frac{d\langle H(t)\rangle}{dt}
\\ &=& \int_{-\infty}^{+\infty} dt \int d^2r \mathbf{A}(\rr,t)\cdot \frac{d}{dt}
\langle \jj(\rr)\rangle (t),
\label{eq:dE_exact2}
\eea
where the second equality comes from a time-dependent Hellmann-Feynman theorem and a temporal integration by parts. Calculating $\langle \jj(\rr)\rangle (t)$ by linear response theory gives 

\begin{multline}
\Delta E \simeq \int_{\mathbb{R}} \frac{d\omega}{2\pi} \int d^2 r \int d^2r'\, 
\omega \\ \mbox{Im} \left[\sum_{\alpha,\beta} \chi^{\rm ex}_{\alpha\beta}(\rr,\rr';\omega) 
A_\alpha(\rr,\omega)^*
A_\beta(\rr',\omega)\right],
\eqname{delta_E_A}
\end{multline}
where $\chi^{\rm ex}$ is the exact current susceptibility in real space taking
into account the spatial inhomogeneity of the trapped cloud. Note that,
contrarily to Eq.(\ref{eq:dE_exact1}), Eqs.(\ref{eq:dE_exact2}),(\ref{eq:delta_E_A}) still
hold when $f(t)=0$ for $t<0$ and has a discontinuous jump in $t=0$.

We now use the particular form (\ref{eq:pf}) for $\mathbf{A}$ and
consider the relevant limiting case $q\sigma \gg 1$,
$q\,\textrm{min}(\xi,\lambda)\ll 1$, where $\xi$ is the healing length
of the gas and $\lambda$ is the thermal de Broglie wavelength. 
We also assume that $\sigma$ is much smaller than the radius of the trapped cloud,
so that the density variation within a region of radius $\sigma$ around $\rr_0$ may be neglected.

Within a local density approximation, we then replace $\chi^{\rm ex}$
with the susceptibility $\chi$ of a spatially homogeneous system with a density
equal to the one of the trapped gas at position $\rr_0$ and with the
same temperature, $\chi^{\rm ex}(\rr,\rr';\omega)\simeq \chi(\rr-\rr';\omega)$.

This local density approximation leads to
\begin{multline}
\Delta E \simeq 
\int_{\mathbb{R}} \frac{d\omega}{2\pi}\omega\,
|f(\omega)|^2
\int \! d^2 R \int\! d^2 u\,
\Big[|c_+|^2 +|c_-|^2 \\ 
+c^*_+ c_- e^{-i\qq\cdot(\RR+\uu/2) }+c_+ c_-^* e^{i\qq\cdot(\RR+\uu/2)} \Big]
\\
\left[|c_+|^2 +|c_-|^2 + 
c^*_+ c_- e^{-i\qq\cdot(\RR-\uu/2) }+c_+ c_-^* e^{i\qq\cdot(\RR-\uu/2)}
\right] \\
\mbox{Im}[\chi_{yy}(\uu;\omega)]\,
e^{-|\RR-\rr_0|^2/\sigma^2}e^{-u^2/(4\sigma^2)}
\end{multline}
where we have performed the change of variables $\rr=\RR+\uu/2,\rr'=\RR-\uu/2$.
As we work in the $q\sigma \gg 1$ regime, we have for example
\begin{multline}
\int d^2R\, e^{-2i\qq\cdot \RR}\, e^{-|\RR-\rr_0|^2/\sigma^2}= \\ 
e^{-2i\qq\cdot \RR_0}\,
\pi \sigma^2\, e^{-q^2\sigma^2} \ll \pi \sigma^2
\end{multline}
so that all the oscillating terms in $\RR$ may be neglected.
Introducing the Fourier transform of $\chi_{yy}(\kk;\omega)$, which is an even function
of $\kk$ due to parity or rotational invariance, we obtain
\bea
\Delta  E &=& \Delta E_1 + \Delta E_2 \\
\Delta E_1 &\simeq &
\int_{-\infty}^{+\infty} \frac{d\omega}{2\pi} \omega |f(\omega)|^2
(2\pi\sigma^2)^2 \int \frac{d^2k}{(2\pi)^2}\nonumber \\ & & \mathrm{Im}\, \chi_{yy}(\kk;\omega)
\left(|c_+|^2+|c_-|^2\right )^2 e^{-k^2\sigma^2} \eqname{deltaE1_app} \\
\Delta E_2 &\simeq &
\int_{-\infty}^{+\infty} \frac{d\omega}{2\pi} \omega |f(\omega)|^2
(2\pi\sigma^2)^2 \int \frac{d^2k}{(2\pi)^2}\nonumber \\
 & & \mathrm{Im}\, \chi_{yy}(\qq+\kk;\omega)
2|c_+|^2 |c_-|^2 e^{-k^2\sigma^2 }.\eqname{deltaE2_app}
\eea
The second contribution $\Delta E_2$ comes from the spatially modulated gauge
field at $\qq$, while the first contribution $\Delta E_1$ is due
to the non-modulated term which follows the broad Gaussian envelope.
The expression \eq{deltaE2} in the main text is obtained from
\eq{deltaE2_app} by noting that the integration over $\kk$ is effectively limited by the Gaussian factor to a small region of radius $1/\sigma$ in which one is allowed to
neglect the $\kk$-dependence of the susceptibility.

Naively, one could guess that a necessary condition for the accuracy of
our local density approximation is that the switch-off time of the
gauge field $\gamma^{-1}$ is short as compared to the characteristic
time $2R/v$ for the induced mechanical perturbation to cross the whole cloud,
to be reflected by its boundaries and to turn back to the excitation zone where
it can interfere with the excitation process,
$v$ being the fastest between the sound and thermal speeds in the
cloud of radius $R$.

This condition is actually sufficient, but not necessary within linear
response theory. We now show for $f(t)=\Theta(t) e^{-\gamma t}$, as in Eq.(\ref{eq:choix_ft}),
that the $\gamma\rightarrow 0$ limit for
the deposited energy scheme exists and coincides with the perturbation
induced by the gauge field in the thermodynamic equilibrium state.
As one can show by inserting the explicit form of the temporal Fourier
transform of the gauge field into \eq{delta_E_A} and performing the integral over
$\omega$, the deposited energy can be written in the form
\begin{multline}
\Delta E\simeq \frac{1}{2}
\int d^2r\,d^2r'\,\sum_{\alpha,\beta}
A_\alpha(\rr,t=0^+)\\ \,A_\beta(\rr',t=0^+)\,
\textrm{Re}
\Big[\chi^{\rm ex}_{\alpha\beta}(\rr,\rr';\omega=i\gamma) 
\Big]
\label{eq:encore_une_forme}
\end{multline}
where we have introduced the Kubo formula for the exact current-current susceptibility
\be
\chi^{\rm ex}_{\alpha\beta}(\mathbf{r},\mathbf{r}';\omega)=
\sum_{\lambda,\lambda'} (\pi_\lambda-\pi_{\lambda'})
\frac{\langle \lambda|j_\alpha(\mathbf{r}) |\lambda'\rangle 
\langle
\lambda'|j_\beta(\mathbf{r}')|\lambda\rangle}{E_{\lambda'}-E_\lambda-\hbar\omega
-i0^+} 
\eqname{chi_ex}
\ee
in terms of the thermal equilibrium population $\pi_\lambda$ of quantum
state $\lambda$. This quantity can be simply related to the
susceptibility at thermodynamic equilibrium,
\begin{multline}
\chi_{\alpha\beta}^{\rm th}(\rr,\rr')=\lim_{\gamma\to 0}
\textrm{Re}[\chi^{\rm ex}_{\alpha\beta}(\rr,\rr';\omega=i\gamma)]\,+ \\
+\frac{1}{k_BT}\sum_{\lambda,\lambda'; E_\lambda=E_{\lambda'}}
\pi_\lambda\,\langle \lambda|j_\alpha(\mathbf{r}) |\lambda'\rangle 
\langle
\lambda'|j_\beta(\mathbf{r}')|\lambda\rangle.
\eqname{thermo_chi}
\end{multline}
We recall that the thermodynamic susceptibility relates the mean current
in a thermal equilibrium state at temperature $T$ to the applied (weak)
static gauge field {\sl via} $\langle \jj \rangle = \chi^{\rm th}*\AAA$,
where $*$ is the spatial convolution product.
In the present case of an interacting gas, one can safely expect that
the second line in \eq{thermo_chi} gives a negligible contribution as
there is no systematic degeneracy and the current operator $\jj$ has no
diagonal matrix elements since the eigenstate wavefunctions may be taken real.
Within linear response theory, we thus obtain
\be
\Delta E \underset{\gamma\to 0}{\to} \frac{1}{2} \int d^2r d^2r'\sum_{\alpha,\beta} 
\chi_{\alpha\beta}^{\rm th}(\rr,\rr') A_\alpha(\rr,0^+)A_\beta(\rr',0^+).
\label{eq:deltae_lim_zero}
\ee
Since local density approximation in the thermodynamic
equilibrium state is a standard procedure, we expect that it can be applied
to evaluate the right-hand side of Eq.(\ref{eq:deltae_lim_zero})
\cite{foot_dF}. 
As a consequence,
within linear response theory, our deposited
energy method (that measures the left-hand side of Eq.(\ref{eq:deltae_lim_zero}))
should be accurately described by the local density approximation down to the $\gamma\to 0$ limit.

\section{Some results of linear response theory and the Bogoliubov
 expression of the deposited energy} 

\label{app:bogo}

A system of time independent Hamiltonian $H_0$ experiences,
at times $t>0$, a time dependent weak perturbation of Hamiltonian $-\epsilon f(t) \mathcal{V}$, 
where $\epsilon\to 0$, the dimensionless time dependent factor $f(t)$ is zero for $t<0$ and tends rapidly
to zero for $t\to +\infty$, and the operator $\mathcal{V}$ is time independent.
At time $t=+\infty$, the system is free again, with a mean energy modified by the perturbation.
The question is to calculate the mean energy change to second order in $\epsilon$.

Suppose first that, at $t=0^-$, the system is prepared in the eigenstate $|\lambda\rangle$ of $H_0$
of eigenenergy $E_\lambda$. The energy change $\delta E$ between time $0$ and time $+\infty$ is
\be
\delta E = \lim_{t\to +\infty} \langle \psi(t)| (H_0-E_\lambda)|\psi(t)\rangle
\ee
where $|\psi(t)\rangle$ is the system state vector at time $t$.
The usual time dependent perturbation theory gives an expansion in powers of $\epsilon$:
\be
|\psi(t)\rangle = |\psi_0(t)\rangle + \epsilon |\psi_1(t)\rangle + \epsilon^2 |\psi_2(t)\rangle + \ldots
\ee
where $|\psi_0(t)\rangle= \exp(-iE_\lambda t/\hbar)|\lambda\rangle$, 
\be
|\psi_1(t)\rangle = -\int_0^t \frac{d\tau}{i\hbar} e^{-iH_0(t-\tau)/\hbar} f(\tau) \mathcal{V} e^{-iE_\lambda \tau/\hbar}
|\lambda\rangle,
\eqname{psi1}
\ee
and the expression of higher order contributions is not needed. Using $(H_0-E_\lambda)|\psi_0(t)\rangle=0$
and $\langle \psi_0(t) | (H_0-E_\lambda)=0$, we find to second order in $\epsilon$:
\be
\delta E \simeq  \lim_{t\to +\infty} \epsilon^2\, \langle \psi_1(t) | (H_0-E_\lambda)|\psi_1(t)\rangle.
\ee
In this paper, $f(t)=\Theta(t) \exp(-\gamma t)$, with $\gamma > 0$ and $\Theta(t)$ is 
the Heaviside step function,
see Eq.(\ref{eq:choix_ft}).
Also, the system is prepared initially in a statistical mixture of eigenstates of $H_0$ with a probability distribution
$\pi_\lambda$. 
After explicit integration of \eq{psi1} over $\tau$ and then average over $|\lambda\rangle$, the expression for the signal to be detected experimentally is
\begin{multline}
\label{eq:signal}
\mbox{Signal}\,(\mathcal{V}) \equiv \lim_{\epsilon\to 0} \frac{\langle \delta
 E\rangle}{\epsilon^2} = \\
\frac{1}{2}\,\textrm{Re}
\sum_{\lambda,\lambda'} 
\frac{(\pi_\lambda-\pi_{\lambda'})}{E_{\lambda'}-E_{\lambda}-i\hbar\gamma}\,
|\langle \lambda'|\mathcal{V}|\lambda\rangle|^2.
\end{multline}
The sum may be restricted to $E_\lambda\neq E_{\lambda'}$ since the contributions with $E_\lambda=E_{\lambda'}$
are zero. This also shows that the signal has a finite limit for $\gamma\to 0^+$.
Note that in a thermal equilibrium state $\pi_\lambda=Z^{-1}\exp(-E_\lambda/k_B T)$, the signal is necessarily
positive.

The calculation of the noise on the experimental signal can be performed along the same lines.
One defines $\delta E^2\equiv\lim_{t\to+\infty} \langle\psi(t)|(H_0-E_\lambda)^2|\psi(t)\rangle$ 
with the initial state vector $|\psi(0)\rangle = |\lambda\rangle$, and one finds after average
over the initial state:
\begin{multline}
[\mbox{Noise}\,(\mathcal{V})]^2 \equiv \lim_{\epsilon\to 0}
 \frac{\langle \delta E^2\rangle}{\epsilon^2} \underset{\gamma\to 0}{=}
\sum_{\lambda,\lambda',E_\lambda\neq E_\lambda'} \pi_\lambda |\langle \lambda |\mathcal{V} |
 \lambda'\rangle|^2 \simeq \\ 
\sum_\lambda \pi_\lambda [\langle \lambda |\mathcal{V}^2 |
 \lambda\rangle- \langle \lambda | \mathcal{V} | \lambda \rangle^2],
\label{eq:defNoise}
\end{multline}
where the approximate equality is based on the assumption that there are no systematic degeneracies in the many-body spectrum.
The $\epsilon^2$ scaling of the variance in Eq.(\ref{eq:defNoise}) shows that the typical value of the energy change at the end of the excitation sequence is of order $\epsilon$. This scaling is to be contrasted with the $\epsilon^2$ one of the expectation value that is suggested by Eq.(\ref{eq:signal}).

We now apply the general formula Eq.(\ref{eq:signal}) to the Bogoliubov analysis of subsection \ref{subsec:bogo}. In this case, 
$\epsilon=\epsilon_{\rm gauge}/2$ and $\mathcal{V}=\mathcal{V}_q + \mathcal{V}_{-q}$ with 
\be
\mathcal{V}_q = \int_{[0,L]^2}  d^2r \, e^{iqx} j_y(\rr) = 
\sum_\kk \frac{\hbar k_y}{m} a^\dagger_{\kk+\qq} a_{\kk}
\ee
where $a_{\kk}$ is the annihilation operator of a particle of the gas of wavevector $\kk$,
and we have set $\qq=q\,\mathbf{e}_x$.
In a translationally invariant system, the eigenstates $|\lambda\rangle$ can be taken of well defined total momentum; as the action of $\mathcal{V}_{\pm q}$ changes this total momentum by $\pm \hbar \qq$, the two operators $\mathcal{V}_{q}$ and $\mathcal{V}_{-q}$ cannot interfere in the signal and thus $\mbox{Signal}\,(\mathcal{V}) = 2\, \mbox{Signal}\,(\mathcal{V}_q)$.
In terms of the annihilation operators $b_\kk$ of Bogoliubov quasiparticles, we can split
\be
\mathcal{V}_q = \mathcal{V}_q^{(0)} + \mathcal{V}_q^{(2)} + \mathcal{V}_q^{(-2)}
\label{eq:splitV}
\ee
in terms of
\bea
\label{eq:vq0}
\mathcal{V}_q^{(0)} = 
\sum_{\kk \neq \mathbf{0}, -\mathbf{q}} 
\frac{\hbar k_y}{m} (U_\kk U_{\kk+\qq}-V_\kk V_{\kk+\qq})\, b_{\kk+\qq}^\dagger b_{\kk}
\\
\label{eq:vq2}
\mathcal{V}_q^{(2)} \!=  \!\!\!\!\!\!\!\sum'_{\kk \neq \mathbf{0}, -\mathbf{q}} 
\frac{\hbar k_y}{m} (V_\kk U_{\kk+\qq} - U_\kk V_{\kk+\qq})\, b_{\kk+\qq}^\dagger b_{-\kk}^\dagger
\\
\label{eq:vq-2}
\mathcal{V}_q^{(-2)}\! =  \!\!\!\!\!\!\!\sum'_{\kk \neq \mathbf{0}, -\mathbf{q}} 
\frac{\hbar k_y}{m} (U_\kk V_{\kk+\qq} - V_\kk U_{\kk+\qq})\,  b_{\kk} b_{-(\kk+\qq)}.
\eea
The primed sum $\sum'$ indicates restriction of the sum over wavevectors such that $k_y>0$. In Bogoliubov theory, the eigenstates $|\lambda\rangle$ may be taken in the form of Fock states of quasiparticles. Since $\mathcal{V}_q^{(n)}$ changes the total number of quasiparticles by the amount $n$, the terms in the right hand side of Eq.(\ref{eq:splitV}) cannot interfere
in the signal and
\be
\mbox{Signal}\,(\mathcal{V}_q) = 
\mbox{Signal}\,(\mathcal{V}_q^{(0)}) + 
\mbox{Signal}\,(\mathcal{V}_q^{(2)}) + 
\mbox{Signal}\,(\mathcal{V}_q^{(-2)}).
\ee
Thanks to the clever writing of $\mathcal{V}_q^{(n)}$ with the constraint $k_y>0$ \cite{lemma}, 
there are no interferences in Eqs.(\ref{eq:vq0},\ref{eq:vq2},\ref{eq:vq-2}) between the different terms of the sums over $\kk$. As a result, the whole signal is the sum over the contribution of the different $\kk$'s. The last trick is to express the ratios $\pi_{\lambda'}/\pi_\lambda$ in terms of the mean occupation numbers $n_\kk$ of the Bogoliubov modes of energy $\epsilon_\kk$ and to make use of the identity
\be
e^{\beta\epsilon_\kk}=\frac{n_\kk+1}{n_\kk}
\label{eq:muoti}
\ee
satisfied by the Bose law.
A little bit of rewriting taking advantage of the relation $\mbox{Signal}(\mathcal{V})=(N/m) f_n^{\rm eff}$ and of the remarks in \cite{lemma}, finally leads to Eq.(\ref{eq:f_n_eff}).

An alternative procedure is to calculate the current-current susceptibility
\eq{chi_ex} for a spatially homogeneous system
within the Bogoliubov theory, which for $\qq\perp \mathbf{e}_y$ gives
in dimension $d$:
\bea
\chi_{yy}(\qq;\omega) = \frac{1}{L^d} \sum_{\kk\neq 0,-\qq}
\frac{\hbar^2 k_y^2}{m^2} \nonumber \\
\Big[  \frac{n_\kk - n_{\kk+\qq}}{\epsilon_{\kk+\qq}-\epsilon_\kk-\hbar\omega -i0^+}
\left(U_{\kk+\qq}^2 U_\kk^2-U_{\kk+\qq}V_{\kk+\qq} U_\kk V_\kk\right) \nonumber\\
-
\frac{1+n_\kk + n_{\kk+\qq}}{-\epsilon_{\kk+\qq}-\epsilon_\kk-\hbar\omega -i0^+}
\left(V_{\kk+\qq}^2 U_\kk^2-U_{\kk+\qq}V_{\kk+\qq} U_\kk V_\kk\right) \nonumber\\
+
\frac{1+n_\kk + n_{\kk+\qq}}{\epsilon_{\kk+\qq}+\epsilon_\kk-\hbar\omega -i0^+}
\left(U_{\kk+\qq}^2 V_\kk^2-U_{\kk+\qq}V_{\kk+\qq} U_\kk V_\kk\right) \nonumber\\
+
\frac{n_{\kk+\qq}- n_\kk}{\epsilon_\kk-\epsilon_{\kk+\qq}-\hbar\omega -i0^+}
\left(V_{\kk+\qq}^2 V_\kk^2-U_{\kk+\qq}V_{\kk+\qq} U_\kk V_\kk\right)
\Big].\nonumber \\
\eea
From Eq.(\ref{eq:delta_E_A}) one then recovers expression Eq.(\ref{eq:f_n_eff}) for the effective normal fraction $f_n^{\rm eff}$.

Another application of Eq.(\ref{eq:signal}) is to calculate the energy deposited by the external potential 
$\mathcal{U}(\rr,t)=(\mathcal{U}_0 e^{i\QQ\cdot\rr}+\mbox{c.c})\Theta(t)
e^{-\eta t}$ in the spatially homogeneous case, for $\eta>0$.
This is useful for Appendix \ref{app:experimental} and Appendix \ref{app:adiab} to estimate the effect of undesired light-shifts.
In second quantized form, and to leading order in Bogoliubov theory, one then has
$\epsilon \mathcal{V} = N^{1/2} (U_Q + V_Q) [\mathcal{U}_0 (b^\dagger_{\QQ}+b_{-\QQ})+\mbox{h.c.}]$. These terms do not interfere in Eq.(\ref{eq:signal}).
For non-zero temperature, using (\ref{eq:muoti}), we then obtain a temperature independent result
\be
\Delta E_{\mathcal{U}} \simeq 2 N |\mathcal{U}_0|^2 (U_Q+V_Q)^2\, \mbox{Re}\, \frac{1}{\epsilon_Q-i\hbar\eta}.
\label{eq:degen}
\ee
Remarkably, this also allows to calculate the energy change $\Delta E_g$ due to the switch-on-and-off of a spatially 
modulated coupling constant, $\delta g(\rr,t)=(\delta g_0 e^{i\qq\cdot\rr}+\mbox{c.c.})\Theta(t) e^{-\gamma t}$.
For a spatially homogeneous system, to leading order of Bogoliubov theory, the relevant operator
is  $\epsilon\mathcal{V}=N^{1/2}(U_q+V_q)[\rho\delta g_0 (b^\dagger_{\qq}+b_{-\qq})+\mbox{h.c.}]$, so that one can formally apply 
Eq.(\ref{eq:degen}) with $\mathcal{U}_0=\rho \delta g_0$.
This can be applied to the variation of the coupling constant due to $g_{aa}\neq g_{ab}$
in Eq.(\ref{eq:g}). In this case $\delta g_0 = 2g [(g_{ab}-g_{aa})/g_{aa}] (\Omega_p^+\Omega_p^{-*})(0^+)/|\Omega_c|^2$ so that
for $q\xi=1$ and $\gamma\to 0$,
\be
\frac{\Delta E_{g}}{\Delta E_2} \simeq \frac{16}{5 f_n} \frac{m\rho g}{(\hbar k_c)^2}  \left(\frac{g_{ab}-g_{aa}}{g_{aa}}\right)^2.
\ee
For the values of Table~\ref{tab:val} and $|g_{ab}-g_{aa}|\lesssim 0.1 |g_{aa}|$ as expected for $^{87}$Rb, this gives 
$\Delta E_{g}/\Delta E_2\approx 7\cdot 10^{-4}/f_n$ which is negligible.

\section{The noise on the deposited energy}
\label{app:noise}

In the section \ref{subsec:num_inv}, while presenting the numerical results on the deposited energy measurement, 
we mentioned that the statistical noise on the deposited energy was larger for smaller values of the 
gauge field amplitude $\epsilon_{\rm gauge}$.

To understand this feature, it can be useful to rewrite the deposited energy for a single realization of the 
classical field simulation in the form
\begin{multline}
\delta E=-\int
d^2r\,\jj(\rr,0)\cdot\AAA(\rr,0^+) \\ 
-\int_{0^+}^{+\infty}\!dt \int d^2r\,\jj(\rr,t)\,\partial_t 
\AAA(\rr,t).
\eqname{deltaE_fl}
\end{multline}
The first term comes from the abrupt switch-on of the gauge field. For each realization, it is of order $\epsilon_{\rm gauge}$ but averages to zero in the limit of an infinite number of realizations of the experiment as $\langle \jj(\rr,0) \rangle=0$. In any actual calculation, an average over a finite number $n_{\rm real}$ of realizations is taken, which gives a non-zero random value 
for $\delta E$ scaling as $\epsilon_{\rm gauge}/\sqrt{n_{\rm real}}$. 

The relevant signal $\langle \delta E\rangle$ is given by the second term in (\ref{eq:deltaE_fl}),
obtained from the classical Hamiltonian identity
$dH/dt=\partial_t H$. For small values of the gauge field switch-off rate $\gamma$, this term is of order $O(\epsilon_{\rm gauge}^2$) as in this limit $\jj$ adiabatically follows the thermal equilibrium value for the instantaneous value of the gauge field. As a result, the number of realizations that are needed to extract the signal out of the statistical noise due to the first term grows as $|\epsilon_{\rm gauge}|^{-2}$, which perfectly explains the numerical observation.

Furthermore, it has been demonstrated by a number of recent cold atom experiments that noise is not always just an hindrance but can be also a source of useful physical information~\cite{noise_exp,noise_th}. 
As a simple example, we consider here the amplitude of the noise on the energy that is deposited in the system at each realization of the experiment. This quantity is quantified by the average $\langle \delta E^2 \rangle$ of the square of the deposited energy in the $\gamma\to 0$ limit. Looking at \eq{deltaE_fl}, it is immediate to see that in the small $\epsilon_{\rm gauge}$ limit the dominant contribution comes from the square of the first term, which suggests that the noise on the deposited energy is related to the variance of the instantaneous fluctuations of the current operator. From the fluctuation-dissipation theorem, this quantity can then be related to the normal fraction of the gas.

This idea can be put on solid grounds by developing a full quantum calculation. The linear response theory calculation performed along the lines of Appendix \ref{app:bogo} leads to the expression in Heisenberg picture 
\begin{multline}
\lim_{\gamma\to 0}\langle
 [H_0(+\infty)-H_0(0)]^2\rangle  \simeq \\
\langle
 [\int\!d^2r\,\jj(\rr)\cdot\AAA(\rr,0^+)]^2 \rangle
\eqname{fullquantumdeltaE}
\end{multline}
where $H_0$ is the unperturbed Hamiltonian, that is the Hamiltonian without coupling to the gauge field.
This relation connects the variance of the quantum equivalent of the deposited energy to the instantaneous fluctuations of the current operator and confirms our expectation based on the classical field model. Moreover, combined with the fluctuation-dissipation theorem, it can be the starting point for another proposal to measure $f_n$.

For a generic hermitian operator $V$ with vanishing diagonal matrix elements in the eigenbasis of $H_0$,
the fluctuation-dissipation theorem of linear response theory relates the imaginary part of the susceptibility $\chi$ to the Fourier transform of the correlation function
$S_{VV}(t)=\langle V(t)\,V(0) \rangle$,
\begin{equation}
\textrm{Im}[\chi_{VV}(\omega)]=\frac{1}{2\hbar}S_{VV}(\omega)\left[1-e^{-\hbar\omega/k_BT}\right].
\end{equation}
The Fourier transform $S_{VV}(\omega)$ of the correlation function in the thermodynamic equilibrium state is defined as usual as
\begin{equation}
S_{VV}(\omega)=\int_{-\infty}^{\infty}\!dt\,e^{i\omega t}\,\langle V(t)\,V(0) \rangle.
\end{equation}
Under the assumption that most of the spectral weight of the $V$ operator lies in the low-energy region $\hbar \omega\ll k_B T$, we can approximate $1-e^{-\hbar\omega/k_BT}\simeq \hbar \omega/k_B T$. This is a quite standard approximation of many-body theory and is generally accurate in the small $q$
limit~\cite{SSLP-book}. After a few manipulations, it leads to the general expression
\begin{multline}
S_{VV}(t=0)=\!\int \!\frac{d\omega}{2\pi}\,S_{VV}(\omega)
 \simeq  
\!\int \!\frac{d\omega}{2\pi}\,\frac{2k_B
 T}{\omega}\,\textrm{Im}[\chi_{VV}(\omega)] \\ = 
k_BT\,
\textrm{Re}[\chi_{VV}(\omega=0)],
\eqname{correl}
\end{multline}
where the equivalent of Eq.(\ref{eq:KK}) was used to obtain the last identity. An application of the fluctuation-dissipation relation \eq{correl} to the susceptibility and the fluctuations of the mass current in liquid He can be found in~\cite{dalfovo}.

The link between the variance of the deposited energy and the normal fraction is immediately obtained by applying \eq{correl} to the 
specific operator 
\begin{equation}
V=-\int d^2r\,\jj(\rr)\cdot\AAA(\rr,0^+)
\end{equation}
and isolating the contribution of the spatially modulated gauge field proportional to $\Omega_p^+ \Omega_p^-$. 
In this way, using Eq.(\ref{eq:encore_une_forme}), one is led to the final expression
\begin{multline}
\frac{\pi w^2}{4}\,
 \frac{\rho}{m}\,\left(\frac{\epsilon_{\rm gauge}}{2}\right)^2 f_n\simeq \\
\simeq  \frac{1}{2k_B T}\,\lim_{\gamma\to 0}\langle 
 [H_0(+\infty)-H_0(0)]^2\rangle,
\eqname{encore_une_autre_forme}
\end{multline}
which demonstrates an alternative way of extracting the value of the normal fraction $f_n$ from a measurement of the statistical variance of the deposited energy in a series of experiments.

It is however crucial to note that a measurement of $f_n$ based on the relation \eq{encore_une_autre_forme} requires taking expectation values of the Hamiltonian operator at different times. This may be experimentally challenging as it requires either a non-destructive measurement of the initial energy of the system at $t=0$ before switching on the gauge field, or a very precise {\em a priori} knowledge of its value in a sort of microcanonical ensemble~\cite{foot_microcan}.


\begin{thebibliography}{99}
\bibitem{foot_theq} 
Experimentally, thermal equilibrium in the rotating frame is not always granted. Most likely, it was not achieved in the experiments of \cite{vortex_lattice_expt} where the vortex lattices did not nucleate at the expected Landau critical rotation frequency, but rather at a higher rotation frequency when a dynamical instability of the condensate was triggered~\cite{Sinha}. This picture was confirmed by numerical simulations \cite{vortex_lattice_simul}.
\bibitem{foot_phys_cur} 
Note that the current operator $\jj(\rr)$ has to be distinguished from the physical current 
 $\jj_{\rm phys}=\jj-\AAA \hat{\phi}^\dagger \hat{\phi}/m$ corresponding to mass transport.
\bibitem{foot_3dgeom_beam} 
The probe beams Gaussian profiles are at angles $O(q/k_p)$ with respect to the $z=0$ plane. 
However, the resulting $z$-dependence of $\Omega_p^\pm$ can be safely neglected as it occurs on a length-scale $O(w\,k_p/q)$ much larger than the 
spatial extension $a_{\rm ho}^z$ of the atomic cloud along $z$.
\bibitem{foot_firstq} 
Note that the $-\int \jj_{3D}\cdot \AAA_{3D}$ gauge Hamiltonian is equivalent to the first-quantized form $-\sum_{j=1}^{N} [\pp_j \cdot \AAA_{3D}(\rr_j) +\AAA_{3D}(\rr_j)\cdot \pp_j]/2m$ 
where $N$ is the particle number. Differently from~\cite{olshanii}, our scalar potential $W_{3D}$ includes the $A^2_{3D}$ term.
\bibitem{foot_order_q2}
For the choice (\ref{eq:magic_choice}), to zeroth order in $q/k_p$ and $1/(k_p w)$, $W$ actually vanishes
to all order in $\Omega_p/\Omega_c$. To order $(q/k_p)^2$, 
it does not vanish and scales as $-\hbar^2 q^2 |\Omega_p/\Omega_c|^2/(4 m)$, which
gives for the parameters of Table~\ref{tab:val} a negligible contribution $\Delta E_W/\Delta E_2\approx 10^{-3}/f_n$ to the deposited energy
scheme.
\bibitem{foot_sptpdep} 
The spatio-temporal dependence of $g_{3D}(\rr,t)= g_{aa}+\Delta g(\rr,t)\simeq g_{aa} + 2(g_{ab}-g_{aa})\,|\Omega_p(\rr,t)/\Omega_c|^2$ is responsible for spurious excitation of the gas {\sl via} two different effects that can be captured by a Bogoliubov treatment. (i) An effective external potential appears of the form $\Delta V_g(\rr,t)=\rho_{3D}\,\Delta g(\rr,t)$: its spatially modulated part at $\pm \qq$ is able to excite phonons. (ii) The spatial modulation of the interaction constant 
that is switched on and off is able to excite pairs of phonons by a parametric down-conversion effect~\cite{DCE}. For suitable atomic species such that $g_{aa}\simeq g_{ab}$ (in particular Rb atoms), the spurious deposited energy turns out to be smaller than the quantity $\Delta E_2$ to be measured (see Appendix \ref{app:bogo} for the analysis of the 
second contribution).
\bibitem{apropos_alpha}
In the configurations shown in Fig.\ref{fig:Rb_levels}, only the case $\alpha=\pi/2$ is realized. 
To obtain $\alpha=0$ with the first choice of level scheme, one may e.g. choose a sublevel $|b\rangle=|F=2,m_F=-1\rangle$ and 
take the coupling beam propagating along $x$ with a linear polarization along $y$.
\bibitem{foot1} The apparent agreement of this value $f_n\simeq 0.22$
	obtained from a quantum Bogoliubov calculation with the result
	of the classical field simulation of Fig.\ref{fig:numeric} for
	$T/T_d=0.1$ is accidental. A Bogoliubov calculation for the
	classical field would give a $40\%$ deviation from this value.
\bibitem{foot_dis_lap}
The discrete Laplacian $\Delta$ admits the plane waves $e^{i\kk\cdot \rr}$ as
eigenmodes on the lattice with eigenvalue $\hbar^2 k^2 /(2m)$. In practice
a Fast Fourier Transform is used to evaluate the action of the Laplacian on the field $\Psi$.
\bibitem{foot_real_time}
For the real time evolution,
the kinetic, external potential, and interaction energy terms are treated with the usual splitting technique, each corresponding
evolution operator being then diagonal either in position space or in Fourier space. 
The evolution operator under the gauge potential \eq{A_y} during the time-step $dt$ involves numerical evaluation of objects of the form
\[
\Psi'=e^{-i dt \,[p_y-h(x,y,t)]^2 }\,\Psi,
\]
where we can assume without loss of generality that the function $h(x,y,t)$ has a zero spatial mean over $y$, $\int_0^L dy\, h(x,y,t)=0, \forall x,\forall t$.
This requires a bit more care as the evolution operator is diagonal neither in the real-space nor in the $\kk$-space,
nor in any mixed representation (real-space along some direction, Fourier-space along
the other).
In a first stage, we performed a local gauge transform on $\Psi$ of the form
\[
U\Psi=e^{-i H(x,y,t)/\hbar} \Psi
\]
with $\partial_y H(x,y,t)=h(x,y,t)$. In this way, the operator that one has to exponentiate to obtain the evolution operator
is  simply proportional to $p_y^2$. Since $h(x,y,t)$ has a zero spatial mean along $y$ axis, 
$H(x,0,t)=H(x,L,t), \forall x,\forall t,$ so that the gauge transform is compatible with the periodic boundary conditions.
In a second stage, we realized however that this gauge transform trick is not exact for our lattice model,
since $p_y$ is not simply $(\hbar/i)\partial_y$ (contrarily to continuous space). As a consequence, relation
(\ref{eq:deltaE_fl}) was not well satisfied numerically. Since the non-zero-spatial-mean version 
$h(x,y,t)$ is factorized in a function $f(x,t)$ of $x$ and
$t$ times $\exp[-y^2/(2\sigma^2)]$, with $w=2\sigma$, we simply diagonalized numerically once the operator
$W=p_y \exp[-y^2/(2\sigma^2)]+\mbox{h.c.}$ (in practice, a $64\times 64$ hermitian matrix for Fig.\ref{fig:numeric2})
and evaluated the evolution operator $\exp[i dt f(x,t) W]$ in the corresponding
eigenbasis (where it is diagonal).
\bibitem{foot_ideal_gas}
Above the critical temperature, there is no superfluid phononic branch, 
and one may expect that the condition $q\xi \lesssim 1$ for $f_n^{\rm eff}$ to reach
the zero-$q$ limit is replaced by the condition that $q \lambda \lesssim 1$, where $\lambda$ is the thermal de Broglie
wavelength, which is less stringent in the high temperature limit.
To test this intuitive idea,
Bogoliubov theory is not applicable, but it is tempting to use the ideal gas model as a first guess.
Then for $q\lambda \lesssim 1$ and $T\ll T_d$, one obtains
for the ideal quantum gas in the thermodynamic limit
\[
\lim_{\gamma\to 0} (f_n^{\rm eff}-f_n) \simeq \frac{2T}{T_d} \left[1-\frac{\mbox{atanh}\, X}{X}\right]
\]
with $X=\epsilon/(4+\epsilon^2)^{1/2}$, 
$\epsilon^2=\hbar^2 q^2/(2m |\mu|)$
and the chemical potential $\mu\simeq - k_B T \exp(-T_d/T)$. The condition that this is much smaller than $f_n$, which
is here equal to unity, thus does not only involve a condition on $q\lambda$,  it also involves 
a comparison of $\hbar^2 q^2/2m$ to the chemical potential, which is much more stringent for $T\ll T_d$. 
This could be expected from the fact that,
for the occupation numbers of the low momenta modes, the Bose-Einstein formula reduces to the Lorentzian shape
$n_k\simeq k_B T/(|\mu|+\hbar^2 k^2/2m)$, which has a narrow momentum width $\propto |\mu|^{1/2}$. 
As an example, to have $f_n^{\rm eff}=0.9$ for the ideal gas in the zero $\gamma$ limit, one has to take
$q \lambda\simeq 0.48$ for $T/T_d=0.18$, and $q\lambda \simeq 0.08$ for $T/T_d=0.1$
(these values of $q$ correspond to $q\xi=0.64$ and $q\xi=0.08$ for the interacting
gas with $g=0.1 \hbar^2/m$, respectively).
\bibitem{foot_evolu}
To avoid the rapid increase of the error bars on $f_n$ for $\epsilon_{\rm gauge}\to 0$,
due to the effect discussed in Appendix~\ref{app:noise}, we performed, for each stochastic field realisation of the thermal ensemble,
two independent temporal evolutions,
one in presence of the gauge field $\AAA$ and the other in presence of the opposite gauge field $-\AAA$.
We then took the average of the two corresponding deposited energies. This in particular exactly eliminates all the noisy terms
in the deposited energy that are linear in the gauge field.
\bibitem{foot_mesurer}
Measuring the gas energy is in general reduced to measuring the gas spatial density profile,
before time of flight (to measure the trapping energy) and after time of flight (to measure the initial kinetic 
plus interaction energy, which is the expansion energy).
These density measurements have reached a precision giving the absolute
equation of state of the gas, the pressure for  example, within five 
percent \cite{Salomon_eos_unit,Dalibard_eos_2D}
or even at the percent level \cite{Zwierlein_eos},
but not yet at the sub-percent level.
Our scheme however requires a measurement of an energy change, not an absolute measurement of the energy,
giving hope for a better precision. 
Recent experiments have actually measured the change of the momentum \cite{Hannaford1}
and of the energy \cite{Hannaford2} of a gas to determine its structure factor with good precision,
not for a gauge field excitation as proposed here, but for a Bragg laser excitation.
\bibitem{foot_finite_tau}
We have checked by performing simulations (for $q\xi=1$, $\gamma/(c_sq)=0.4$ and $\epsilon_{\rm gauge}/(m k_B T_d)^{1/2}=0.02$)
with a duration $\tau=6/\gamma$ that the truncation of the time evolution
at $\tau=3/\gamma$ as done in Fig.\ref{fig:numeric2} introduces an error on the normal fraction which is within
the statistical error bars.
\bibitem{precision}
Eqs.(\ref{eq:EQQz},\ref{eq:scatt},\ref{eq:scatt3},\ref{eq:relat_scatt_int},\ref{eq:relat_scatt_int2}) have been obtained from the operatorial
form of (\ref{eq:transm_var}), where the atomic field operators have already been reduced to 2D.
This is an approximation.
The exact mean intensity of the light field involves the expectation value of products of $\hat{\phi}_{3D}$ and
$\hat{\phi}_{3D}^\dagger$ that are not normally ordered, and one has to put the atomic field operators
in normal order before the 2D reduction. In the operatorial form of (\ref{eq:transm_var}), this amounts to replacing
$\kk_c\cdot[\mathbf{j}+\frac{\hbar}{2im}\nabla n]$ with $(\kk_c-\kk_p)\cdot\mathbf{J}$, where $\mathbf{J}(x,y)=\int_{-\infty}^{+\infty} dz \frac{\hbar}{im} \hat{\phi}_{3D}^\dagger(x,y,z) 
\nabla \hat{\phi}_{3D}(x,y,z)$. Then one finds additive corrections to the above mentioned equations. For example,
for $k_p\simeq k_c$,
in Eq.(\ref{eq:relat_scatt_int}) one has to add to the expression $
\frac{k_B T \rho(\rr_0)}{m} [\cos^2\phi_{\rm sc}+f_n\sin^2\phi_{\rm sc}]$ the correction
$\hbar\omega_z \rho(\rr_0)/(2m)$. This correction, corresponding to spontaneous emission due to the motional
coupling between $|NC\rangle$ and $|C\rangle$ along $z$, is a flat function of $\phi_{\rm sc}$, 
that can thus be distinguished experimentally from the relevant $f_n\sin^2\phi_{\rm sc}$ component.
\bibitem{foot_petitQ}
As a consequence of the continuity equation, the fluctuations of the density and of the longitudinal current are 
related by $\omega n_\qq = \qq\cdot \jj_\qq$. For small wavevectors, fluctuations are mostly sound-like with $\omega\simeq c_s\,q$. 
Inserting this result into the continuity equation, one finds that $\hbar q n_\qq/2m \ll j_{L,\qq}$ if $q\xi/2\ll 1$.
\bibitem{foot_un_un}
Note that, for the second choice of laser configuration to come, atoms with a $J=1\to J'=1$ transition and a large fine structure
are quite favorable, because the spurious light-shift $U^p$ is strongly suppressed. Also, since $|a\rangle=|J=1,m=-1\rangle$ and $|b\rangle=|J=1,m=0\rangle$,
one finds $g_{aa}=g_{ab}$ from rotational symmetry arguments \cite{Hoetal,Cargese}. A potential problem to take care of is the partial
inelastic nature of the $b-b$ collisions.
\bibitem{foot_effet_ft}
We have checked that the contribution to $\langle C|-\partial_t|NC\rangle$ of the switch-off function $f(t)$ in $\Omega_p$ 
(that was not included in \eq{champ_chi}) gives a negligible correction to the estimate of the fluorescence rate.
\bibitem{foot_si_petite_diff}
At the lowest order in $\Omega_p/\Omega_c$, the effect of a non-zero $g_{aa}-g_{ab}$ 
for the measurement scheme of section \ref{subsec:eftl} is to introduce a 
smooth polarization profile $\mathcal{P}_p=[4|d_{ae}|^2/(\hbar^2|\Omega_c|^2)]\,(g_{aa}-g_{ab})\,
\langle \phi_{3D}^{\dagger 2}\phi_{3D}^2\rangle \,\mathcal{E}_p$. 
As the corresponding phase shift does not depend on $\Omega_p/\Omega_c$, it can be isolated in the experiment. 
The contribution of higher order terms $O[|\Omega_p/\Omega_c|^3]$ have a non-trivial spatial structure and 
may interfere with the signal to be measured. 
Their relative value as compared to $\Delta\phi_2$ is however a factor 
roughly $[(g_{aa}-g_{ab})/g_{aa}]\,\tilde{g}\,(\rho\lambda_c^2)/(2\pi^2f_n)$ weaker.
\bibitem{foot_dF}
We have checked that the right-hand side of Eq.(\ref{eq:deltae_lim_zero}) is simply $-\Delta F$ 
(within linear response theory), where $\Delta F$ is the free-energy  of
the system at thermal equilibrium in presence of the static gauge
field $\AAA(\rr,0^+)$ minus the free energy in the absence of gauge field, for fixed temperature and atom number.
This establishes a formal link between our proposal and the one of \cite{Ho}.
\bibitem{lemma} 
We used the fact that, for a generic function $F$ satisfying $F[-(\kk+\qq)]=F(\kk)$ for all $\kk$, one has $\sum_{\kk\neq  \mathbf{0}, -\mathbf{q}; k_y>0} F(\kk) =\sum_{\kk\neq  \mathbf{0}, -\mathbf{q}; k_y<0} F(\kk).$
\bibitem{foot_microcan}
A sufficient condition is that the uncertainty in the initial total energy is $\ll \hbar \gamma$.
\end{thebibliography}
\end{document}